\documentclass[a4paper,11pt]{article}
\pdfoutput=1 

\usepackage{jheppub}
\usepackage{amsmath,amssymb,amscd}
\usepackage{bbm}
\usepackage{graphicx,xcolor}
\usepackage{url}
\usepackage[export]{adjustbox}
\usepackage{accents}
\usepackage[]{algorithm2e}
\usepackage{verbatim}

\graphicspath{ {fig/}{../fig} }

\makeatletter
\def\hlinewd#1{%
  \noalign{\ifnum0=`}\fi\hrule \@height #1 \futurelet
   \reserved@a\@xhline}
\makeatother

\makeatletter
\renewcommand\@fpheader{}
\renewcommand\@journal{}
\makeatother

\newcommand\thickbar[1]{\accentset{\rule{.4em}{.8pt}}{#1}}

\definecolor{darkgreen}{rgb}{0.,.3,0}
\definecolor{darkblue}{rgb}{0.0,0.0,0.5}

\newcommand{\Reduze}{\texttt{Reduze\;2}}
\newcommand{\ud}{\mathrm{d}}

\newcommand{\hide}[1]{}

\title{
{Multiple polylogarithms with algebraic arguments
and the two-loop EW-QCD Drell-Yan master integrals
}}
\preprint{MITP/19-043, MSUHEP-19-012}

\author[a]{Matthias Heller,}
\author[b]{Andreas von Manteuffel,}
\author[\,b]{and Robert M. Schabinger}

\affiliation[a]{Institut f\"ur Kernphysik und PRISMA$^+$ Cluster of Excellence, Johannes-Gutenberg Universit\"{a}t,\\
55099 Mainz, Deutschland}
\affiliation[b]{Department of Physics and Astronomy, Michigan State University, \\
East Lansing, Michigan 48824, USA}

\emailAdd{maheller@students.uni-mainz.de}
\emailAdd{vmante@msu.edu}
\emailAdd{schabing@msu.edu}

\abstract{
We consider Feynman integrals with algebraic leading singularities
and total differentials in $\epsilon\, \ud\ln$ form.
We show for the first time that it is possible to evaluate integrals with singularities involving unrationalizable roots in terms of conventional multiple polylogarithms, by either parametric integration or matching the symbol. As our main application, we evaluate the two-loop master integrals relevant to the $\alpha \alpha_s$ corrections to Drell-Yan lepton pair production at hadron colliders. We optimize our functional basis to allow for fast and stable numerical evaluations in the physical region of phase space.
}

\begin{document}
\maketitle

\allowdisplaybreaks[4]

\section{Introduction} 
\label{sec:intro}

The Drell-Yan process~\cite{Drell:1970wh} is one of the most important and basic processes measured at the Large Hadron Collider at CERN. It is used for precision measurements of the $W^{\pm}$ mass, the $Z$ mass, and the weak mixing angle, as well as for new physics searches. In order to interpret the increasingly precise data, higher-order corrections must be included in the theory predictions. In fixed-order perturbation theory, the pure Quantum Chromodynamic (QCD) corrections to the cross section are known through to next-to-next-to-leading order \cite{Hamberg:1990np,Harlander:2002wh,Anastasiou:2003ds,Melnikov:2006di}, the pure Quantum Electrodynamics (QED) corrections to neutral gauge boson production and decay are known at next-to-next-to-leading order \cite{Berends:1987ab,Blumlein:2019srk}, whereas the exact electroweak (EW) corrections are known only at next-to-leading order \cite{Baur:2001ze,Dittmaier:2001ay,Baur:2004ig}. Since these corrections turn out to be significant, the inclusion of exact mixed EW-QCD corrections is well-motivated; at the present time, results are known only for the  mixed Quantum Electrodynamic (QED)-QCD corrections \cite{Kilgore:2011pa} and in the pole approximation for the full EW-QCD corrections \cite{Dittmaier:2014qza,Dittmaier:2015rxo}.

\begin{figure}[b]
\centering
\begin{align}
\includegraphics[scale=0.5]{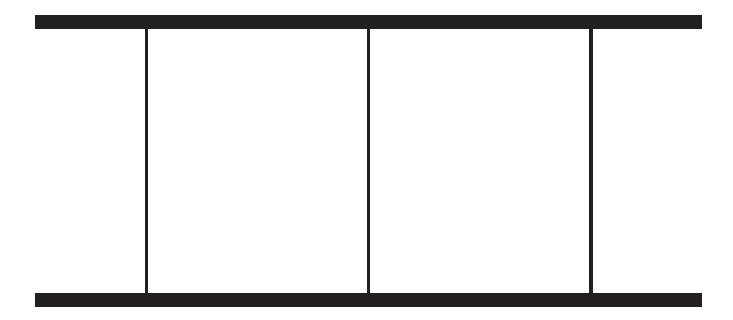} \qquad \qquad \includegraphics[scale=0.5]{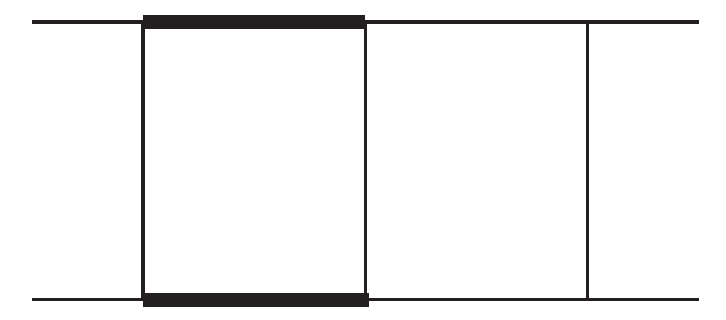}\nonumber
\end{align}
\caption{Planar top-level topology for the two-loop QED corrections to Bhabha electron-positron scattering
(left) and the most complicated top-level topology for the mixed EW-QCD two-loop  corrections to Drell-Yan lepton production (right).
Thick lines denote massive propagators and massive legs, thin lines are used in the massless case.
Both of the integral topologies depicted above admit $\epsilon\,\ud \ln$ differential equations.
}
\label{fig:toplevel}
\end{figure}

\begin{figure}[t]
\centering
\begin{align}
\includegraphics[valign = m, raise = .1 cm, height = 1 cm,keepaspectratio]{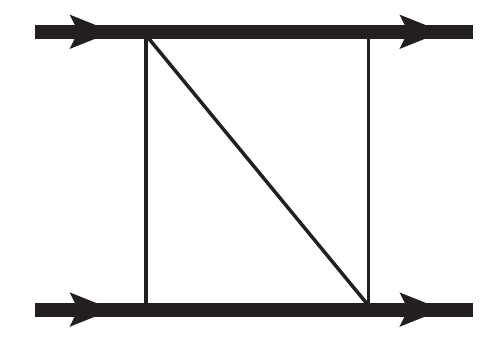}\big(s,t,m^2\big) \qquad \qquad \includegraphics[valign = m, raise = .05 cm, height = 1 cm,keepaspectratio]{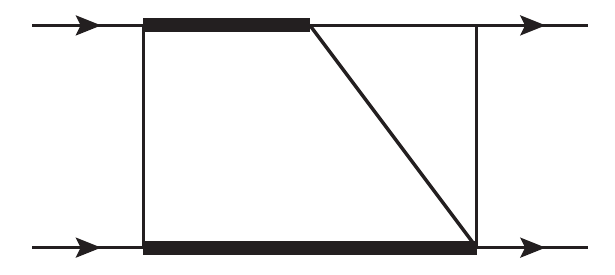}\big(s,t,m^2\big)\nonumber
\end{align}
\caption{Feynman integrals relevant to the two-loop QED corrections to Bhabha electron-positron scattering (left) and the two-loop mixed EW-QCD corrections to Drell-Yan lepton production (right),
which have problematic root-valued leading singularities.
}
\label{fig:algLS}
\end{figure}

The two-loop master integrals relevant to the virtual part of the mixed EW-QCD corrections to the Drell-Yan process were studied in \cite{Bonciani:2016ypc,vonManteuffel:2017myy} and all of them
admit $\epsilon$-decoupled and, subsequently, $\epsilon\, \ud \ln$ differential equations \cite{Henn:2013pwa}. All of the double-box integrals are actually planar, and for most of them, no intrinsically non-rational prefactors appear in the  $\epsilon$-decoupled differential equations written with respect to suitable kinematic variables.
It has come to light, however, that the two most complicated two-loop integral topologies (right panels of Figures \ref{fig:toplevel} and \ref{fig:algLS}) involve three square roots at once which can {\it not} be rationalized simultaneously~\cite{mainz,Besier:2019hqd}.
Here and in the following, we will refer to a set of roots as \emph{unrationalizable} if no locally invertible rational variable transformation exists, which turns all of them into rational functions at the same time.
The presence of non-rational {\it symbol letters} \cite{Goncharov:2010jf} in the $\epsilon\, \ud \ln$ differential equations makes the standard approach to the integration of the differential equations in terms of multiple polylogarithms impossible.

For most Feynman integrals considered in the literature so far,
kinematic square roots could be rationalized by a suitable
choice of variables, {\it e.g.}\ by using a Landau variable
for Feynman integrals with a two-mass threshold \cite{Barbieri:1972as}, by using momentum twistor variables \cite{Hodges:2009hk} for multi-leg Feynman integrals \cite{Gehrmann:2015bfy,Bourjaily:2018aeq,Caron-Huot:2018dsv}, or by using diophantine equations to construct suitable variables \cite{JohannesAmplitudes2015talk}.
However, in several processes studied in the literature, no simultaneous rationalization of the square roots which appear could be found.
In addition to the Drell-Yan process mentioned above, this is an issue for a subset of the planar two-loop master integrals for Bhabha electron-positron scattering \cite{Henn:2013woa,Festi:2019} and for a subset of the planar two-loop master integrals for the next-to-leading order QCD corrections to Higgs plus jet production with full heavy top quark mass dependence \cite{Bonciani:2016qxi}. 

So far, master integrals satisfying $\epsilon\, \ud \ln$ differential equations which involve non-rational symbol letters have typically been treated as generic Chen iterated integrals \cite{Chen:1977oja}, with a focus on the Euclidean region of phase space. Given our experience from other processes \cite{Bonciani:2013ywa,Gehrmann:2015ora}, one may expect significantly faster and also more stable numerical evaluations for results directly expressed in terms of suitably chosen multiple polylogarithms.
A constructive algorithm to obtain such a solution in the presence of root-valued symbol letters was used in \cite{vonManteuffel:2017hms} to calculate Feynman integrals.
However, the latter case is special since the symbol alphabet is univariate and simple, and a rational parametrization does in fact exist.
In other words, it was unclear until now whether it should generically be possible to find linear combinations of multiple polylogarithms which solve $\epsilon\, \ud \ln$ differential equations with symbol letters involving unrationalizable algebraic functions.

In this article, we show for the first time that it is indeed possible to integrate Feynman integrals of current phenomenological interest with unrationalizable roots in their symbol letters in terms of multiple polylogarithms, focusing  primarily on the most complicated two-loop mixed EW-QCD Drell-Yan master integrals.
Our method allows us to derive results written in terms of multiple polylogarithms well-suited for the physical region of phase space.
The two-loop mixed EW-QCD Drell-Yan master integrals with two massive internal lines exhibit a rich structure of thresholds and pseudo-thresholds, which means that one must think carefully about the analytic structure of the involved functions.
We discuss in detail a procedure to systematically filter the functions in a given region of phase space, such that no explicit Feynman $+i\,0$ prescriptions are required for the kinematic variables and spurious singularities of individual multiple polylogarithms at pseudo-thresholds of the Feynman integrals are avoided.

The outline of the article is as follows. In Section \ref{sec:linred} we discuss the linear reducibility of examples which are well-known in the literature to frustrate the standard machinery of Feynman integral calculus. 
Specifically, we prove that the integral topologies of Figure \ref{fig:algLS}, a five-line master integral for Bhabha scattering and a six-line master integral for Drell-Yan production, are linearly reducible for the first time. Linear reducibility is a technical criterion which is of interest here because linearly reducible Feynman integrals are guaranteed to be integrable in terms of multiple polylogarithms to all orders in $\epsilon$.
In Section \ref{sec:dydefs}, we define a normal form basis with $\epsilon\,\ud \ln$ differential equations for the two-loop master integrals for the mixed EW-QCD corrections to Drell-Yan production with two massive internal lines (see the right panel of Figure \ref{fig:toplevel}) and discuss a partial rationalization of the roots appearing in our integral basis definition.
In Section \ref{sec:diffeqint}, we show how to integrate the differential equations directly in terms of multiple polylogarithms even in the presence of root-valued symbol letters. In Section \ref{sec:analycontandi0}, we review the analytic continuation of multiple polylogarithms and outline our procedure to filter multiple polylogarithms with undesirable analytic properties out of our ans\"{a}tze. In Section \ref{sec:DYresult}, we present results for the most complicated two-loop mixed EW-QCD Drell-Yan master integrals. In particular, we highlight the notable analytic features of our solution for the six-line integral from the right panel of Figure \ref{fig:algLS}.
We conclude in Section \ref{sec:outlook} and, for clarity, we give the set of complete $\epsilon\, \ud \ln$ differential equations for the two-loop master integrals considered in this paper in Appendix \ref{sec:dydeq}.
In Appendix \ref{sec:multiroot} we give an example for the construction of algebraic letters in the presence of multiple root-valued leading singularities.

\section{Linear reducibility for algebraic symbol letters}
\label{sec:linred}

In this section, we discuss the direct integration of a five-line master integral for the two-loop QED corrections to massive Bhabha scattering (left panel of Figure \ref{fig:algLS}). This particular five-line integral is of special interest because its symbol letters are not simultaneously rationalizable \cite{Festi:2019}. It has been known for quite some time that the planar master integrals for the two-loop QED corrections to  Bhabha scattering (see the left panel of Figure \ref{fig:toplevel}) satisfy $\epsilon$-decoupled differential equations \cite{Henn:2013woa}, but, even at leading order in the $\epsilon$ expansion, it is not at all clear that integral \eqref{eq:FPBalgLS} below may be expressed as a linear combination of standard
multiple polylogarithms~\cite{LDanilevsky,Goncharov:1998kja,Borwein:1999js}.
In fact, it was suggested in \cite{ClaudeAmplitudestalk} that elliptic multiple polylogarithms \cite{BrownLevin} might actually be required for such cases.  It is therefore of some importance to demonstrate that one can actually integrate it to all orders in $\epsilon$ in terms of Goncharov or multiple polylogarithms using a {\tt HyperInt}-like approach \cite{Brown:2008um,Brown:2009ta,Panzer:2014caa,Bogner:2014mha,Panzer:2015ida,vonManteuffel:2013uoa,Gehrmann:2013cxs,vonManteuffel:2013vja,Gehrmann:2014bfa,Duhr:2019tlz,Becchetti:2019tjy}.

 In fact, this may be readily achieved by studying the polynomials which appear in the polynomial reduction \cite{Brown:2008um,Brown:2009ta,Panzer:2015ida} of the Symanzik polynomials at intermediate stages. The idea is to make changes of variables which factorize totally quadratic polynomials to allow for further integrations without producing square root-valued functions of the remaining Feynman parameters. Ultimately, the goal is to find a complete sequence of integration variables or, in the language introduced in \cite{Brown:2009ta} and refined in \cite{Panzer:2014caa}, to show that \eqref{eq:FPBalgLS} is a linearly reducible Feynman integral in the chosen variables. Practically speaking, a Feynman integral is  linearly reducible if it can be evaluated in terms of Goncharov polylogarithms, starting from the Feynman parametric representation but allowing for arbitrary variable changes along the way. In this case, the Feynman parametric representation is
\begin{equation}
\label{eq:FPBalgLS}
     \includegraphics[valign = m, raise = .4 cm, height = 1 cm,keepaspectratio]{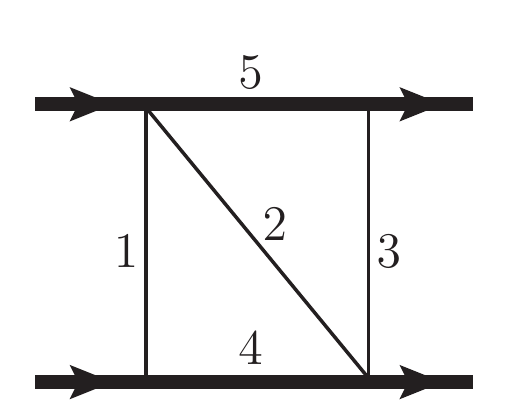} \left(s,t,m^2\right) = -e^{2 \gamma_E \epsilon}\Gamma(1+2\epsilon)\Bigg[ \prod_{i=1}^5 \int_0^{\infty} {\rm d} \alpha_i \Bigg] 
	\delta(1-\alpha_5)\,\mathcal{U}^{-1+3\epsilon}\mathcal{F}^{-1-2\epsilon},
\end{equation}
where
\begin{align}
    \mathcal{U} &= \alpha_1 \alpha_2+\alpha_1 \alpha_3+\alpha_1 \alpha_5+\alpha_2 \alpha_3+\alpha_2 \alpha_4+\alpha_2 \alpha_5+\alpha_3 \alpha_4+\alpha_4 \alpha_5\,,
\\
    \mathcal{F} &= -s\, \alpha_2 \alpha_4 \alpha_5-t\, \alpha_1 \alpha_2 \alpha_3 + m^2 \left(\alpha_4+\alpha_5\right)\mathcal{U}
    \\&
     -m^2 \left(\alpha_1 \alpha_2 \alpha_4+\alpha_1 \alpha_2 \alpha_5+\alpha_1 \alpha_3 \alpha_4+\alpha_1 \alpha_3 \alpha_5+\alpha_1 \alpha_4 \alpha_5+\alpha_2 \alpha_3 \alpha_4+\alpha_2 \alpha_3 \alpha_5+\alpha_3 \alpha_4 \alpha_5\right)\nonumber 
\end{align}
are the integral's Symanzik polynomials. We implement the delta function constraint at the very beginning by setting $\alpha_5 = 1$ inside $\mathcal{U}$ and $\mathcal{F}$. Our normalization for the integral is chosen to facilitate comparison with {\tt FIESTA 4} \cite{Smirnov:2015mct}.

At the outset, the polynomial reduction of $\mathcal{U}$ and $\mathcal{F}$ indicates that just one integration can safely be performed, with respect to either $\alpha_1$ or $\alpha_3$. The irreducible polynomials in the remaining Feynman parameters which could potentially appear in the result after integrating out $\alpha_1$ are \cite{Panzer:2014caa}:
\begin{align}
\label{eq:polyredalp1}
   &{\rm L}_{\alpha_1} = \Big\{\vphantom{\left(1+\alpha_2+\alpha _3\right)^2}1+\alpha _2+\alpha _3,m^2-t\,\alpha _2\, \alpha _3,\alpha _2 \left(1+\alpha _3\right)+\left(1+\alpha _2+\alpha _3\right) \alpha_4,
   \\ &
   m^2 \alpha _2+ \left(m^2+\left(2\, m^2-s\right)\alpha _2\right)\alpha _4+m^2\left(1+\alpha _2+\alpha _3\right) \alpha _4^2,
    \nonumber \\ &
    \alpha_2^2 \left(m^2+ t \,\alpha _3\left(1+\alpha_3\right)\right)+ \left(2\, m^2-s+ t\,\alpha _3\right)\left(1+\alpha _2+\alpha _3\right)\alpha _2\, \alpha _4 + m^2 \left(1+\alpha_2+\alpha _3\right)^2 \alpha _4^2\Big\}.\nonumber
\end{align}

The final totally quadratic polynomial in ${\rm L}_{\alpha_1}$ is conveniently dealt with by using the first non-trivial variable change from the discussion of the period $P_{7, 11}$ in \cite{Panzer:2015ida}. It is readily apparent from the form of the polynomial that two powers of $\alpha_2$ factorize from it once the variable change 
\begin{equation}
    \alpha_4 = \frac{x_4 \,\alpha_2}{1+\alpha_2+\alpha_3}
\end{equation}
is applied:
\begin{align}
\label{eq:alp2hardpoly3}
    &\alpha_2^2 \left(m^2+ t \,\alpha _3\left(1+\alpha_3\right)\right)+ \left(2\, m^2-s+ t\,\alpha _3\right)\left(1+\alpha _2+\alpha _3\right)\alpha _2\, \alpha _4 + m^2 \left(1+\alpha_2+\alpha _3\right)^2 \alpha _4^2 
    \nonumber \\ 
    &= \alpha_2^2 \left(m^2+t\, \alpha_3 \left(1+\alpha_3\right) + \left(2\, m^2-s+ t\,\alpha _3\right) x_4 + m^2 x_4^2\right).
\end{align}
Quite remarkably, the other polynomials in ${\rm L}_{\alpha_1}$ which depend on $\alpha_4$ still give rise exclusively to irreducible polynomials linear in $\alpha_2$ after changing variables:
\begin{align}
\label{eq:alp2hardpoly1}
  \alpha _2 \left(1+\alpha _3\right)+\left(1+\alpha _2+\alpha _3\right) \alpha_4 &= \alpha_2 \left(1+\alpha_3+x_4\right),
    \\
    \label{eq:alp2hardpoly2}
     m^2 \alpha _2+ \left(m^2+\left(2\, m^2-s\right)\alpha _2\right)\alpha _4+m^2\left(1+\alpha _2+\alpha _3\right) \alpha _4^2 &=
    \\ 
 \frac{\alpha_2}{1+\alpha_2+\alpha_3}\Big(m^2\left(1+\alpha_2+\alpha_3\right) + \big(  m^2+ \big(2\, m^2-& s\big)  \alpha _2\big)x_4+m^2\alpha_2\, x_4^2\Big).\nonumber
\end{align}
From Eqs. \eqref{eq:alp2hardpoly3}, \eqref{eq:alp2hardpoly1}, and \eqref{eq:alp2hardpoly2}, it is clear that we can now safely integrate out $\alpha_2$. 

Rerunning the polynomial reduction with {\tt HyperInt} after changing variables shows that the irreducible polynomials
\begin{align}
\label{eq:polyredalp1alp2}
    &{\rm L}_{\alpha_1 \alpha_2} = \Big\{1+\alpha _3,1+\alpha _3+x_4,m^2+t\,\alpha _3+t\,\alpha _3^2,m^2+\left(2\, m^2-s\right) x_4+m^2 x_4^2,
    \\&
    m^2\left(1+x_4\right)-s+\left(m^2(2+x_4) - s\right) \alpha _3,m^2+\left(2\, m^2-s\right) x_4+m^2 x_4^2+t(1+x_4)\alpha _3 +t\,\alpha _3^2\Big\}\nonumber
\end{align}
could appear in the result after integrating out $\alpha_1$ followed by $\alpha_2$. ${\rm L}_{\alpha_1 \alpha_2}$ is more non-trivial to treat than ${\rm L}_{\alpha_1}$ and requires us to think carefully, both about what sort of result we expect to find and how the {\tt Mathematica}-based private direct integration code written by one of us actually operates in practice. As should by now be clear, our goal is to delay the appearance of square roots until the final variable of integration is moved into the arguments of the Goncharov polylogarithms which depend on it. With the final integration of the most complicated weight-three functions in mind, it is useful to aim for a final integration domain of $[0,1]$. This will naturally produce Goncharov polylogarithms of argument $1$ which contain a complicated square root in the weights. Actually, due to the way that our integration script is written, it is most natural for us to integrate complicated finite Feynman parameter integrals such as the one considered in this section over the unit hypercube. This is because it is easier for our integration code to take a definite integral on $[0,1]$ than on $[0,\infty)$, due to the fact that no complicated argument inversion formulae for non-trivial Goncharov polylogarithms need to be derived if the upper integration endpoint is always set to $1$.

In this spirit, we initiate our treatment of ${\rm L}_{\alpha_1 \alpha_2}$ by trivially mapping one of the remaining integration domains onto $[0,1]$ via the transformation
\begin{equation}
\label{eq:trivalp3}
    \alpha_3 = \frac{x_3}{1-x_3}.
\end{equation}
It is also useful to note that two of the quadratic polynomials in Eq. \eqref{eq:polyredalp1alp2} depend on just a single Feynman parameter each, a situation which strongly suggests that a change of kinematic variables would be advantageous. Indeed, with Euclidean $s$ and $t$ in mind, we see that the variable changes
\begin{equation}
\label{eq:partratB}
    s = -\frac{4\, m^2 (v_2-v_1)^2}{(1-v_1)(1+v_1)(1-v_2)(1+v_2)}\qquad {\rm and}\qquad t = -\frac{m^2 (v_2-v_1)^2}{v_1 v_2}
\end{equation}
rationalize two square roots which would otherwise arise from the third irreducible polynomial of ${\rm L}_{\alpha_1 \alpha_2}$,
\begin{equation}
\label{eq:rootrat1}
    m^2+t\,\alpha _3+t\,\alpha _3^2 = \frac{m^2 \left(\vphantom{v^2}v_1 - (v_2 - v_1) \alpha_3\right) \left(\vphantom{v^2}v_2 + (v_2- v_1) \alpha_3\right)}{v_1 v_2},
\end{equation}
and the fourth irreducible polynomial of ${\rm L}_{\alpha_1 \alpha_2}$,
\begin{align}
\label{eq:rootrat2}
    & m^2+\left(2\, m^2-s\right) x_4+m^2 x_4^2 = 
    \\&
    \qquad \frac{m^2 \left(\vphantom{v^2}(1 + v_1) (1 - v_2) + (1 - v_1) (1 + v_2) x_4\right) \left(\vphantom{v^2}(1 - v_1) (1 + v_2) + (1 + v_1) (1 - v_2) x_4\right)}{(1 - v_1) (1 + v_1) (1 - v_2) (1 + v_2)}.\nonumber
\end{align}
Curiously, transformation \eqref{eq:partratB} was also found to be useful in the context of the multi-loop QED corrections to light-by-light scattering \cite{Caron-Huot:2014lda}.

At this stage, it turns out that just a single non-trivial algebraic function of $x_4$ is generated if one attempts to naively continue the calculation by integrating out $x_3$:
\begin{equation}
\label{eq:finalsqrtBx4}
    \sqrt{1+\frac{2 (4\, m^2 - 2\, s- t)}{4\, m^2-t}x_4 + x_4^2}\,.
\end{equation}
To our knowledge, a situation such as this was first discussed in the Feynman integral literature in the context of the fully-massless (three-loop) $K_4$ integral \cite{Panzer:2014gra}. Using the method of parametrization by lines \cite{Panzer:2014gra,Besier:2018jen},
it is completely straightforward to derive a final variable change for $x_4$ which rationalizes \eqref{eq:finalsqrtBx4} and produces an integration domain of $[0,1]$. As usual, one begins by identifying a suitable rational point of the algebraic variety
\begin{equation}
    \label{eq:varietyB}
    1+\frac{2 (4 \,m^2 - 2 \,s- t)}{4 \,m^2-t}x_4 + x_4^2 = \rho^2.
\end{equation}
For our purposes, the point
\begin{equation}
\label{eq:ratpointB}
    \left(x_4^{(0)}, \rho^{(0)}\right) = \left(-\frac{2 (4 \,m^2 - 2\, s- t)}{4\, m^2-t},1\right)
\end{equation}
is ideal.\footnote{We avoid the other obvious rational point, $\left(x_4^{(0)}, \rho^{(0)}\right) = (0,1)$, because the resulting variable change does not map the $x_4$ integration domain onto the unit interval.}
Eq. \eqref{eq:varietyB} defines a hyperbola which can be rationally parametrized by a family of lines passing through \eqref{eq:ratpointB},
\begin{equation}
\label{eq:linefamB}
    \rho = y_4 \left(x_4 - x_4^{(0)}\right)+\rho^{(0)}.
\end{equation}
We combine Eqs. \eqref{eq:varietyB} and \eqref{eq:linefamB} to determine $x_4$ as a function of $y_4$:
\begin{equation}
\label{eq:finalvarchangeB}
x_4 = \frac{2 \,y_4 \left(1+\frac{4 \,m^2 - 2 \,s - t}{4 \,m^2 - t} y_4\right)}{(1-y_4)(1+y_4)}.
\end{equation}
As desired, we see from the above that the integration domain for $y_4$ is the unit interval. 

By combining together the various variable changes given above, we obtain the complete sequence of integration variables $\{\alpha_1, \alpha_2, x_3, y_4\}$ and establish that, up to an overall normalization factor, integral \eqref{eq:FPBalgLS} may be evaluated as a rational linear combination of Goncharov polylogarithms to all orders in $\epsilon$. As a sanity check, we explicitly evaluate all Feynman parameter integrals analytically using our direct integration code at $\mathcal{O}\left(\epsilon^0\right)$ and then evaluate the result obtained numerically to high precision using {\tt GiNaC} \cite{Vollinga:2004sn,Bauer:2000cp} at the randomly chosen Euclidean phase space point
\begin{equation}
\label{eq:vPSptB}
    \left(v_1,v_2,m^2\right) = \left(1/7, 1/5, 2\right).
\end{equation}
We find
\begin{equation}
\label{eq:BalgLSnum}
    \includegraphics[valign = m, raise = .2 cm, height = .8 cm,keepaspectratio]{BalgLSunmarked} \left(-1/36,-8/35,2\right) \approx -6.317550089475753330169497\ldots \,+\, \mathcal{O}\left(\epsilon\right),
\end{equation}
which we could rapidly confirm agrees to five significant digits with an independent {\tt FIESTA 4} evaluation of the integral.

Let us stress once more that the unrationalizable square root
\begin{equation}
\label{eq:finalsqrtB}
\sqrt{p_1(v_1,v_2)p_2(v_1,v_2)p_3(v_1,v_2)p_4(v_1,v_2)}\,,
\end{equation}
where
\begin{align}
    p_1(v_1,v_2) &= 1 + v_1 - v_2 (1 - v_1),\nonumber\\
    p_2(v_1,v_2) &= 1 + v_2 - v_1 (1 - v_2),\nonumber\\
    p_3(v_1,v_2) &= v_1 (1 - v_1) + v_2 (1 - v_2) + v_1 v_2 (2 - v_1 - v_2),
    \nonumber \\
    p_4(v_1,v_2) &= v_1 (1 + v_1) + v_2 (1 + v_2) - v_1 v_2 (2 + v_1 + v_2),
\end{align}
appears in our analytic integrations only at the very end, once all Feynman parameters have been integrated out. The most complicated, weight-four functions which appear in the final result have argument unity and weights which are non-trivial functions of the square root above. For the final integration, our integration code actually utilizes the generalized weights of \cite{vonManteuffel:2013vja}. The idea is that, for the final integration over $y_4$, one can employ non-linear integrating factors of the form $\mathrm{d}y_4(\partial f(y_4;v_1,v_2)/\partial y_4)/f(y_4;v_1,v_2)$, for non-linear irreducible polynomials $f(y_4;v_1,v_2)$ in $y_4$. This allows for a more concise representation of the final result without giving up access to the well-tested numerical routines provided by {\tt GiNaC}. If desired, a representation written strictly in terms of Goncharov polylogarithms may also be derived by fully factoring all of the generalized weights.

Of course, the direct integration approach described in this section is not without its limitations. For one thing, it would at first seem quite non-trivial to find an explicit integration order for {\it e.g.} the two-loop double box with two massive internal lines from the two-loop mixed EW-QCD corrections to Drell-Yan lepton production (right panel of Figure \ref{fig:toplevel}). Actually, with the help of {\tt HyperInt}, it is comparatively easy to find a complete sequence of integration variables; for a fixed number of loops and legs, one gets the impression that the number of internal lines is a less important factor in determining the difficulty of a linear reducibility analysis than the total number of massive lines. Although we could directly integrate through to weight four the six-line basic scalar integral (right panel of Figure \ref{fig:algLS}) which first inherits unrationalizable symbol letters from its unrationalizable leading singularity, we find the direct integration of the top-level two-massive-line integrals to be very cumbersome at the technical level. Furthermore, the produced Goncharov polylogarithms are very complicated, not minimal, and suboptimal for numerical evaluations with {\tt GiNaC}.

An alternative approach which can avoid these issues is to apply the differential equation method. We shall see that the differential equation method has the very appealing feature that one can fit an ansatz of multiple polylogarithmic functions appropriate for each physical kinematic region separately. In the following sections, we discuss in some detail how this is achieved for the mixed EW-QCD Drell-Yan master integrals with two massive lines, including, in particular, the six- and seven-line integrals with an unrationalizable square root in their symbol letters.

\section{An \boldmath{$\epsilon$} basis for the Drell-Yan master integrals with two massive lines}
\label{sec:dydefs}

In this work, we are primarily concerned with neutral-current lepton pair production\footnote{The two-loop mixed EW-QCD corrections to the charged-current Drell-Yan process could be accessed by expanding in $1-m_W^2/m_Z^2$, as this would allow one to make effective use of the equal-mass integrals relevant to the neutral-current process.}  in quark-antiquark annihilation,
\begin{align}
\label{eq:ourprocess}
 q(p_1) \, \bar q (p_2) &\to \ell^- (p_3) \, \ell^+ (p_4),
\end{align}
where all external particles are taken massless and on their mass shell. The most complicated master integrals for the two-loop mixed EW-QCD corrections to above process are those with two internal lines of mass $m$, where, depending on the parent Feynman diagram, $m$ may refer to either $m_W$ or $m_Z$. In particular, as mentioned above, it has been known to us for some time that the integral topologies from the right panels of Figures \ref{fig:toplevel} and \ref{fig:algLS} actually contain master integrals with unrationalizable symbol letters in their $\epsilon$-decoupled differential equations. To index the seventeen master integrals with two massive internal lines, it suffices to consider a single integral family based on the two-loop planar double box with two massive internal lines from Figure \ref{fig:toplevel} (Family $\mathrm{E}$ from Table \ref{tab:famsmixedDY}). For completeness, we will also give definitions for those normal form integrals with zero or one massive internal lines which appear on the right-hand sides of the $\epsilon$-decoupled differential equations for the master integrals with two massive internal lines. To index these auxiliary integrals, we reintroduce two additional integral families which were already studied in the physical region by two of us some time ago \cite{vonManteuffel:2017myy} (Family $\mathrm{A}$ and Family $\mathrm{C}$ from Table \ref{tab:famsmixedDY}).

\begin{table}[t]
\centering
\begin{tabular}{llll}
Family $\mathrm{A}$\hspace{18mm} & Family $\mathrm{C}$ & Family $\mathrm{E}$\\[1mm]
\hlinewd{2pt}                                                                                                            
\rule{0pt}{2ex}~$k_1^2$                 & $k_1^2$                   & $k_1^2$\\
\rule{0pt}{2ex}~$k_2^2$                 & $k_2^2$                   & $k_2^2 - m^2$\\
\rule{0pt}{2ex}~$(k_1-k_2)^2$           & $(k_1-k_2)^2$             & $(k_1-k_2)^2$\\
\rule{0pt}{2ex}~$(k_1-p_1)^2$           & $(k_1-p_1)^2$             & $(k_1-p_1)^2$\\
\rule{0pt}{2ex}~$(k_2-p_1)^2$           & $(k_2-p_1)^2$             & $(k_2-p_1)^2$\\
\rule{0pt}{2ex}~$(k_1-p_1-p_2)^2$       & $(k_1-p_1-p_2)^2$         & $(k_1-p_1-p_2)^2$\\
\rule{0pt}{2ex}~$(k_2-p_1-p_2)^2$       & $(k_2-p_1-p_2)^2 - m^2$   & $(k_2-p_1-p_2)^2 - m^2$\\
\rule{0pt}{2ex}~$(k_1-p_3)^2$           & $(k_1-p_3)^2$             & $(k_1-p_3)^2$\\
\rule{0pt}{2ex}~$(k_2-p_3)^2$           & $(k_2-p_3)^2$             & $(k_2-p_3)^2$
\end{tabular}
\caption{Integral families for the master integrals which appear in the differential equations for the two-loop mixed EW-QCD Drell-Yan master integrals with two massive internal lines.}
\label{tab:famsmixedDY}
\end{table}

In order to obtain a closed system of differential equations for the masters with two massive internal lines, we need to consider thirty-six integrals in total. Our notation for Feynman integrals in this section is exactly that of \cite{vonManteuffel:2017myy} ({\it i.e.} dots for doubled propagators, heavy lines for massive propagators, numerator insertions written in square brackets, and $\mathrm{\thickbar{F}}$:$x$ for the crossed version of sector $x$ from family $\mathrm{F}$).
For the kinematic invariants we use
\begin{align}
s = (p_1+p_2)^2,\qquad t = (p_1-p_3)^2,\qquad u = (p_2-p_3)^2,\qquad
p_1^2 = p_2^2 = p_3^2 = p_4^2 = 0\,.
\end{align}
In the following, we keep the dependence on the internal mass parameter $m$ implicit for the sake of brevity and since it is anyway clear from the thick-line notation.
We build up our normal form basis out of the following thirty-six Feynman integrals:
\begin{align}
\mathbf{f}_{1}^{\mathrm{\thickbar{A}:38}} = \includegraphics[valign = m, raise = .3 cm, height = .125\linewidth, width = .125\linewidth,keepaspectratio]{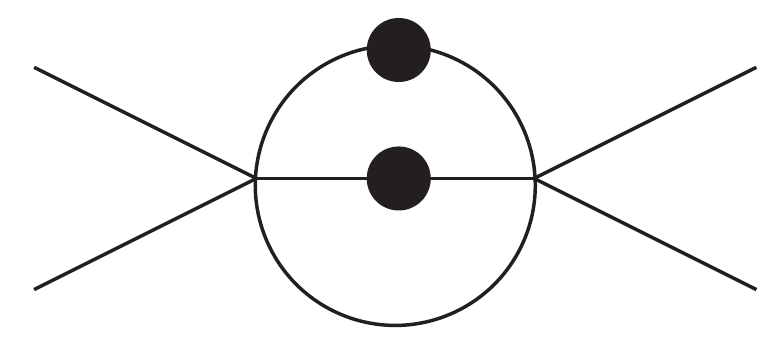}(t)\quad
\mathbf{f}_{2}^{\mathrm{C:97}} = \includegraphics[valign = m, raise = .3 cm, height = .125\linewidth, width = .125\linewidth,keepaspectratio]{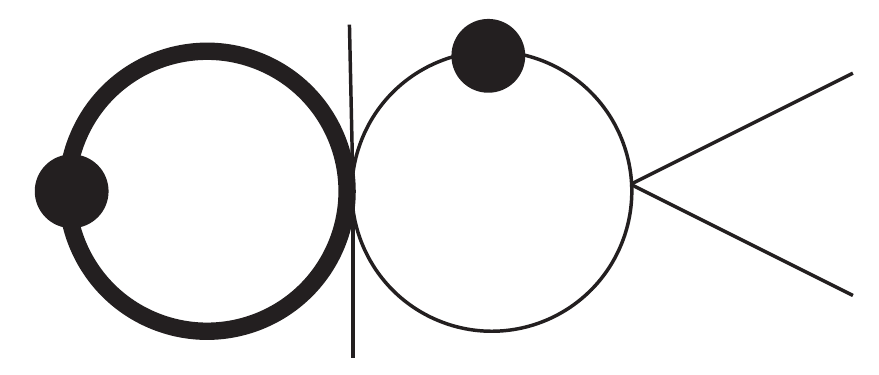}(s)\quad
\mathbf{f}_{3}^{\mathrm{C:76}} = \includegraphics[valign = m, raise = .3 cm, height = .05\linewidth, width = .05\linewidth,keepaspectratio]{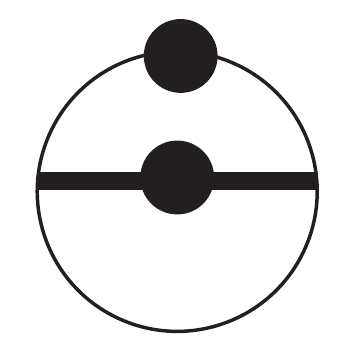} \nonumber
\end{align}

\begin{align}
\mathbf{f}_{4}^{\mathrm{C:69}} = \includegraphics[valign = m, raise = .3 cm, height = .125\linewidth, width = .125\linewidth,keepaspectratio]{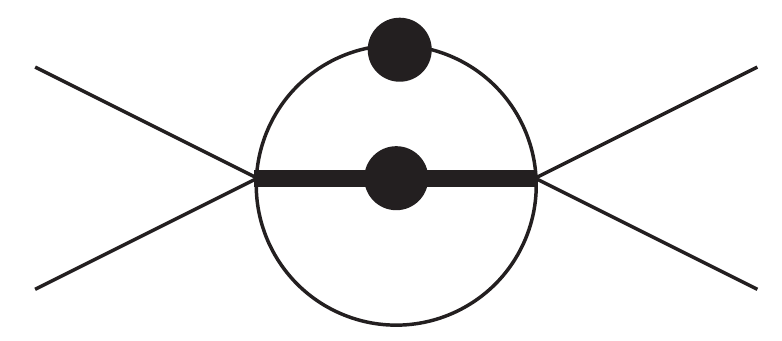}(s)\quad
\mathbf{f}_{5}^{\mathrm{C:69}} = \includegraphics[valign = m, raise = .3 cm, height = .125\linewidth, width = .125\linewidth,keepaspectratio]{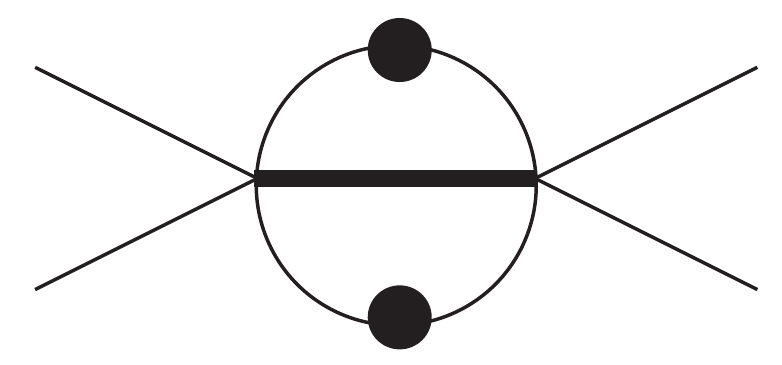}(s)\quad
\mathbf{f}_{6}^{\mathrm{A:53}} = \includegraphics[valign = m, raise = .2 cm, height = .135\linewidth, width = .135\linewidth,keepaspectratio]{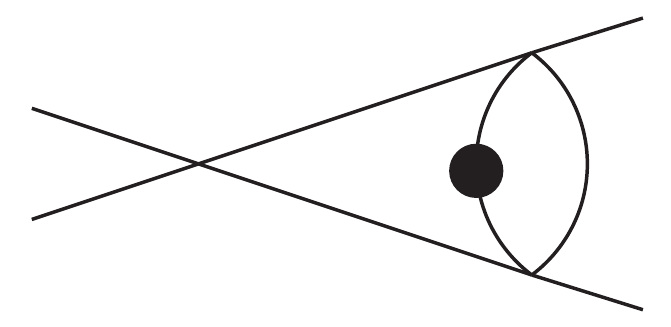}(s) \nonumber
\end{align}

\begin{align}
\mathbf{f}_{7}^{\mathrm{\thickbar{A}:53}} = \includegraphics[valign = m, raise = .2 cm, height = .135\linewidth, width = .135\linewidth,keepaspectratio]{int6a7}(t)\quad
\mathbf{f}_{8}^{\mathrm{C:212}} = \includegraphics[valign = m, raise = .1 cm, height = .15\linewidth, width = .15\linewidth,keepaspectratio]{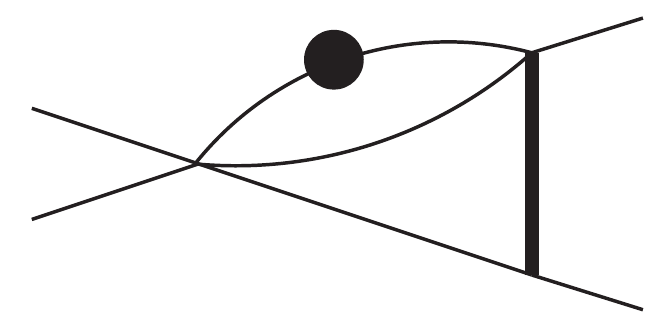}(t)\quad
\mathbf{f}_{9}^{\mathrm{E:99}} = \includegraphics[valign = m, raise = .3 cm, height = .175\linewidth, width = .175\linewidth,keepaspectratio]{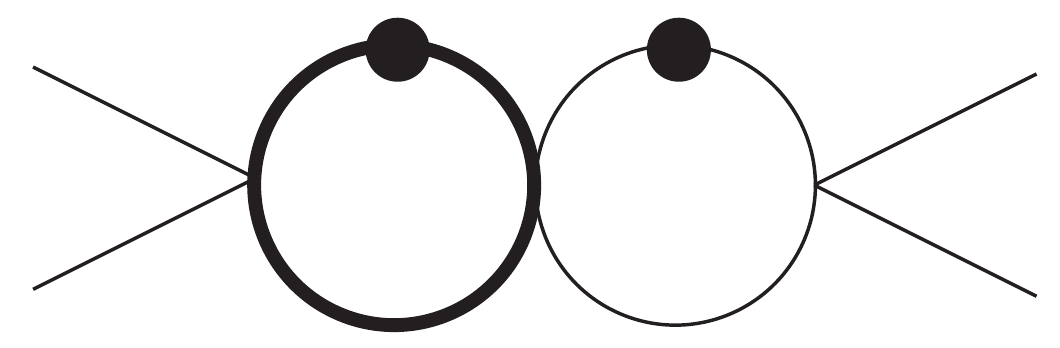}(s) \nonumber
\end{align}

\begin{align}
\mathbf{f}_{10}^{\mathrm{E:71}} = \includegraphics[valign = m, raise = .6 cm, height = .16\linewidth, width = .16\linewidth,keepaspectratio]{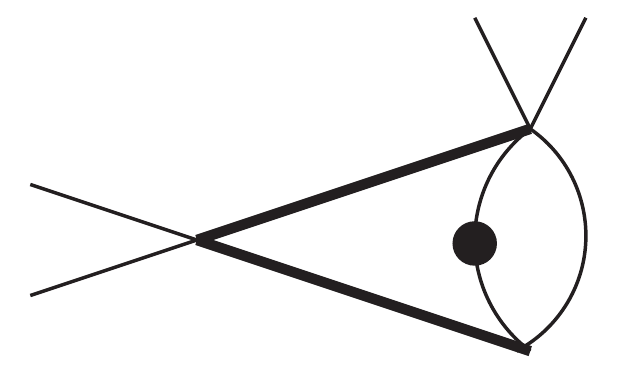}(s)\quad
\mathbf{f}_{11}^{\mathrm{E:78}} = \includegraphics[valign = m, raise = .1 cm, height = .16\linewidth, width = .16\linewidth,keepaspectratio]{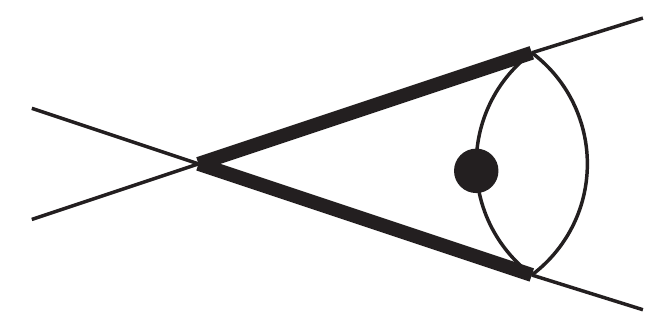}(s)\quad
\mathbf{f}_{12}^{\mathrm{E:78}} = \includegraphics[valign = m, raise = .1 cm, height = .16\linewidth, width = .16\linewidth,keepaspectratio]{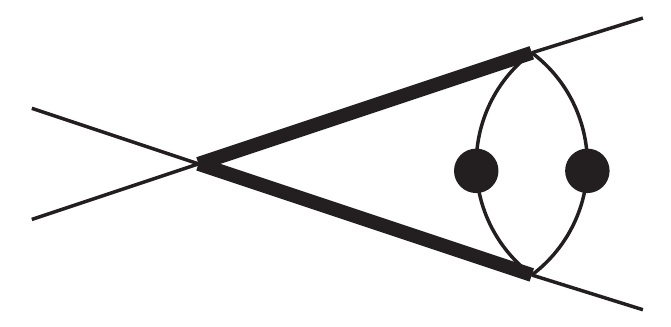}(s) \nonumber
\end{align}

\begin{align}
\mathbf{f}_{13}^{\mathrm{\thickbar{A}:174}} = \includegraphics[valign = m, raise = .2 cm, height = .125\linewidth, width = .125\linewidth,keepaspectratio]{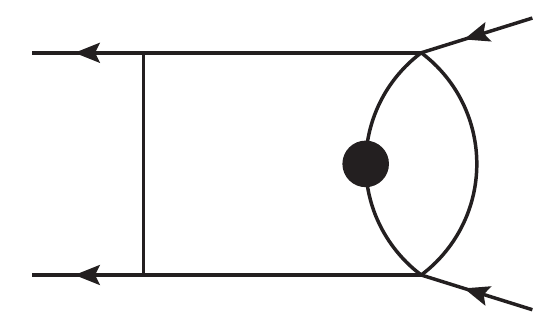}(t,s)\quad
\mathbf{f}_{14}^{\mathrm{C:372}} = \includegraphics[valign = m, raise = .2 cm, height = .15\linewidth, width = .15\linewidth,keepaspectratio]{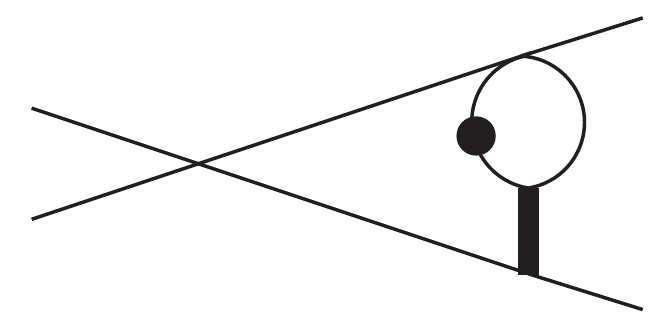}(t)\quad
\mathbf{f}_{15}^{\mathrm{C:244}} = \includegraphics[valign = m, raise = .7 cm, height = .175\linewidth, width = .175\linewidth,keepaspectratio]{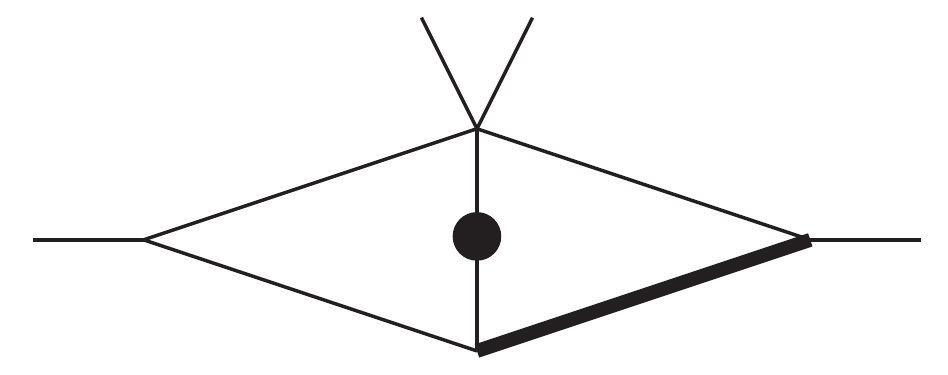}(t) \nonumber
\end{align}

\begin{align}
\mathbf{f}_{16}^{\mathrm{C:117}} = \includegraphics[valign = m, raise = .7 cm, height = .175\linewidth, width = .175\linewidth,keepaspectratio]{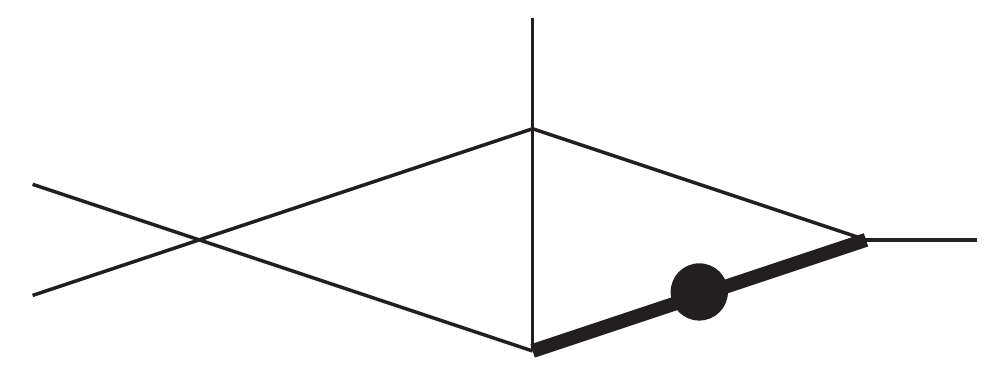}(s)\quad
\mathbf{f}_{17}^{\mathrm{C:117}} = \includegraphics[valign = m, raise = .7 cm, height = .175\linewidth, width = .175\linewidth,keepaspectratio]{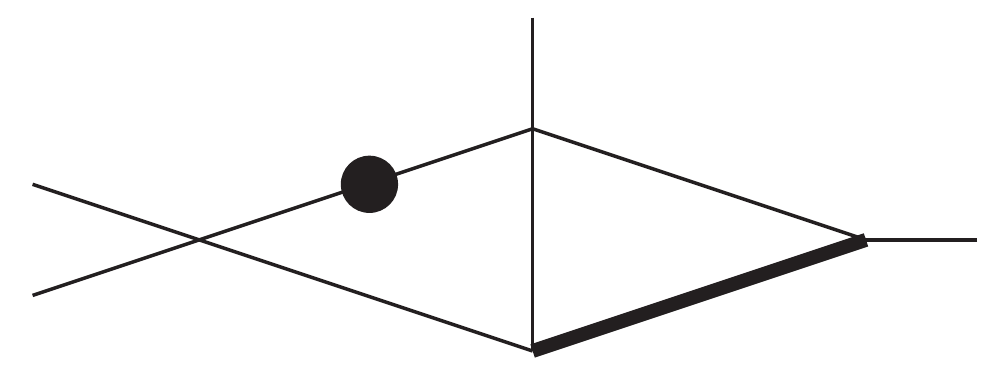}(s)\quad
\mathbf{f}_{18}^{\mathrm{C:341}} = \includegraphics[valign = m, raise = .2 cm, height = .125\linewidth, width = .125\linewidth,keepaspectratio]{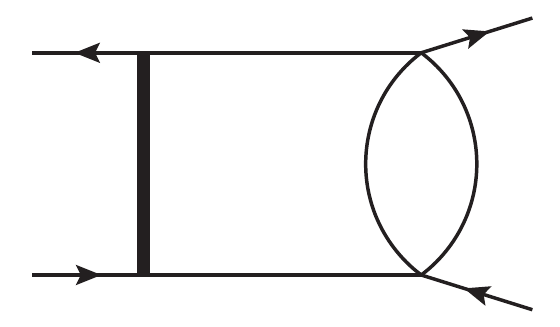}(s,t) \nonumber
\end{align}

\begin{align}
\mathbf{f}_{19}^{\mathrm{C:341}} = \includegraphics[valign = m, raise = .2 cm, height = .125\linewidth, width = .125\linewidth,keepaspectratio]{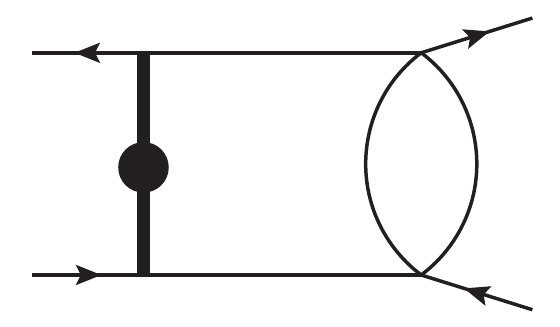}(s,t)\quad
\mathbf{f}_{20}^{\mathrm{C:213}} = \includegraphics[valign = m, raise = .2 cm, height = .135\linewidth, width = .135\linewidth,keepaspectratio]{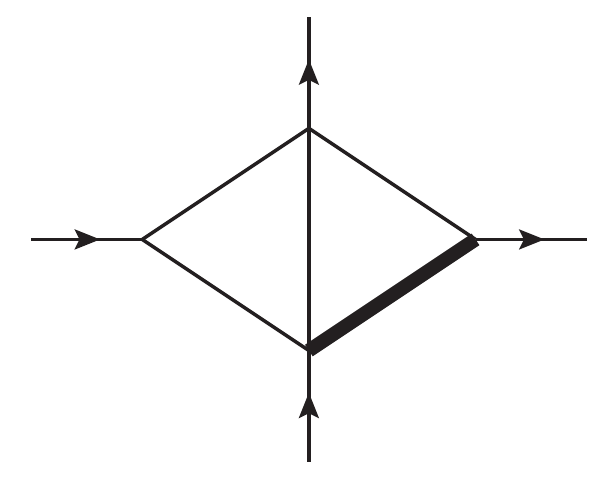}(s,t)\quad
\mathbf{f}_{21}^{\mathrm{C:213}} = \includegraphics[valign = m, raise = .2 cm, height = .135\linewidth, width = .135\linewidth,keepaspectratio]{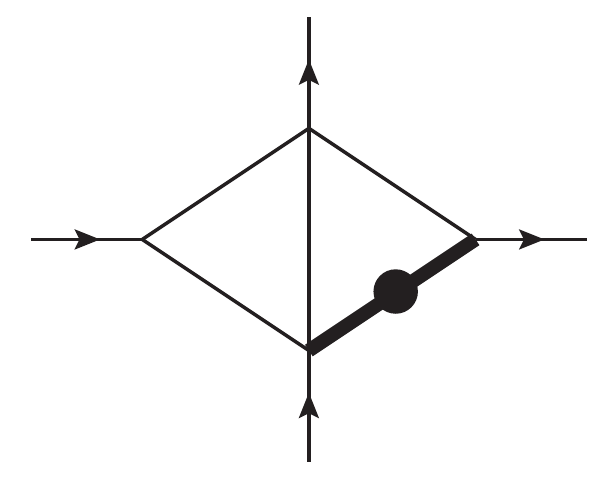}(s,t)\quad \nonumber
\end{align}

\begin{align}
\mathbf{f}_{22}^{\mathrm{E:115}} = \includegraphics[valign = m, raise = .3 cm, height = .175\linewidth, width = .175\linewidth,keepaspectratio]{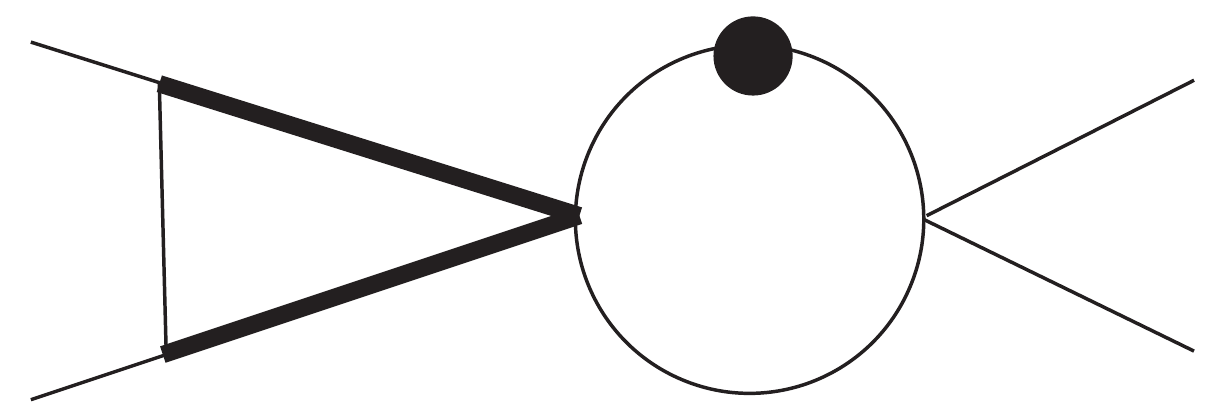}(s)\quad
\mathbf{f}_{23}^{\mathrm{E:103}} = \includegraphics[valign = m, raise = .3 cm, height = .175\linewidth, width = .175\linewidth,keepaspectratio]{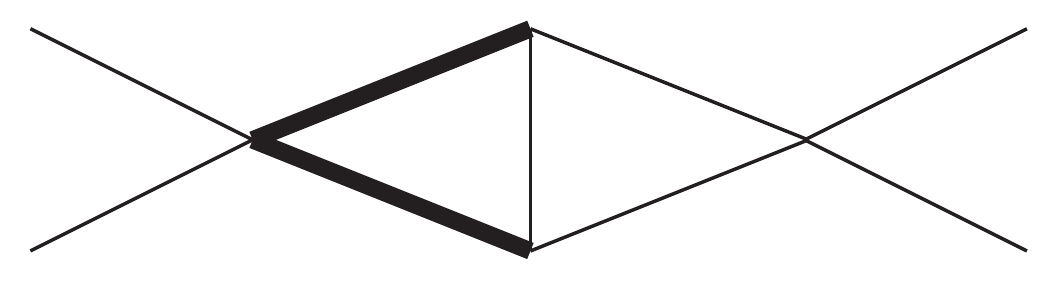}(s)\quad
\mathbf{f}_{24}^{\mathrm{E:87}} = \includegraphics[valign = m, raise = .1 cm, height = .15\linewidth, width = .15\linewidth,keepaspectratio]{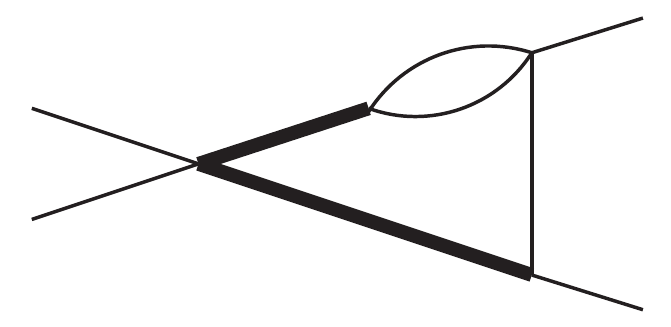}(s) \nonumber
\end{align}

\begin{align}
\mathbf{f}_{25}^{\mathrm{E:79}} = \includegraphics[valign = m, raise = .7 cm, height = .175\linewidth, width = .175\linewidth,keepaspectratio]{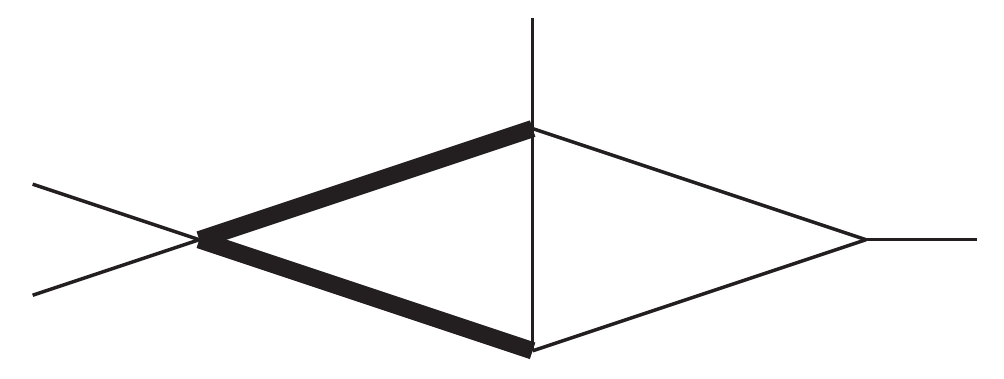}(s)\quad
\mathbf{f}_{26}^{\mathrm{E:214}} = \includegraphics[valign = m, raise = .2 cm, height = .125\linewidth, width = .125\linewidth,keepaspectratio]{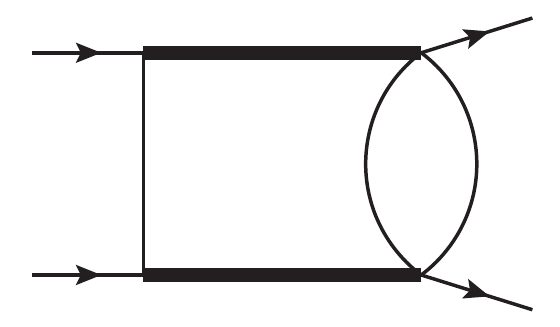}(s,t)\quad
\mathbf{f}_{27}^{\mathrm{E:214}} = \includegraphics[valign = m, raise = .2 cm, height = .125\linewidth, width = .125\linewidth,keepaspectratio]{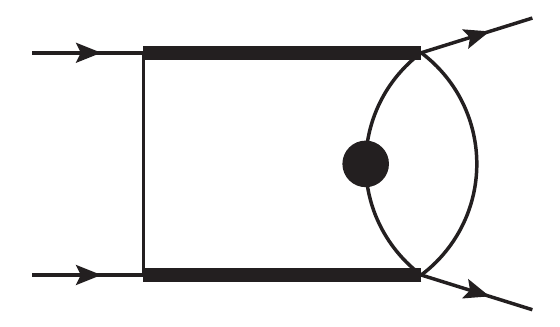}(s,t) \nonumber
\end{align}

\begin{align}
\mathbf{f}_{28}^{\mathrm{C:245}} = \includegraphics[valign = m, raise = .45 cm, height = .15\linewidth, width = .15\linewidth,keepaspectratio]{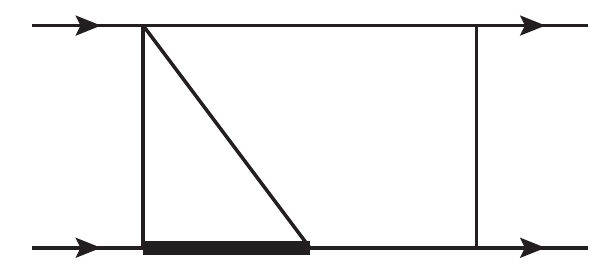}(s,t)\quad
\mathbf{f}_{29}^{\mathrm{C:245}} = \includegraphics[valign = m, raise = .45 cm, height = .15\linewidth, width = .15\linewidth,keepaspectratio]{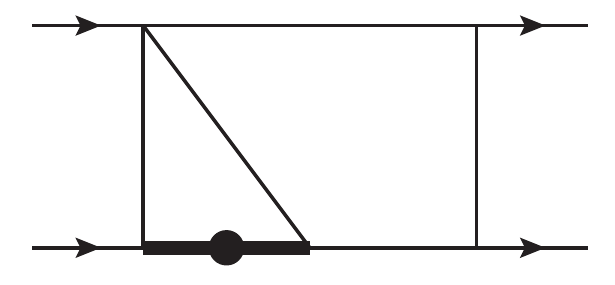}(s,t)\quad
\mathbf{f}_{30}^{\mathrm{E:119}} = \includegraphics[valign = m, raise = .2 cm, height = .135\linewidth, width = .135\linewidth,keepaspectratio]{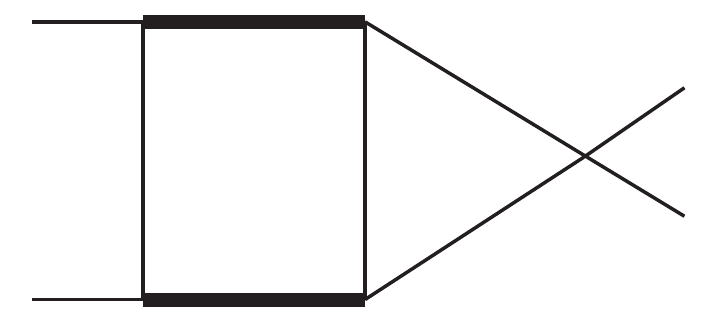}(s) \nonumber
\end{align}

\begin{align}
\mathbf{f}_{31}^{\mathrm{E:343}} = \includegraphics[valign = m, raise = .45 cm, height = .15\linewidth, width = .15\linewidth,keepaspectratio]{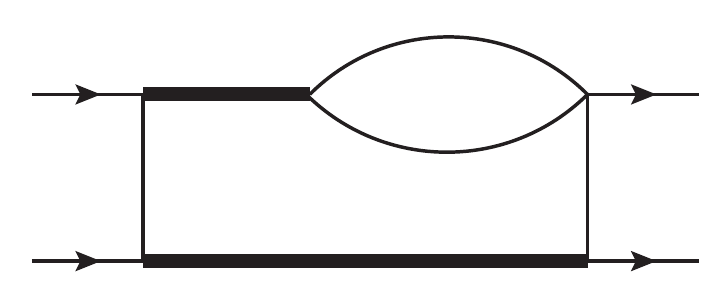}(s,t)\quad
\mathbf{f}_{32}^{\mathrm{E:215}} = \includegraphics[valign = m, raise = .2 cm, height = .55 cm,keepaspectratio]{int32}(s,t)\quad \mathbf{f}_{33}^{\mathrm{E:247}} = \includegraphics[valign = m, raise = .3 cm, height = .15\linewidth, width = .15\linewidth,keepaspectratio]{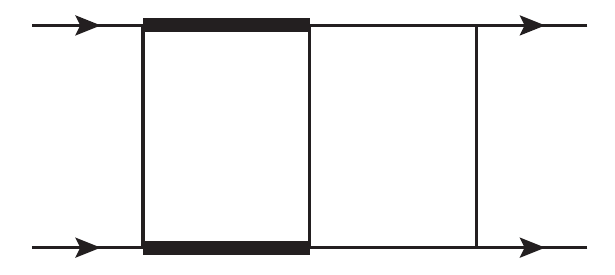}(s,t) \nonumber
\end{align}

\begin{align}
\mathbf{f}_{34}^{\mathrm{E:247}} = \includegraphics[valign = m, raise = .3 cm, height = .15\linewidth, width = .15\linewidth,keepaspectratio]{int33a34a35a36}\left[(k_2 - p_3)^2\right](s,t)\quad 
\mathbf{f}_{35}^{\mathrm{E:247}} = \includegraphics[valign = m, raise = .3 cm, height = .15\linewidth, width = .15\linewidth,keepaspectratio]{int33a34a35a36}\left[(k_1 - p_1)^2\right](s,t) \nonumber
\end{align}

\begin{align}
\label{eq:rawints}
    \mathbf{f}_{36}^{\mathrm{E:247}} = \includegraphics[valign = m, raise = .3 cm, height = .15\linewidth, width = .15\linewidth,keepaspectratio]{int33a34a35a36}\left[(k_1 - p_1)^2(k_2 - p_3)^2\right](s,t).
\end{align}
From the general principles discussed in \cite{Gehrmann:2014bfa}, one can readily cast \eqref{eq:rawints} above into a normal form basis for the integrals of interest.\footnote{Due to the presence of unrationalizable square roots, the available public software packages for the construction of a normal form basis of integrals \cite{Prausa:2017ltv,Gituliar:2017vzm,Meyer:2017joq} are not applicable to the problem at hand.} Abbreviating the three square roots which appear as
\begin{align}
\label{eq:root1}
    r_1 &= \sqrt{s\, \big(s - 4 \, m^2\big)}\,,
\\
\label{eq:root2}
    r_2 &= \sqrt{-s\, t\, \big(4 m^2 (m^2 + t) - s\, t\big)}\,,
\\
\label{eq:root3}
    r_3 &= \sqrt{s\, \big(t^2 (s - 4\, m^2) + m^2\, s\, (m^2 - 2\, t)\big)}\,,
\end{align}
we find:
\begin{align}
\mathbf{m}_{1} &= \epsilon^2 \, t \,\mathbf{f}_{1}^{\mathrm{\thickbar{A}:38}},
\nonumber\\
\mathbf{m}_{2} &= \epsilon^2 \, s \,\mathbf{f}_{2}^{\mathrm{C:97}},
\nonumber\\
\mathbf{m}_{3} &= \epsilon (1-\epsilon) \, m^2 \, \mathbf{f}_{3}^{\mathrm{C:76}},
\nonumber\\
\mathbf{m}_{4} &= \epsilon^2 \, s \,\mathbf{f}_{4}^{\mathrm{C:69}},
\nonumber\\
\mathbf{m}_{5} &= 2 \,\epsilon^2 (s - m^2) \mathbf{f}_{4}^{\mathrm{C:69}} + \epsilon^2 (s - m^2) \mathbf{f}_{5}^{\mathrm{C:69}},
\nonumber\\
\mathbf{m}_{6} &= \epsilon^3 \, s \, \mathbf{f}_{6}^{\mathrm{A:53}},
\nonumber\\
\mathbf{m}_{7} &= \epsilon^3 \, t \, \mathbf{f}_{7}^{\mathrm{\thickbar{A}:53}},
\nonumber\\
\mathbf{m}_{8} &= \epsilon^3 \, t \, \mathbf{f}_{8}^{\mathrm{C:212}},
\nonumber\\
\mathbf{m}_{9} &= \epsilon^2 \, s \, r_1 \mathbf{f}_{9}^{\mathrm{E:99}},
\nonumber\\
\mathbf{m}_{10} &= \frac{\epsilon (1-\epsilon)m^2\, s}{4\, r_1}\mathbf{f}_{3}^{\mathrm{C:76}}-\frac{\epsilon^2\,s(s - 3\, m^2)}{2\,r_1}\mathbf{f}_{4}^{\mathrm{C:69}}
-\frac{\epsilon^2\,s(s - m^2)}{4\,r_1}\mathbf{f}_{5}^{\mathrm{C:69}}
 +\frac{\epsilon^2 (1-2\,\epsilon)m^2\, s}{r_1}\mathbf{f}_{10}^{\mathrm{E:71}},
\nonumber\\
\mathbf{m}_{11} &= \epsilon^3 \, s\, \mathbf{f}_{11}^{\mathrm{E:78}},
\nonumber\\
\mathbf{m}_{12} &= -\frac{\epsilon (1-\epsilon) m^2\, s}{r_1}\mathbf{f}_{3}^{\mathrm{C:76}}+\frac{3\,\epsilon^3\, s (s-2\, m^2)}{r_1}\mathbf{f}_{11}^{\mathrm{E:78}}+\frac{\epsilon^2\,m^4\, s}{r_1}\mathbf{f}_{12}^{\mathrm{E:78}},
\nonumber\\
\mathbf{m}_{13} &= \epsilon^3 \, s\, t\, \mathbf{f}_{13}^{\mathrm{\thickbar{A}:174}},
\nonumber\\
\mathbf{m}_{14} &= \epsilon^3 \, m^2\, t\, \mathbf{f}_{14}^{\mathrm{C:372}},
\nonumber\\
\mathbf{m}_{15} &= \epsilon^3 \, m^2\, t\, \mathbf{f}_{15}^{\mathrm{C:244}},
\nonumber\\
\mathbf{m}_{16} &= \epsilon^3 \, m^2\, s\, \mathbf{f}_{16}^{\mathrm{C:117}},
\nonumber\\
\mathbf{m}_{17} &= \epsilon^3 \, s^2\, \mathbf{f}_{17}^{\mathrm{C:117}},
\nonumber\\
\mathbf{m}_{18} &= \epsilon^3(1-2\,\epsilon)\, t\, \mathbf{f}_{18}^{\mathrm{C:341}},
\nonumber\\
\mathbf{m}_{19} &= 3\,\epsilon^3(1-2\,\epsilon)\, m^2\, \mathbf{f}_{18}^{\mathrm{C:341}}
+\epsilon^2(1-2\,\epsilon)\, m^2\, (m^2 + t)\mathbf{f}_{19}^{\mathrm{C:341}},
\nonumber\\
\mathbf{m}_{20} &= \epsilon^4\, (s + t) \mathbf{f}_{20}^{\mathrm{C:213}},
\nonumber\\
\mathbf{m}_{21} &= \epsilon^3\, m^2\, (s + t)\mathbf{f}_{21}^{\mathrm{C:213}},
\nonumber\\ 
\mathbf{m}_{22} &= \epsilon^3\, s^2\, \mathbf{f}_{22}^{\mathrm{E:115}},
\nonumber\\
\mathbf{m}_{23} &= \epsilon^3(1-2\,\epsilon)\, s\, \mathbf{f}_{23}^{\mathrm{E:103}},
\nonumber\\
\mathbf{m}_{24} &= \epsilon^3(1-2\,\epsilon)\, s\, \mathbf{f}_{24}^{\mathrm{E:87}},
\nonumber\\
\mathbf{m}_{25} &= \epsilon^4\, s\,\mathbf{f}_{25}^{\mathrm{E:79}},
\nonumber\\
\mathbf{m}_{26} &= \epsilon^3(1-2\,\epsilon)\, s\, \mathbf{f}_{26}^{\mathrm{E:214}},
\nonumber\\
\mathbf{m}_{27} &= \epsilon^3\, r_2\, \mathbf{f}_{27}^{\mathrm{E:214}},
\nonumber\\
\mathbf{m}_{28} &= \epsilon^4\, s\, t\, \mathbf{f}_{28}^{\mathrm{C:245}},
\nonumber\\
\mathbf{m}_{29} &= 2\, \epsilon^4\, m^2\, s\, \mathbf{f}_{28}^{\mathrm{C:245}}+\epsilon^3\, m^2\, s\,(m^2+t) \mathbf{f}_{29}^{\mathrm{C:245}},
\nonumber\\
\mathbf{m}_{30} &= \epsilon^4\, s\, r_1\, \mathbf{f}_{30}^{\mathrm{E:119}},
\nonumber\\
\mathbf{m}_{31} &= \epsilon^3(1-2\,\epsilon)\, r_2\, \mathbf{f}_{31}^{\mathrm{E:343}},
\nonumber\\
\mathbf{m}_{32} &= \epsilon^4\, r_3\, \mathbf{f}_{32}^{\mathrm{E:215}},
\nonumber\\ 
\mathbf{m}_{33} &= \epsilon^4\, s\, r_2\, \mathbf{f}_{33}^{\mathrm{E:247}},
\nonumber\\
\mathbf{m}_{34} &= \epsilon^4\, s^2\, \mathbf{f}_{34}^{\mathrm{E:247}},
\nonumber\\
\mathbf{m}_{35} &= 2\, \epsilon^4\, t\, r_1\, \mathbf{f}_{32}^{\mathrm{E:215}}-\epsilon^4\, s\, t\, r_1\, \mathbf{f}_{33}^{\mathrm{E:247}}+\epsilon^4\, s\, r_1\, \mathbf{f}_{35}^{\mathrm{E:247}},
\nonumber\\
\label{eq:normformints}
    \mathbf{m}_{36} &= 2\,\epsilon^4\, t\, \mathbf{f}_{20}^{\mathrm{C:213}}+\epsilon^3\, s\, t\, \mathbf{f}_{22}^{\mathrm{E:115}}-\frac{1}{2}\epsilon^3\, s\, \mathbf{f}_{23}^{\mathrm{E:103}}-\frac{1}{2}\epsilon^4\, s\, (s - 2\, m^2) \mathbf{f}_{30}^{\mathrm{E:119}}+\epsilon^4\, s\, t\, \mathbf{f}_{32}^{\mathrm{E:215}}
    \nonumber \\
    &\quad -\frac{1}{2}\epsilon^4\, s\, t\, (s - 2\, m^2) \mathbf{f}_{33}^{\mathrm{E:247}} +\frac{1}{2}\epsilon^4\, s\, (s - 2\, m^2) \mathbf{f}_{35}^{\mathrm{E:247}}+\epsilon^4\, s\,\mathbf{f}_{36}^{\mathrm{E:247}}.
\end{align}

For $\mathbf{m}_{1},\ldots,\mathbf{m}_{36}$, we employ the integration measure
\begin{equation}
\label{eq:norm2}
    \left(\frac{\Gamma(1-\epsilon) s^{\epsilon}}{i \pi^{2-\epsilon}}\right)^2 \int {\rm d}^{4-2\epsilon} k_1 \int {\rm d}^{4-2\epsilon} k_2
\end{equation}
with $\epsilon=(4-d)/2$, which allows us to consider our integrals to be functions of two dimensionless kinematic variables.\footnote{
The normal form basis \eqref{eq:normformints} closely resembles the one given in \cite{Bonciani:2016ypc}, but we correct typos and employ a different overall normalization.
Note that $r_1^2 = s (s - 4\, m^2)$ is positive for both $s > 4\, m^2$ and $s < 0$. However, one must take care when attempting to simplify $r_1$ to make contact with the Euclidean-region analysis of \cite{Bonciani:2016ypc}; $r_1 = s \sqrt{1-4\, m^2/s}$ for $s >  4\, m^2$, but $r_1 = -s \sqrt{1-4\, m^2/s}$ for $s < 0$.
} For instance, 
\begin{equation}
\label{eq:m1}
    \mathbf{m}_{1} = \frac{\Gamma^5(1-\epsilon)\Gamma(1+2\epsilon)}{\Gamma(1-3\epsilon)}\left(-\frac{t}{s}\right)^{-2\epsilon}.
\end{equation}
One achieves a rationalization of two of our three roots, $r_1$ and $r_2$, with the parametrization~ \cite{Bonciani:2016ypc}
\begin{equation}
\label{eq:DYpartrat}
    s=-\frac{m^2(1-w)^2}{w}\qquad{\rm and}\qquad t=-\frac{m^2 w\, (1+z)^2}{z\, (1+w)^2},
\end{equation}
and we hereafter work primarily with the variables $w$ and $z$. It is immediate from \eqref{eq:normformints} that, in the $(w,z)$ representation, the {\it only} normal form integral which involves a square root in its definition is $\mathbf{m}_{32}$ ({\it i.e.} the integral from the right panel of Figure \ref{fig:algLS}). Indeed, we see from \eqref{eq:DYpartrat} and Appendix A of \cite{Bonciani:2016ypc} that
\begin{align}
\label{eq:r1simp}
    r_1 &= -\frac{m^2(1-w)(1+w)}{w},
\\
\label{eq:r2simp}
    r_2 &= -\frac{m^4(1-w)(1-z)(1+z)}{z (1+w)},
\\
\label{eq:r3simp}
    r_3 &= \frac{m^4(1-w)}{w\,z(1+w)}\sqrt{(1 + w^2 z^2) (w + z)^2 + 2 w\,z (w - z)^2 + 4 w\,z^2 (1 + w^2)}
\end{align}
in the region $s > 4\,m^2$ (see also Section \ref{sec:analycont} for a detailed discussion of the $(w,z)$ parametrization).

In the following, we will replace $r_1$, $r_2$ and $r_3$ according to \eqref{eq:r1simp}-\eqref{eq:r3simp} in the normal form definitions $\mathbf{m}_{1},\ldots,\mathbf{m}_{36}$ and use these partially rationalized expressions for the entire physical region of phase space.
That is, only the definition of $\mathbf{m}_{32}$ involves a root
\begin{equation}
    \label{eq:rdef}
    r \equiv \sqrt{(1 + w^2 z^2) (w + z)^2 + 2 w\,z (w - z)^2 + 4 w\,z^2 (1 + w^2)}
\end{equation}
in the prefactor, which requires a non-trivial analytic continuation from
$s>4\,m^2$ to other regions of phase space on its own.
Using these definitions, it is straightforward to obtain the differential equations in the $\epsilon$-decoupled form discussed in the next section.
We used \Reduze\, \cite{vonManteuffel:2012np,Studerus:2009ye,Bauer:2000cp,fermat} to compute the integration-by-parts identities required to derive the differential equations.

\section{Integrating root-valued symbols in terms of multiple polylogarithms}
\label{sec:diffeqint}

Our normal form basis for the mixed EW-QCD Drell-Yan integrals with two massive internal lines, \eqref{eq:normformints}, is chosen to bring the associated differential equations into an
$\epsilon$-decoupled form \cite{Kotikov:2010gf,Henn:2013pwa}:
\begin{equation}\label{EqDENormal}
\ud\mathbf{m}_i = \epsilon\; \sum_{j,k} \ud \ln(l_k) \big(A^{(k)}\big)_{ij} \, \mathbf{m}_j\,,
\end{equation}
where $l_k$ are the symbol letters,
$A^{(k)}$ are matrices of rational numbers,
and $i,j=1,\ldots,36$.
It was demonstrated already in \cite{Bonciani:2016ypc} that a $\epsilon\, \ud \ln$ form does in fact exist, but their choice of the symbol letters is not optimal for our purposes.
For now, we proceed with the understanding that {\it some} algebraic symbol letters appear in Eq. \eqref{EqDENormal}, but leave their number and precise form to be determined by the analysis below.

Using $\vec{x} = (w,z)$ and $\vec{\mathbf{m}} = (\mathbf{m}_i)$, $i=1,\ldots,36$, one can give a formal solution of Eq. \eqref{EqDENormal} in terms of Chen iterated integrals as
\begin{equation}
\label{eq:formalChen}
    \vec{\mathbf{m}}(\epsilon,\vec{x})=\mathcal{P}\exp\left[\epsilon\int_{\gamma}\ud \mathbb{A}\right] \vec{\mathbf{m}}(\epsilon,\vec{x}_0),
\end{equation}
where
\begin{equation}
\ud \mathbb{A} = \sum_{k} \ud \ln(l_k) A^{(k)}\,,
\end{equation}
$\vec{\mathbf{m}}(\epsilon,\vec{x}_0)$ are the boundary constants of the master integrals at the point $\vec{x}=\vec{x}_0$, $\gamma$ is a piece-wise smooth path connecting $\vec{x}_0$ and $\vec{x}$, and the path-ordered exponential in Eq. \eqref{eq:formalChen} is defined in the usual way as an infinite series of integral operators acting to the right,
\begin{equation}
\label{eq:Pexpdef}
    \mathcal{P}\exp\left[\epsilon\int_{\gamma}\ud \mathbb{A}\right]\equiv\mathbf{1}+\epsilon\int_{\gamma}\ud \mathbb{A}+\epsilon^2\int_{\gamma}\ud \mathbb{A} \ud \mathbb{A}+\cdots\,.
    \end{equation}
In \eqref{eq:Pexpdef}, the product of two or more $\ud\ln$ terms is understood as an instruction to take the corresponding iterated integral of the kernel along the path $\gamma$. As a concrete example, let us consider a straight-line path $\tau$ on the real axis from $0$ to $x$. One can write, {\it e.g.}:
\begin{align}
    \int_\tau\ud\ln(x)\ud\ln(x+1) &= \int_{0}^x \frac{\mathrm{d}x_2}{x_2+1}\left[\int_{0}^{x_2} \frac{\mathrm{d}x_1}{x_1}\right] 
    \\
    &= {\rm Li}_2(-x) + \ln(x)\ln(x+1). \nonumber
\end{align}
Note that, in the above example, one could equally well identify the iterated integral as a Goncharov polylogarithm,
\begin{equation}
    \int_\tau \ud\ln(x)\ud\ln(x+1)=G(-1,0;x).
\end{equation}

If all symbol letters are linear, it is always possible to integrate Chen iterated integrals in closed form using $G$ functions. This is no longer true, however, if one encounters non-linear or non-rational letters. In the former case, it is often possible to choose an appropriate integration order for which the non-linear letters appear only in the final integration kernel. At this stage, one may employ the generalized weights used in Section \ref{sec:linred} to obtain concise results.
If one encounters non-rational symbol letters, the standard integration algorithms for $G$ functions cannot be applied. In many cases, a transformation can be found which simultaneously rationalizes all letters of the alphabet. Nevertheless, it can be proven that the symbol alphabet of the five most complicated integrals from \eqref{eq:normformints}, $\{\mathbf{m}_{32},\ldots,\mathbf{m}_{36}\}$, cannot be rationalized \cite{mainz,Besier:2019hqd}. Of course, it is {\it a priori} not obvious whether the functional basis for $\{\mathbf{m}_{32},\ldots,\mathbf{m}_{36}\}$ consists solely of multiple polylogarithms; at the outset, it is certainly possible that one could have to deal with a more involved space of functions.

Fortunately, our linear reducibility analysis from Section \ref{sec:linred} guarantees that standard multiple polylogarithms suffice; what we need is a better way to integrate $\epsilon\, \ud\ln$ differential equations. A clear alternative is to proceed by matching the symbol of the Chen iterated integrals to a suitable ansatz built out of logarithms and ${\rm Li}$ functions.
For rational alphabets, a method was provided in \cite{Duhr:2011zq} by Duhr, Gangl, and Rhodes to construct suitable ${\rm Li}$ function arguments, such that the functional basis contains no spurious letters. At weight one, it is clear that logarithms of the letters are admissible functions. The non-trivial step is to find a sufficiently large set of admissible ${\rm Li}_n$ arguments; once these arguments have been found, it is straightforward to construct all admissible arguments for the ${\rm Li}$ functions of depth greater than one. By considering the symbol of these functions,
\begin{equation}
\mathcal{S}\big({\rm Li}_n(f)\big)=-(1-f)\otimes \underbrace{f\otimes...\otimes f}_{(n-1)~\text{times}},
\end{equation}
we see that it is desirable to admit only those function arguments $f$ which have the property that both $f$ and $1-f$ may also be written as a power product of the symbol letters. In practice, one therefore forms power products $f$ out of the letters and tests if $1-f$ factorizes over the alphabet. The symbols of higher-depth {\rm Li} functions are more complicated and lead to additional constraints. To treat ${\rm Li}_{2,1}$, ${\rm Li}_{3,1}$, and ${\rm Li}_{2,2}$, let $\mathcal{F}$ be the union of the set of admissible ${\rm Li}_n$ function arguments and the set $\{1\}$. Then, the symbol dictates that a possible pair of arguments for ${\rm Li}_{n_1,n_2}$, $(f_i,f_j)$ such that $f_i,f_j \in \mathcal{F}$, is admissible if $1-f_i f_j$ factorizes over the alphabet.
It has proven useful for practical applications to impose further constraints on the functional basis to ensure real-valuedness and good numerical performance.
An implementation of this method written by one of us in {\tt Mathematica} has been applied successfully to various processes~\cite{vonManteuffel:2013uoa,Gehrmann:2013cxs,Gehrmann:2014bfa,Gehrmann:2015ora,vonManteuffel:2017myy,Becchetti:2019tjy}.

In the presence of square roots, we use a heuristic factorization algorithm to detect admissible function arguments.
For a given expression $g$ we are interested in factorizations of the form
\begin{equation}
\label{powerproduct}
g = c^{a_0} l_1^{a_1} l_2^{a_2} \cdots,
\end{equation}
with a rational number $c$ and $a_n\in \mathbbm{Q}$.
It is non-trivial to find such factorizations using standard computer algebra systems due to the presence of the root $r$ in the symbol letters.
We observe that the factorization \eqref{powerproduct} implies
\begin{equation}
\label{logrelations}
\ln(g) - a_0 \ln(c) - a_1 \ln(l_1) - a_2 \ln(l_2) - \ldots = 0\,.
\end{equation}
Replacing the variables by numerical samples allows us to find these relations using \emph{heuristic} integer relation finders.
To find the required factorizations, we employ the Lenstra-Lenstra-{Lov\'asz} (LLL)
algorithm~\cite{Lenstra1982} implemented in {\tt PARI/GP}~\cite{PARI2} for a parallelized {\tt C++} code written by one of us.

We would like to stress that the definition of the symbol letters is not unique. One can replace a letter by power products of letters and it is {\it a priori} unclear which choice is optimal for practical purposes.
For example, we find that out of the 17 letters presented in \cite{Bonciani:2016ypc}, only 16 combinations are actually required for the integration of the Drell-Yan integrals.
Furthermore, since we consider actual derivatives of Feynman integrals, we are not sensitive to numerical letters like 2.
In principle, one needs to include also {\it ad-hoc letters} like $-1$, $2$, etc.\ in the construction of function arguments $f$.
This is a problem occurring also in the purely rational case and does not seem to be a major obstacle in practice.
In the applications we have considered so far, including $-1$ and $2$ was sufficient.
For the case of the Drell-Yan integrals, we were able to absorb the letter $2$ by a redefinition of our symbol alphabet, as will be explained below.

In practice, we encounter two main problems specific to the case of non-rational symbol alphabets:
\begin{enumerate}
    \item[(i)] In general, one needs to allow for non-integer powers, {\it e.g.} $1/2$, $1/4$, etc.\ when forming power products $f$. Consequently, one may have to test many more expressions than in the rational case in order to construct enough function arguments to successfully ``integrate the symbol.'' In fact, without additional constraints, the inherent combinatorial complexity makes this extension of the Duhr-Gangl-Rhodes method too costly for two-loop calculations of current phenomenological interest.
    \item[(ii)] Factorization over algebraic functions is not unambiguously defined; in principle, it is not clear where to ``stop'' factorizing. Consider, for instance, the set of letters $\{\sqrt{x},\sqrt{y},x-y\}$ and note that it is {\it a priori} unclear whether one ought to factor the third letter further, {\it i.e.} $x-y=(\sqrt{x}+\sqrt{y})(\sqrt{x}-\sqrt{y})$.
\end{enumerate}
In the case of the Drell-Yan integrals, both of these problems can be tackled by the following observations: given an alphabet, we introduce the subset of rational letters, $\mathcal{L}_R$, and the subset of intrinsically non-rational letters, $\mathcal{L}_A$. Let us assume that in $\mathcal{L}_A$ we encounter a square root, $r$. We find it natural to take $r$ itself to be an element of $\mathcal{L}_A$. For a given algebraic letter with a non-trivial rational part, $l_a$, we define the conjugated letter, $\bar{l}_a$, by making the replacement $r\rightarrow -r$ in $l_a$. For each $l_a\in \mathcal{L}_A$ we observe that $\bar{l}_a$ can be written as a power product of other symbol letters; that is to say, one could exchange any letter for its conjugate without affecting the singularity structure of the alphabet. Furthermore, we observe that the product $l_a \bar{l}_a$ always factorizes over the rational part of the alphabet, $\mathcal{L}_R$. We find these observations to be quite compelling and therefore conjecture that these structural constraints hold also for other symbol alphabets of this type. 

Our conjecture is a strong one, since it implies that one can predict the form of the remaining unrationalizable symbol letters which appear in the presence of the square root $r$: knowing only the rational part of the alphabet, $\mathcal{L}_R$, and the unrationalizable square root which appears, it allows one to {\it construct} the algebraic part of the alphabet, $\mathcal{L}_A$, unambiguously. 
As we shall see, our new
insight allows for a drastic simplification of the algebraic part of the symbol alphabet relative to what was presented in Eqs. (5.13)-(5.19) of \cite{Bonciani:2016ypc} and what we were able to derive ourselves initially using the above-mentioned heuristic factorization code. 

We want to show this in some detail for the EW-QCD Drell-Yan master integrals with two massive internal lines. At the outset, after carefully considering various possibilities, we find the rational alphabet
\begin{align}
\label{eq:Rletters}
    \mathcal{L}_R&=\{1 - w, -w, 1 + w, 1 - w + w^2, 1 - z, -z, 1 + z,\nonumber\\
    &\qquad 1 - w\,z, 1 + w^2\,z, -z-w^2, z-w\}
\end{align}
and the intrinsically algebraic alphabet
\begin{align}
\label{eq:rawAletters}
    &\mathcal{L}_A=\{r, -(1 - w)(z - w)(1 - w\,z)+r\,(1 + w),
    \\
    &-(1 - w)\left(\vphantom{w^2}4 w\,z+(w + z) (1 + w\,z)\right) -r\,(1 + w), r^2-2 w\,z^2(1 - w)^2 + r\,(w + z) (1 + w\,z),
    \nonumber\\
    &r^2 (1 - z)^2 + 2 z^2 (z + w^2) (1 + w^2 z) + r\,(1 - z)(1 + z)\left(\vphantom{w^2}2 w\,z - (w + z) (1 + w\,z)\right)\},\nonumber
\end{align}
where $r=\sqrt{(1 + w^2 z^2) (w + z)^2 + 2 w\,z (w - z)^2 + 4 w\,z^2 (1 + w^2)}$, using our heuristic factorization code.
Note that $r$ already appears as a letter in Eq. \eqref{eq:rawAletters}. However, there are also two rather complicated-looking symbol letters with terms involving $r^2$.

As stated above, knowing only $r$ and $\mathcal{L}_R$, we can also construct an improved representation of the algebraic part of the alphabet by making an ansatz of the form
\begin{equation}
    l_a=q_a + r,
\end{equation}
where $q_a$ is a rational function in $w$ and $z$, and then requiring that $l_a \bar{l}_a$ factorize over $\mathcal{L}_R$. In practice, it is more convenient to directly make an ansatz of the form
\begin{align}
    (q_a+r)(q_a-r)=q_a^2-r^2=\sum_{ij} a_{ij} w^i z^j-r^2,
\end{align}
and then solve for the unknowns $a_{ij}$.
The algorithm to construct (simple) algebraic letters then reads:
\newline\newline
\begin{algorithm}[H]
 \KwIn{rational part of alphabet and square root $r$ depending on $x_i$}
 \KwResult{simplified letters}
 initialization:
 monlist=monomials of rational alphabet up to degree $n$\;
 \For{$f$ in monlist}{
  $d_i=deg(f,x_i$)\;
  polynomial ansatz: $p=\sum_i\sum_{n_i}^{d_i} a_{n_i} x_i^{n_i}$\;
  solve $f= p-r^2$ for unknown coefficients $a_{n_i}$\;
  \If{$p$ is perfect square}{
  set polynomial $q\equiv \sqrt{p}$ \;
  add $l=q+r$ to new alphabet\;}
 }
\end{algorithm}
In order to also reproduce the original letters we would need
to allow for a polynomial prefactor in front of the $r^2$ term,
which we deliberately avoid here to simplify our construction.
In practice, we run the algorithm with a degree $n$ at least
sufficiently large to be able to express the differential equations
in terms of the new alphabet.

By using this procedure up to degree $4$, we find the following
algebraic letters for the Drell-Yan integrals:
\begin{align}
\label{eq:niceAletters}
    \tilde{\mathcal{L}}_A &=\bigg\{r, \frac{1}{2}\big(2 + z - w + w\,z (w + z) + r\big), \frac{1}{2}\big(2 w^2 + z - w + w\,z (w + z) + r\big), \nonumber\\
    &\qquad\frac{1}{2}\big(-(w + z)(1 - w\,z)+r\big), \frac{1}{2}\big(-(z - w)(1 + w\,z)+r\big)\bigg\}.
\end{align}
The overall factors of $1/2$ in Eq. \eqref{eq:niceAletters} are judiciously chosen after the fact to prevent the appearance of explicit factors of $2$ in our {\rm Li} function arguments; as alluded to above, our original construction of the functional basis appended $2$ as an auxiliary letter.
The presence of this factor can also be understood in light of the factorization property: one can easily check that all letters in $\tilde{\mathcal{L}}_A$ after multiplication with their conjugate, indeed factorize over the rational part of the alphabet without the presence of the factor $2$ in this product. In order to find the correct normalization, one can run the above algorithm first adding the letter $2$ in the rational part of the alphabet.

Using our heuristic approach to algebraic function factorization, we immediately find that the non-trivial elements of $\mathcal{L}_A$ can indeed be expressed as power products of letters drawn from the improved alphabet:
\begin{align}
\label{eq:nicelettermaps2}
    &-(1 - w)(z - w)(1 - w\,z)+r\,(1 + w) =
    \\
    &\qquad \qquad \qquad\frac{2 \big(-w\big) \big(1+z\big) \big(-z-w^2\big) \big(2 + z - w + w\,z (w + z) + r\big)}{2 w^2 + z - w + w\,z (w + z) + r},\nonumber
\end{align}
\begin{align}
\label{eq:nicelettermaps3}
    &-(1 - w)\left(\vphantom{w^2}4 w\,z+(w + z) (1 + w\,z)\right) -r\,(1 + w) =
    \\
    &\quad \qquad \frac{8 \big(-w\big)^2 \big(-z\big) \big(1+z\big)^3 \big(1+w^2\, z\big) \big(2 w^2 + z - w + w\,z (w + z) + r\big)}{\big(2 + z - w + w\,z (w + z) + r\big) \big(-(w + z)(1 - w\,z)+r\big)
   \big(-(z - w)(1 + w\,z)+r\big)},\nonumber
\end{align}
\begin{align}
\label{eq:nicelettermaps4}
    &r^2-2 w\,z^2(1 - w)^2 + r\,(w + z) (1 + w\,z) =
    \\
    &\quad \frac{\big(-z\big)^2 \big(2 + z - w + w\,z (w + z) + r\big)^2 \big(2 w^2 + z - w + w\,z (w + z) + r\big)^2}{8 \big(1+z\big)^2 \big(1+w^2\, z\big)^2 \big(-(w + z)(1 - w\,z)+r\big)^2\big(-(z - w)(1 + w\,z)+r\big)^{-2}},\nonumber
\end{align}
\begin{align}
\label{eq:nicelettermaps5}
    &r^2 (1 - z)^2 + 2 z^2 (z + w^2) (1 + w^2 z) + r\,(1 - z)(1 + z)\left(\vphantom{w^2}2 w\,z - (w + z) (1 + w\,z)\right)
    \\
    &\qquad \qquad \qquad \qquad \qquad = \frac{2 \big(-z\big)^2 \big(1+w^2\, z\big)^2 \big(-(w + z)(1 - w\,z)+r\big)^2}{\big(-(z - w)(1 + w\,z)+r\big)^2}.\nonumber
\end{align}

In summary, we define the full symbol alphabet for the two-massive-line Drell-Yan integrals, $\mathcal{L} = \mathcal{L}_R \cup \tilde{\mathcal{L}}_A$, in terms of positive definite letters for physical phase space points which satisfy $-1 < w < 0$ and $w < z < -w^2$ (this component of $(w,z)$ space corresponds to part of the $s>4\,m^2$ region, see Section \ref{sec:analycont} for details). From Eqs. \eqref{eq:Rletters} and \eqref{eq:niceAletters}, we have:
\begin{align}
\label{eq:fullalphabet}
\mathcal{L} &=
    \{l_1,\ldots,l_{16}\} 
    \notag \\
    &= \bigg\{1 - w, -w, 1 + w, 1 - w + w^2, 1 - z, -z, 1 + z,1 - w\,z, 1 + w^2\,z, -z-w^2,
    \nonumber\\
    &\quad z-w, r, \frac{1}{2}\big(2 + z - w + w\,z (w + z) + r\big), \frac{1}{2}\big(2 w^2 + z - w + w\,z (w + z) + r\big), 
    \nonumber\\
    &\quad \frac{1}{2}\big(-(w + z)(1 - w\,z)+r\big), \frac{1}{2}\big(-(z - w)(1 + w\,z)+r\big)\bigg\}.
\end{align}
Most importantly, our improved representation of the alphabet effectively solves the problem of finding power products of high degree, since we observe in practice that we do not need to consider square roots of any letter in $\mathcal{L}$. For the integration of the EW-QCD Drell-Yan master integrals with two massive internal lines through to weight four, it was enough to consider power products of symbol letters derived above up to total degree $9$.

The construction above employed a representation with only
a single root-valued leading singularity.
In Appendix \ref{sec:multiroot}, we consider a one-loop integral which involves five different root-valued leading singularities. We demonstrate that the algebraic part of the symbol alphabet can be constructed without any reparametrization by generalizing the procedure above to the case of multiple roots.

\section{Analytic continuation and optimization of the functional bases}
\label{sec:analycontandi0}

In this section, we review the salient features of the $(w,z)$ representation for the mixed EW-QCD corrections to Drell-Yan production introduced in Section \ref{sec:dydefs} above, as well as subtleties one encounters when analytically continuing multiple polylogarithms. Our primary goal in Section \ref{sec:analycont} is to motivate the analysis of Section~\ref{sec:i0}, where we show how we avoid all explicit analytic continuations and $+i\,0$ prescriptions by partitioning the physical phase space and finding several solutions to the differential equations in terms of well-behaved {\rm Li} functions valid in judiciously chosen regions.
However, we also find it useful to review the fundamentals of analytic continuation of Feynman integrals as well as some details relevant to our specific representation.

\subsection{Analytic continuation}
\label{sec:analycont}

Feynman's $+i\,0$ prescription for the propagators determines the value of a given Feynman integral in a specific region of phase space unambiguously.
In principle, one could imagine to solve the Feynman integral in each region of phase space separately.
Alternatively, one can try to solve the integral in one region and then use the solution to obtain a result for it in a neighboring region by analytic continuation.
The latter method is of particular interest for the method of differential equations, since it typically involves regularity conditions in some regions of phase space, and one needs to transport this knowledge to the region of interest.

If we want to continue a specific representation of the solution based on just the solution itself (without reference to the original Feynman integral), we need to make sure Feynman's $+i\,0$ prescription is maintained by appropriate complex values of the kinematic parameters. It is is essential to observe that the analytic continuation is along a path in complex phase space and that the $+i\,0$ prescriptions must be respected for all points along this path, not just for the start and endpoint.
This is in general non-trivial and needs to be checked for the representation at hand.

What is commonly referred to as ``analytic continuation'' in the physics literature should really be regarded as a two-step procedure in general:
\begin{enumerate}
    \item[(i)] The actual analytic continuation in the mathematical sense: given a solution for one region, derive a solution for a connected region in some representation.
    \item[(ii)] A possible change of functional representation in the new region such that no explicit $+i\,0$ prescription is necessary.
\end{enumerate}
We will now work out the details for our current application.

From the second Symanzik polynomials of our Drell-Yan master integrals, we see that their Euclidean region is given by $s<0$, $t<0$ and $m^2>0$.\footnote{For non-planar topologies with four massless external legs, also cuts in $u$ must be taken into account, which may actually prevent the existence of a Euclidean region with $s+t+u=0$~\cite{Tausk:1999vh}.}
Considering other regions of phase space, we observe that the Feynman propagator prescription can effectively be implemented by the replacements $s \rightarrow s + i\, 0$, $t \rightarrow t + i\, 0$, since these are external scales, and $m^2 \rightarrow m^2 - i\, 0$, since this is an internal scale.
The $+i\,0$ prescription is relevant for $s$ ($t$) whenever $s$ ($t$) is positive.
For the discussion of $w$ and $z$, we will see that it is sufficient to view $m^2$ as being normalized to 1 without any imaginary part.

Let us focus on the $(w,z)$ parametrization of our integrals
in the physical region of phase space, $s>0$, $t<0$ and $m^2>0$.
It was noted in \cite{Bonciani:2016ypc} that the $(w,z)$ representation has a point of non-analyticity at the physical two-mass threshold $s = 4\,m^2$ and a rather different character depending on whether $s$ is above or below this threshold. We find it natural to adopt the definitions
\begin{align}
\label{eq:wdefabove}
    w &= -\frac{\sqrt{s}-\sqrt{s-4\, m^2}}{\sqrt{s}+\sqrt{s-4\, m^2}},
\\
\label{eq:zdefabove}
    z &= -\frac{\sqrt{4 \,m^4-t\,(s-4\, m^2)}-\sqrt{-t(s-4\, m^2)}}{\sqrt{4\, m^4-t\,(s-4\, m^2)}+\sqrt{-t(s-4\, m^2)}}\,.
\end{align}

\begin{figure}[!t]
\centering
\begin{align}
\includegraphics[scale=0.52]{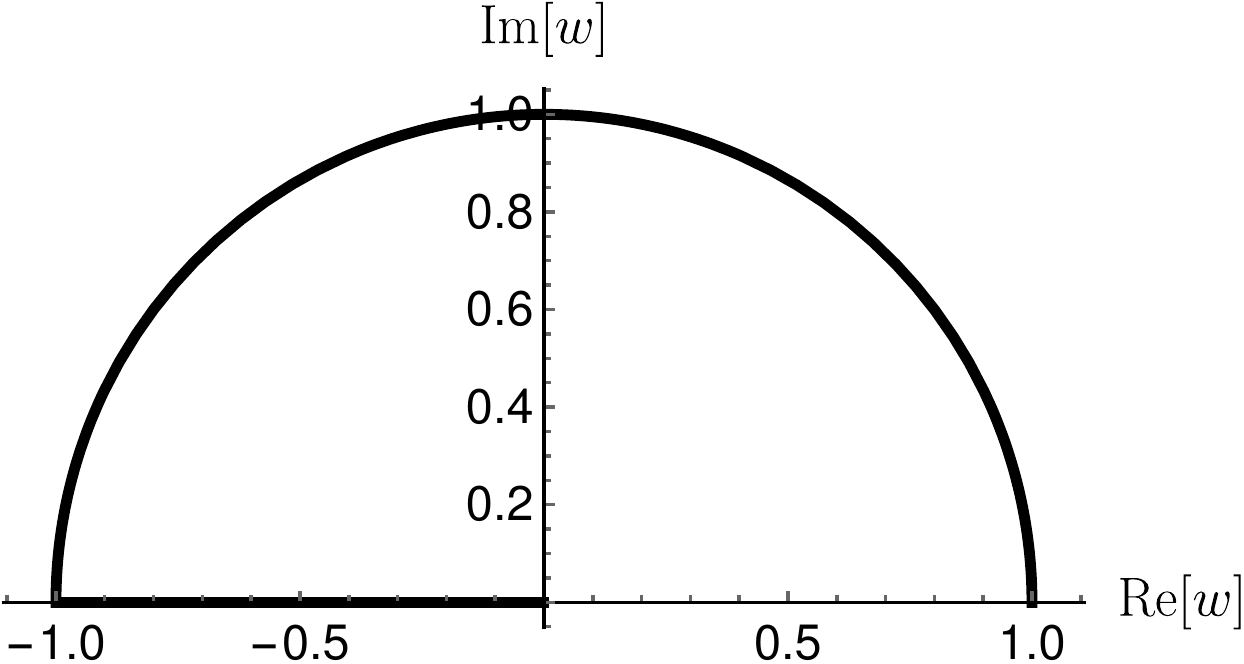} \qquad \qquad \includegraphics[scale=0.52]{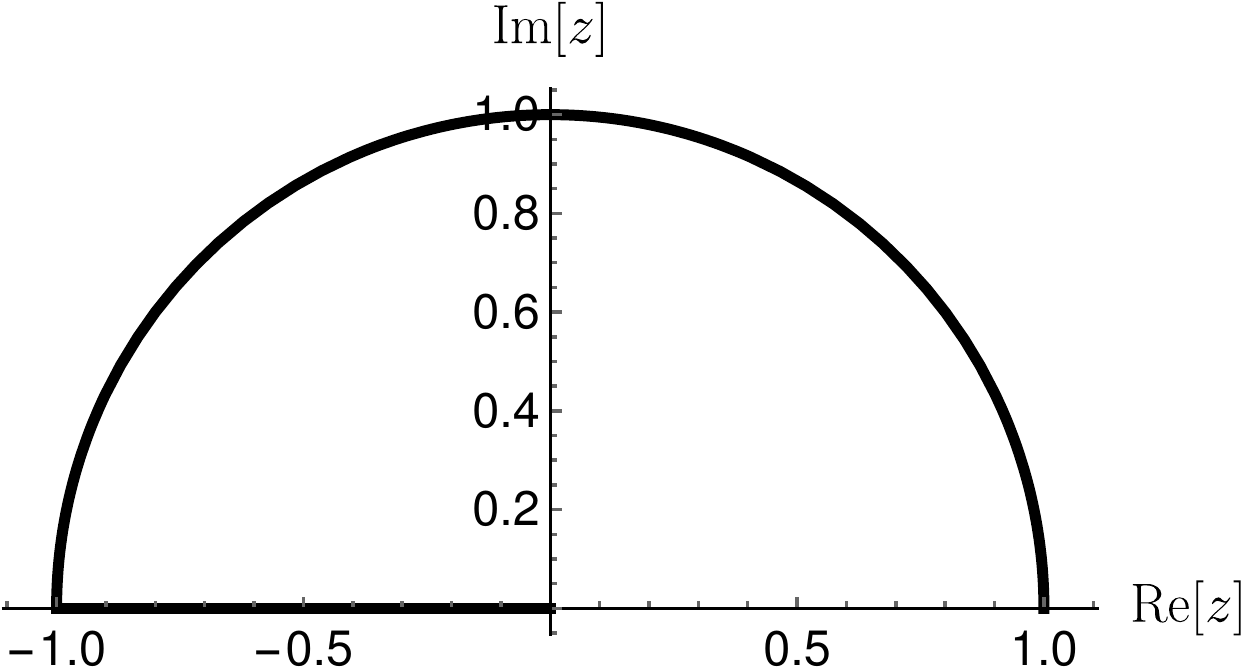}\nonumber
\end{align}
\caption{The $(w,z)$ representation of the physical phase space has two main components which merge at the point of non-analyticity $w = z = -1$, corresponding to the two-mass threshold at $s = 4\,m^2$. The line segments on the negative real axes of $w$ space (left) and $z$ space (right) are half-open intervals which correspond to the above-threshold region; the points $w = 0$ and $z = 0$ are approached in the $s \rightarrow \infty$ limit. The semi-circular domains in the upper $w$ and $z$ half-planes correspond to the below-threshold region; the point $w = 1$ corresponds to the phase space boundary point $s = 0$ and $z = 1$ corresponds to the phase space boundary point $s = -t = 2\,m^2$. Note that the upper endpoint of $z$  depends parametrically on $w$ both above and below the two-mass threshold.
}
\label{fig:wzdomains}
\end{figure}

We can see from the above that $w$ and $z$ are real variables which satisfy $-1 < w < 0$ and $-1 < z < -w^2$ in the region $s>4\,m^2$.
With our choices \eqref{eq:wdefabove} and \eqref{eq:zdefabove}, we find that the Feynman prescription implies $w \rightarrow w + i\, 0$ and $z \rightarrow z + i\, 0$ in this region.
On the other hand, both $w$ and $z$ become pure phases in the region $0<s<4\,m^2$. Therefore, it makes sense to explicitly extract their real and imaginary parts,
\begin{align}
\label{eq:wdefbelow}
    w &= 1 - \frac{s}{2\, m^2} + \sqrt{1-\left(1 - \frac{s}{2\, m^2}\right)^2}\,i\,,
\\
\label{eq:zdefbelow}
 z &= -1 - \frac{t\, (4\, m^2 - s)}{2\, m^4} + \sqrt{1-\left(-1 - \frac{t\, (4\, m^2 - s)}{2\, m^4}\right)^2}\,i\,,
\end{align}
to emphasize that, in the below-threshold region, the imaginary parts of $w$ and $z$ are fixed in terms of the real parts of $w$ and $z$.  In particular, we can deduce from Eqs. \eqref{eq:wdefbelow} and \eqref{eq:zdefbelow} that $\mathrm{Re}[w]$ and $\mathrm{Re}[z]$ satisfy $-1 < \mathrm{Re}[w] < 1$ and $-1 < \mathrm{Re}[z] < 1-2\left(\mathrm{Re}[w]\right)^2$ in the below-threshold region. A visualization of the full physical phase space in the $(w,z)$ representation is given in Figure \ref{fig:wzdomains}.

\begin{figure}[!t]
\centering
\begin{align}
\includegraphics[scale=0.5]{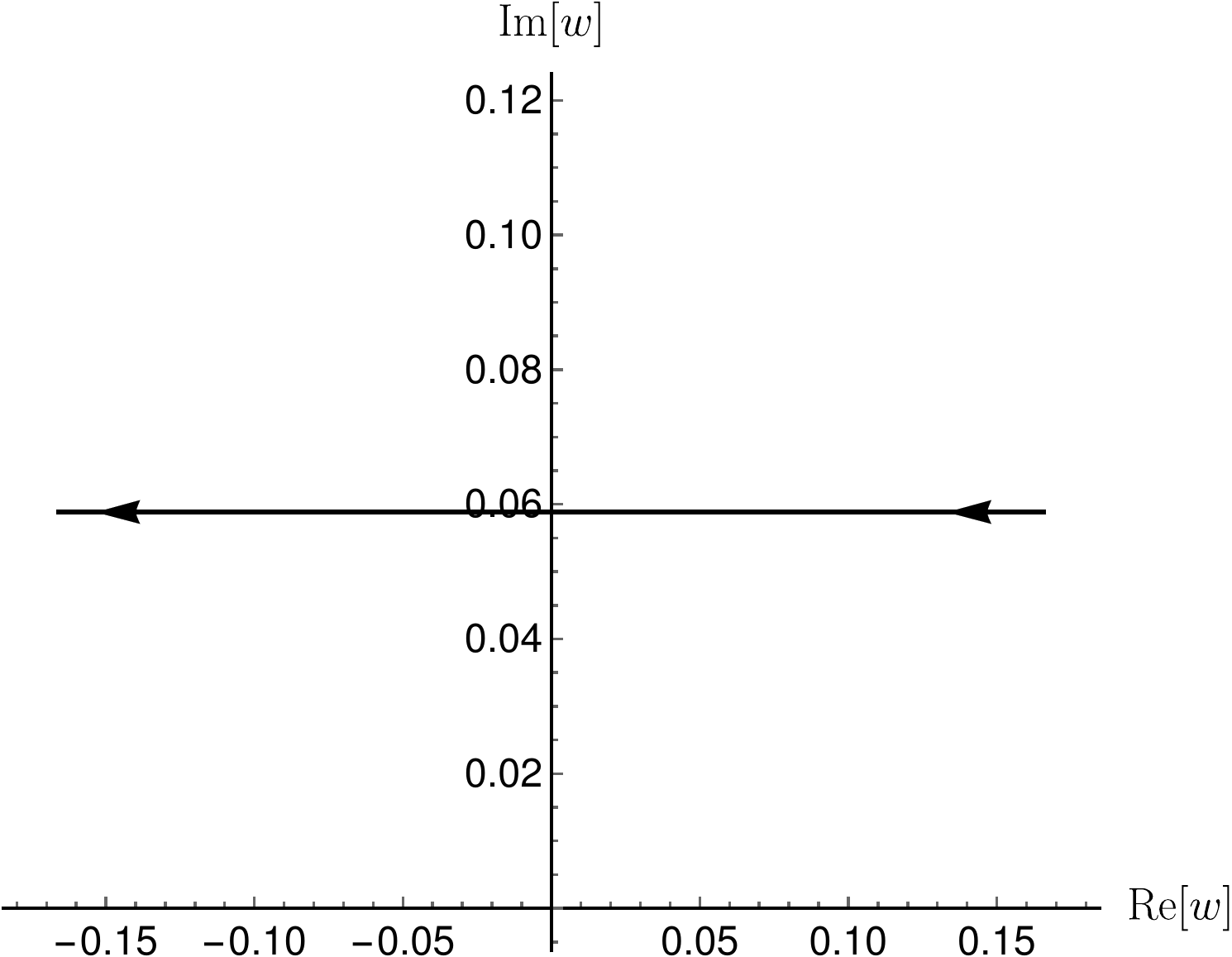} \qquad \qquad \includegraphics[scale=0.5]{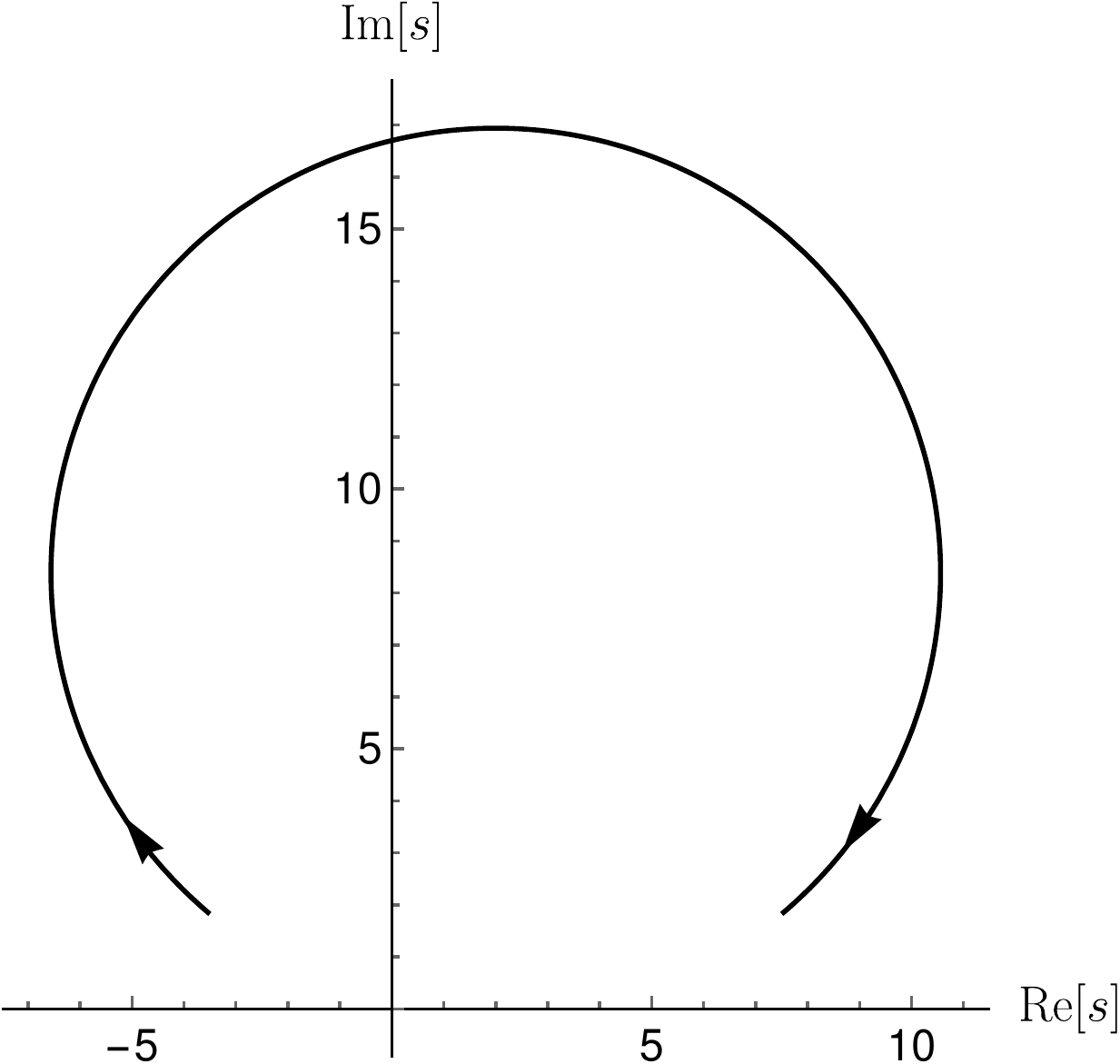}\nonumber
\end{align}
\caption{The path in $w$ space prescribed by Eq. \eqref{eq:analycontpath} for $w_{(0)} = 1/6$ and $\delta = 1/17$ (left) induces a nearly semi-circular path in $s$ space (right) which respects the $+i\,0$ prescription for $s$. Note that, for fixed $t$ and $m^2$, this straight-line path in $w$ determines a nearly straight-line path in $z$ as well.
}
\label{fig:spath}
\end{figure}

In the following, we will study how to analytically continue solutions between different regions in $s$.
In practice, we work with simple straight-line paths in the complex $(w,z)$ space, checking after the fact that the chosen paths of analytic continuation always preserve the $+i\, 0$ prescription for the original kinematic variables. In what follows, we carefully go through typical elementary examples of analytic continuation in order to clearly illustrate the subtleties for our ${\rm Li}$ functions which one must be wary of to avoid introducing errors.
\begin{figure}[!t]
\centering
\begin{align}
\includegraphics[scale=0.45]{wpath} \qquad \includegraphics[scale=0.45]{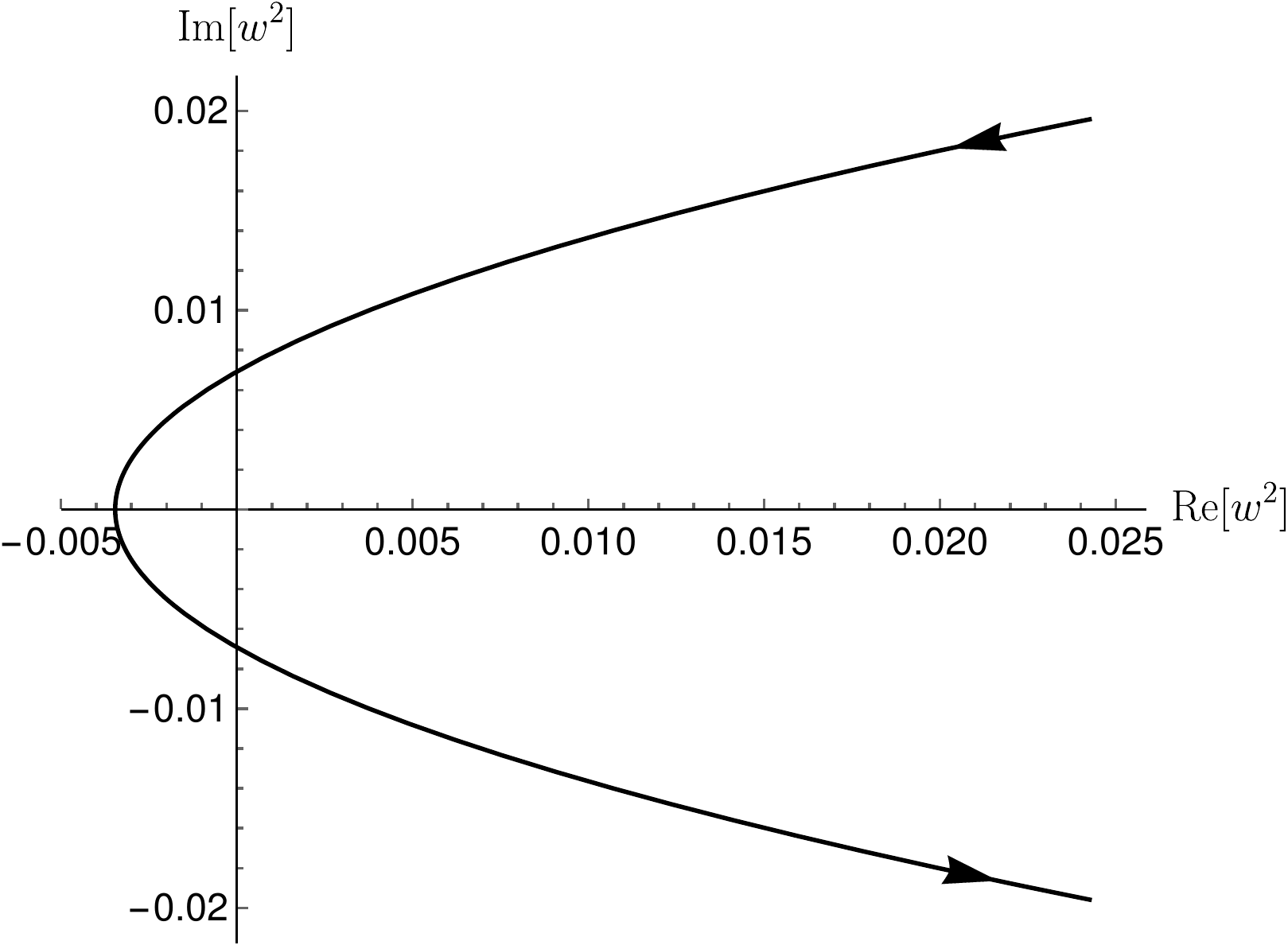}\nonumber
\end{align}
\caption{The path in $w$ space prescribed by Eq. \eqref{eq:analycontpath} for $w_{(0)} = 1/6$ and $\delta = 1/17$ (left) induces an essentially hyperbolic path in $w^2$ space (right) which crosses the negative real axis at $-\delta^2$.
}
\label{fig:w2path}
\end{figure}

First, let us illustrate the importance of taking into account the complete path of analytic continuation rather than just the start and endpoint.
Consider the analytic continuation of $\ln\big(w^2\big)$ along a straight-line path from $w_i = w_{(0)}+i\,\delta$ to $w_e = -w_{(0)}+i\,\delta$, for real $w_{(0)}$ and $\delta$ such that $0 < w_{(0)} < 1$ and $0 < \delta < w_{(0)}$ (see Figure \ref{fig:spath}).
Such a path allows one to connect solutions for $s<0$ with solutions for $s>4\,m^2$.
Naively, one might be tempted to erroneously implement the analytic continuation as $\ln\big(w_i^2\big) = \ln\big(w_{(0)}^2\big) = \ln\big((-w_{(0)})^2\big) = \ln\big(w_e^2\big)$ and incorrectly conclude that the analytic continuation is trivial along the chosen path. The problem here is that the logarithm is a multi-valued function and one must therefore carefully check whether or not the specified path of analytic continuation forces the polynomial function argument to cross the branch cut of the logarithm on the negative real axis, $(-\infty,0)$. In the absence of any branch cut crossings, knowledge of the endpoint of the path is sufficient. However, in the presence of one or more branch cut crossings, the function leaves its principal Riemann sheet and one must add an appropriate {\it monodromy} contribution, taking into account the details of the path of analytic continuation.

Let us spell out in detail how to correctly analytically continue solutions valid for different values of $s$.
We parametrize our chosen path in $w$ space as
\begin{equation}
\label{eq:analycontpath}
    w(v) = (1 - 2 v)\,w_{(0)}+\delta\, i,
\end{equation}
with a parameter $v \in [0,1]$. We then have
\begin{equation}
\label{eq:paramw2}
    w^2(v) = (1-2\,v)^2 w_{(0)}^2-\delta^2+2\,(1-2\,v)\, \delta \,w_{(0)}\, i
\end{equation}
for the argument of $\ln\big(w^2\big)$. As depicted in Figure \ref{fig:w2path}, our path in $w$ space takes $w^2(v)$ from just above the point $w_{(0)}^2-\delta^2$ on the positive real axis to just below the point $w_{(0)}^2-\delta^2$ and, crucially, it passes through the negative real axis at $w(1/2) = \delta\, i$. The monodromy contribution to $\ln\big(w^2\big)$ in this case is well-known to be simply $2 \pi i$, due to the fact that our path induces just one counter-clockwise branch cut crossing. Therefore, the analytically continued function is $\ln\big(w^2\big) + 2 \pi i$ in a neighborhood of the endpoint of our chosen path.
This conclusion may be quickly checked by rewriting the function $\ln\big(w^2\big)$ as $2 \ln(w)$ in the Euclidean region where $0 < \mathrm{Re}[w] < 1$ before carrying out the analytic continuation along our chosen path. When the function is viewed in this alternative way, no branch cut crossing is induced and one can simply rewrite the function to explicitly extract its imaginary part: for $\mathrm{Im}[w] > 0$, $2 \ln(w) = 2\big(\ln(-w)+\pi i\big) = 2 \ln(-w)+2\pi i = \ln\big(w^2\big)+2\pi i$.
In the two-step picture mentioned above we thus have
\begin{alignat}{2}
\ln\big(w^2\big)
 & \underset{\scriptscriptstyle \text{(i)}\,w(v)}{\xrightarrow{\hspace{.8 cm}}}
   \ln\big(w^2\big) + 2 \pi i
 &&\underset{\scriptscriptstyle \text{(ii)}}{=}
   \ln\big(w^2\big) + 2 \pi i
   \\
\intertext{or}
\ln\big(w^2\big) = 2\ln\big(w\big)
 &\underset{\scriptscriptstyle \text{(i)}\,w(v)}{\xrightarrow{\hspace{.8 cm}}}
   2\ln(w)
 &&\underset{\scriptscriptstyle \text{(ii)}}{=}
   2 \ln(-w) + 2 \pi i = \ln\big(w^2\big) + 2 \pi i
\end{alignat}
Depending on the type of representation we consider, either the analytic continuation (i) or the rewriting of the function to be independent of $+i\,0$ prescriptions (ii) is trivial in this example.

The relevant monodromy contributions become more complicated for higher-weight ${\rm Li}$ functions. Already for ${\rm Li}_2$, a new feature emerges: moving across its branch cut on the positive real axis, $(1,\infty)$, onto a Riemann sheet other than the principal one actually exposes the existence of a hidden branch point at 1. To see this, consider Euler's identity for the dilogarithm of $1-w^2$,
\begin{equation}
\label{eq:Eulerdilog}
    {\rm Li}_2\big(1-w^2\big) = -{\rm Li}_2\big(w^2\big) - \ln\big(w^2\big)\ln\big(1-w^2\big)+\frac{\pi^2}{6}.
\end{equation}
The representation furnished by the right-hand side of \eqref{eq:Eulerdilog} and the above discussion of $\ln\big(w^2\big)$ make it clear that ${\rm Li}_2\big(1-w^2\big)$ has non-trivial monodromy for the path $w(v)$ defined above in Eq.\ \eqref{eq:analycontpath}. For the path $w(v)$, we find:
\begin{align}
\label{eq:analycontdilog}
    {\rm Li}_2\big(1-w^2\big) &= -{\rm Li}_2\big(w^2\big) -\ln\big(w^2\big)\ln\big(1-w^2\big)+\frac{\pi^2}{6}
    \nonumber \\
    &\underset{\scriptscriptstyle w(v)}{\longrightarrow} -{\rm Li}_2\big(w^2\big) -\Big(\ln\big(w^2\big)+2\pi i\Big)\ln\big(1-w^2\big)+\frac{\pi^2}{6}
    \nonumber \\
    &\qquad = {\rm Li}_2\big(1-w^2\big) - 2\pi i\ln\big(1-w^2\big),
\end{align}
This implies that ${\rm Li}_2\big(1-w^2\big)$ picks up a monodromy contribution of $- 2\pi i\ln\big(1-w^2\big)$ due to the branch point at $w=1$, despite the fact the function is continuous at that point on the principal sheet.

When considering general analytic continuations of ${\rm Li}_{n_1,\ldots, n_k}$ functions, one must take into account all function arguments to obtain the monodromy contributions as one moves along the path of analytic continuation. Building on the work of Goncharov \cite{Goncharov:2001iea}, it was shown in \cite{Duhr:2012fh} how one can easily compute the monodromy of an arbitrary ${\rm Li}$ function in terms of monodromies of simple logarithms using the coproduct.
In particular, the presence of a monodromy contribution can be detected by studying the first entry of the symbol.
For our master integrals, the sheer number of distinct branch cut crossings which can arise from the arguments of the various ${\rm Li}$ functions renders the analytic continuation between different regions rather involved in practice.
While our final goal is to obtain representations in terms of well-behaved ${\rm Li}$ functions for kinematic regions, we find it convenient to also employ auxiliary representations for the purpose of analytic continuations.

If available, a representation in terms of $G$ functions with the kinematic variables in the argument avoids many of the subtleties involved in the continuation described above.
For the integrals with unrationalizable alphabets, however, this is not an option.
For such cases, we find it useful to use expansions around regular and singular points~\cite{Lee:2017qql}  for the continuation.
For this part of the analysis, we do not include the unrationalizable root of $\mathbf{m}_{32}$ in \eqref{eq:dlogform} in the definition of our basis in order to work with rational differential equations.
Using high-precision numerical evaluations for two expansion points with an overlapping region of convergence and the PSLQ algorithm \cite{PSLQ}, one can effectively transport analytic integration constants.

What has been discussed so far allows us to construct a functional basis for a given region of phase space, integrate the symbol in terms of these functions, and relate solutions for different regions by analytic continuation in order to fix the integration constants.
We will now discuss how to construct domain-restricted but well-behaved ans{\"a}tze of multiple polylogarithms, which don't require an explicit $+i\,0$ prescription and perform well numerically.

\subsection{Optimizing the bases of multiple polylogarithms
}
\label{sec:i0}

Starting from a Duhr-Gangl-Rhodes basis of multiple polylogarithms for the integrals $\vec{\mathbf{m}}$, we wish to remove ${\rm Li}$ functions which have suboptimal analytic properties for physical kinematics ($s > 0$).
In fact, one finds that even the purely rational function arguments allowed by the letters \eqref{eq:Rletters} are non-trivial to treat systematically and, we will therefore restrict the discussion in this section to this rational subset.
The main idea is to make a partition of the physical phase space into regions $\mathcal{D}_i$ such that, inside each region, a solution to the differential equations may be constructed out of ${\rm Li}$ functions which never diverge or move off of their principal Riemann sheets for arbitrary phase space trajectories contained in $\mathcal{D}_i$. Although our primary goal is to show how to one can avoid supplying explicit $+i\,0$ prescriptions for $w$ and $z$, we also find it convenient to impose further aesthetic criteria on our polylogarithmic bases in order to simplify our results. For example, we find it useful in each ansatz to give precedence to those functions which do not involve the symbol letter $1 - w + w^2$, since this ensures that the master integrals which do not depend on $1 - w + w^2$ are manifestly free of $1 - w + w^2$ at the level of functions.\footnote{The presence or absence of the letter $1 - w + w^2$ for particular master integrals is linked to the presence or absence of a one-mass threshold at $s = m^2$.}

The first step of our analysis is to study the letters of the rational alphabet, $\mathcal{L}_R$,
above and below the physical two-mass threshold at $s = 4\, m^2$. To proceed, we must determine under what conditions the logarithms of the letters either diverge or move off of their principal Riemann sheets. We find that, in practice, it is simplest to use Eqs. (\ref{eq:wdefabove}) and (\ref{eq:zdefabove}) above threshold, Eqs. (\ref{eq:wdefbelow}) and (\ref{eq:zdefbelow}) below threshold, and the {\tt Mathematica} function {\tt Reduce} to work out how various constraints on polynomials of $w$ and $z$ map back to conditions on the original and more familiar kinematic variables, $s$, $t$, and $m^2$. We find that {\tt Reduce} works in an efficient way when the system of inequalities to be reduced is formulated in terms of real-valued parameters and the solution to the system does not involve roots of high-degree polynomials. Fortunately, these assumptions are always satisfied for the ${\rm Li}$ functions which have symbols built out of letters from $\mathcal{L}_R$.\footnote{ In Section \ref{sec:DYresult}, the method described in this section is also used to prepare an ansatz involving ${\rm Li}$ functions which have symbols built out of letters from both $\mathcal{L}_R$ and $\mathcal{L}_A$.}
To check whether it is possible to live without a $+i\,0$ prescription, we must first understand for what values of $s$, $t$, and $m^2$ both the real and imaginary parts of the letters vanish simultaneously and for what values of $s$, $t$, and $m^2$ the imaginary parts of the letters vanish while their real parts happen to be negative:
\begin{itemize}
    \item $\ln(l_1) = \ln(1-w)$ diverges at the phase space boundary point $s = 0$.
    \item $\ln(l_2) = \ln(-w)$ has no issues for $m^2 > 0$, except at the phase space boundary point $s = 0$ where it becomes ill-defined.
    \item $\ln(l_3) = \ln(1+w)$ diverges at the two-mass threshold $s = 4\, m^2$.
    \item $\ln(l_4) = \ln(1-w+w^2)$ diverges at the one-mass threshold $s = m^2$.
    \item $\ln(l_5) = \ln(1-z)$ diverges at the
    phase space boundary point $s = -t = 2\, m^2$.
    \item $\ln(l_6) = \ln(-z)$ has no issues for $m^2 > 0$, except at the phase space boundary
    point $s = -t = 2\, m^2$ where it becomes ill-defined.
    \item $\ln(l_7) = \ln(1+z)$ diverges at the phase space boundary where $t = 0$ and at the two-mass threshold $s = 4\, m^2$.
    \item $\ln(l_8) = \ln(1-w\, z)$ diverges at the two-mass threshold $s = 4\, m^2$.
    \item $\ln(l_9) = \ln(1+w^2\, z)$ diverges at the phase space boundary $s = -t$ for $0 \leq s \leq 2\, m^2$ and at the two-mass threshold $s = 4\, m^2$.
    \item $\ln(l_{10}) = \ln(-z-w^2)$ diverges at the phase space boundary $s = -t$ for $s \geq 2\, m^2$ and, in particular, at $s = 4\, m^2$. 
    \item $\ln(l_{11}) = \ln(z-w)$ diverges when $t = -m^2$ for $s \geq m^2$ and, in particular, at $s = 4\, m^2$. $\ln(l_{11})$ is also not analytic at $s = -\frac{4\, m^2 t}{m^2 - t}$ for $-m^2 \leq t \leq 0$.
\end{itemize}

The next step is to determine from the above data how many regions of the physical phase space it makes sense to consider separately. In a sense, this step is the most non-trivial because different choices besides the one we ultimately make are possible.
Our choices are guided by imposing analytic properties of the integrals onto the basis functions themselves.
Due to the absence of $u$ dependence in our planar double boxes, we expect them to have a regular limit as $u$ approaches $0$. We see from the above that, depending on whether $s$ is less than or greater than $2\,m^2$, regularity as $u \rightarrow 0$ suggests the absence, respectively, of $\ln(l_9)$ or $\ln(l_{10})$ at weight one. Of course, we wish to achieve that also the higher weight basis functions lack logarithmic singularities as $l_9 \rightarrow 0$ for $0 < s < 2\, m^2$ and $l_{10} \rightarrow 0$ for $s > 2\, m^2$. This can be achieved by choosing the point $s = 2\, m^2$ as a region boundary and then imposing an appropriate first-entry condition \cite{Gaiotto:2011dt,Dixon:2011ng,CaronHuot:2011ky,Duhr:2012fh} on the symbol to select suitable functions. In other words, we remove from consideration all ${\rm Li}$ functions which have the letter $l_9$ in the first entry of their symbols in $0 < s < 2\, m^2$ regions and all functions which have the letter $l_{10}$ in the first entry of their symbols in $s > 2\, m^2$ regions. All of the other letters exhibit problems only at the physical one- and two-mass thresholds or at other exceptional points on the boundary of the physical phase space like $s = 0$. The only logarithm with a spurious singularity in the physical region is $\ln(l_{11})$, which can be dealt with by imposing an additional first-entry condition on $l_{11}$ in all regions.

Ultimately, we find it natural to partition the physical phase space into four regions:
\begin{align}
   \mathcal{D}_a &= \left\{(s,t) ~|~ s > 4\, m^2,\,-s < t < 0\right\}
   \nonumber \\
\label{eq:regiona}
   &= \left\{(w,z) ~|~ -1 < w < 0,\,-1 < z < -w^2\right\},
   \\
   \mathcal{D}_{b_1} &= \left\{(s,t) ~|~ 2\, m^2 < s < 4\, m^2,\,-s < t < 0\right\}
   \nonumber \\
\label{eq:regionb1}
   &= \left\{(w,z) ~|~ -1 < \mathrm{Re}[w] < 0,\,-1 < \mathrm{Re}[z] < 1-2\left(\mathrm{Re}[w]\right)^2\right\},
   \\
   \mathcal{D}_{b_2} &= \left\{(s,t) ~|~ m^2 < s < 2\, m^2,\,-s < t < 0\right\},
   \nonumber \\
\label{eq:regionb2}
   &= \left\{(w,z) ~|~ 0 < \mathrm{Re}[w] < 1/2,\,-1 < \mathrm{Re}[z] < 1-2\left(\mathrm{Re}[w]\right)^2\right\},
   \\
   &{\rm and}\nonumber
   \\
   \mathcal{D}_{b_3} &= \left\{(s,t) ~|~ 0 < s < m^2,\,-s < t < 0\right\}
   \nonumber \\
\label{eq:regionb3}
   &= \left\{(w,z) ~|~ 1/2 < \mathrm{Re}[w] < 1,\,-1 < \mathrm{Re}[z] < 1-2\left(\mathrm{Re}[w]\right)^2\right\},
\end{align}
where, as suggested by Eqs. \eqref{eq:wdefbelow} and \eqref{eq:zdefbelow}, we have eliminated $\mathrm{Im}[w]$ and $\mathrm{Im}[z]$ below the two-mass threshold by exploiting the fact that $w$ and $z$ become pure phases:
\begin{equation}
\label{eq:imasre}
    \mathrm{Im}[w] = \sqrt{1-\left(\mathrm{Re}[w]\right)^2}\qquad{\rm and}\qquad \mathrm{Im}[z] = \sqrt{1-\left(\mathrm{Re}[z]\right)^2}\,.
\end{equation}
Even for the twelve two-mass master integrals which have rational symbols in the $(w,z)$ representation, it is not obvious that it suffices to consider a partition of the physical phase space into just four separate regions. Fortunately, no further subdivisions are necessary\footnote{This conclusion remains unchanged if one passes from $\mathcal{L}_R$ to the full alphabet.} and it is even possible to consistently impose stronger constraints in each region on the subset of ${\rm Li}$ functions which survive our first-entry cuts.

Our basis of ${\rm Li}$ functions in region $\mathcal{D}_a$ may be further refined along the lines described in \cite{vonManteuffel:2017myy}, where the EW-QCD Drell-Yan master integrals with a single massive line were evaluated in the physical region for the first time. Consider, for example, the set of 192 ${\rm Li}_n$ arguments consistent with our first-entry conditions in this region:
\begin{align}
\label{eq:linargsforslarge}
    \Bigg\{&l_1,l_2,-l_2,l_3,l_4,l_5,l_6,-l_6,l_7,l_8,l_9,\frac{1}{l_9},\frac{1}{l_8},\frac{1}{l_7},\frac{1}{l_6},-\frac{1}{l_6},\frac{1}{l_5},\frac{1}{l_4},\frac{1}{l_3},\frac{1}{l_2},-\frac{1}{l_
   2},\frac{1}{l_1},-l_1 l_2,l_1 l_3,\frac{l_1}{l_8},
   \nonumber\\
   &\frac{l_1}{l_4},\frac{l_1}{l_2},l_2^2,l_2 l_6,-\frac{l_2}{l_4},-\frac{l_2}{l_3},l_3 l_4,\frac{l_3}{l_9},\frac{l_3}{l_8},l_5
   l_7,\frac{l_5}{l_8},\frac{l_5}{l_6},l_6^2,-\frac{l_6}{l_7},\frac{l_7}{l_9},\frac{l_7}{l_8},-\frac{l_{11}}{l_8},\frac{l_{11}}{l_8},-\frac{l_{10}}{l_9},\frac{l_8}{l_9},\frac{l_9}{l_8},\frac{l_{11}}{l_7},
   \nonumber\\
   &-\frac{l_{10}}{l_7},\frac{l_8}{l_7},\frac{l_9}{l_7},-\frac{l_7}{l_6},-\frac{l_{11}}{l_6},\frac{l_{10}}{l_6},\frac{1}{l_6^2},\frac{l_6}{l_5},-\frac{l_{11}}{l_5},\frac{l_8}{l_5},\
   \frac{1}{l_5 l_7},-\frac{l_{11}}{l_3},\frac{l_8}{l_3},\frac{l_9}{l_3},\frac{1}{l_3 l_4},-\frac{l_3}{l_2},-\frac{l_4}{l_2},\frac{l_{11}}{l_2},\frac{1}{l_2 l_6},
   \nonumber\\
   &\frac{1}{l_2^2},\frac{l_2}{l_1},\frac{l_4}{l_1},\frac{l_{11}}{l_1},\frac{l_8}{l_1},\frac{1}{l_1 l_3},-\frac{1}{l_1 l_2},\frac{l_1^2}{l_4},\frac{l_1^2}{l_2},\frac{l_1
   l_2}{l_4},\frac{l_1 l_3}{l_9},-\frac{l_1 l_6}{l_8},-\frac{l_1 l_{11}}{l_9},\frac{l_1 l_8}{l_9},\frac{l_1 l_6}{l_5},\frac{l_1}{l_2 l_5},-\frac{l_1}{l_2^2},l_2^3,l_2^2
   l_6,
   \nonumber\\
   &\frac{l_2^2}{l_4},-\frac{l_2 l_5}{l_8},-\frac{l_2 l_6}{l_8},-\frac{l_2 l_7}{l_9},\frac{l_2 l_7}{l_8},\frac{l_2 l_{11}}{l_8},-\frac{l_2 l_{10}}{l_8},\frac{l_2 l_8}{l_9},-\frac{l_2
   l_7}{l_3},-\frac{l_2 l_8}{l_3},\frac{l_3 l_6}{l_8},-\frac{l_3 l_6}{l_7},-\frac{l_6 l_{11}}{l_8},\frac{l_6 l_{10}}{l_9},
   \nonumber\\
   &-\frac{l_{11}}{l_6 l_8},\frac{l_{10}}{l_6 l_9},-\frac{l_{10}}{l_4
   l_7},-\frac{l_7}{l_3 l_6},-\frac{l_{11}}{l_3 l_6},\frac{l_8}{l_3 l_6},-\frac{l_{11}}{l_3 l_5},-\frac{l_3}{l_2 l_8},-\frac{l_3}{l_2 l_7},\frac{l_{11}}{l_2 l_9},\frac{l_{11}}{l_2
   l_8},\frac{l_9}{l_2 l_8},\frac{l_{11}}{l_2 l_7},\frac{l_8}{l_2 l_7},
   \nonumber\\
   &-\frac{l_9}{l_2 l_7},-\frac{l_8}{l_2 l_6},\frac{l_{11}}{l_2 l_5},-\frac{l_8}{l_2
   l_5},\frac{l_4}{l_2^2},-\frac{l_{10}}{l_2^2},\frac{1}{l_2^2 l_6},\frac{1}{l_2^3},-\frac{l_2^2}{l_1},\frac{l_2 l_5}{l_1},\frac{l_5}{l_1 l_6},\frac{l_{10}}{l_1 l_8},\frac{l_9}{l_1
   l_8},\frac{l_{11}}{l_1 l_7},-\frac{l_{11}}{l_1 l_6},-\frac{l_8}{l_1 l_6},
   \nonumber\\
   &\frac{l_{10}}{l_1 l_3},\frac{l_9}{l_1 l_3},\frac{l_4}{l_1 l_2},\frac{l_2}{l_1^2},\frac{l_4}{l_1^2},\frac{l_1 l_3
   l_6}{l_9},-\frac{l_1 l_3 l_6}{l_7},\frac{l_1 l_{11}}{l_4 l_7},\frac{l_1 l_8}{l_2 l_7},-\frac{l_1 l_3}{l_2^2},-\frac{l_2^2 l_6}{l_9},\frac{l_2^2 l_7}{l_9},\frac{l_2^2 l_{11}}{l_9},\frac{l_2
   l_3 l_6}{l_9},-\frac{l_2 l_3 l_6}{l_8},
   \nonumber\\
   &\frac{l_2 l_{11}}{l_4 l_8},\frac{l_2 l_{10}}{l_3 l_4},-\frac{l_5 l_7}{l_6^2},\frac{l_7 l_9}{l_8^2},\frac{l_{11}^2}{l_8^2},-\frac{l_{10} l_{11}}{l_8
   l_9},\frac{l_8^2}{l_7 l_9},-\frac{l_6^2}{l_5 l_7},\frac{l_6 l_{11}}{l_5 l_7},-\frac{l_6 l_{10}}{l_5 l_7},-\frac{l_{11}}{l_3 l_4 l_6},\frac{l_{10}}{l_3 l_4 l_6},\frac{l_{11}}{l_2 l_5
   l_7},
   \nonumber\\
   &-\frac{l_8}{l_2 l_3 l_6},\frac{l_9}{l_2 l_3 l_6},-\frac{l_{10}}{l_2^2 l_8},-\frac{l_{10}}{l_2^2 l_7},\frac{l_9}{l_2^2 l_7},-\frac{l_9}{l_2^2 l_6},-\frac{l_2^2}{l_1 l_3},\frac{l_2
   l_7}{l_1 l_8},-\frac{l_2 l_{11}}{l_1 l_9},-\frac{l_2 l_{11}}{l_1 l_3},-\frac{l_7}{l_1 l_3 l_6},-\frac{l_{11}}{l_1 l_3 l_6},
   \nonumber\\
   &\frac{l_{10}}{l_1 l_3 l_6},\frac{l_9}{l_1 l_3
   l_6},\frac{l_{10}}{l_1 l_3 l_5},-\frac{l_1 l_3}{l_2^2 l_7},-\frac{l_2^3}{l_3 l_4},-\frac{l_2^2 l_{11}}{l_3 l_4},\frac{l_3^2 l_6}{l_8^2},-\frac{l_3^2 l_6}{l_7 l_9},-\frac{l_7 l_9}{l_3^2
   l_6},-\frac{l_8 l_{11}}{l_3^2 l_6},\frac{l_8^2}{l_3^2 l_6},\frac{l_9 l_{11}}{l_2 l_7 l_8},\frac{l_8 l_{11}}{l_2 l_7^2},
   \nonumber\\
   &\frac{l_8 l_{11}}{l_2 l_5^2},-\frac{l_{10}}{l_2^2 l_5 l_7},-\frac{l_3
   l_4}{l_2^3},-\frac{l_2^2 l_7}{l_1 l_3},-\frac{l_8 l_{11}}{l_1^2 l_6},\frac{l_{10} l_{11}}{l_3 l_4 l_5 l_7},-\frac{l_9 l_{10}}{l_2^2 l_7^2},-\frac{l_{11}^2}{l_1 l_3 l_5 l_7},-\frac{l_9
   l_{11}}{l_1 l_3^2 l_6},\frac{l_9 l_{10}}{l_1^2 l_3^2 l_6}\Bigg\}.
\end{align}

Above threshold, the letters are real-valued functions and it turns out that the ${\rm Li}_n$ functions themselves may be chosen to be real-valued in $\mathcal{D}_a$. By using the {\tt Reduce} function of {\tt Mathematica}, we can rapidly check whether any of the above function arguments assume values larger than one for $(w,z)$ in $\mathcal{D}_a$. As one might expect, this is a non-trivial constraint to impose on \eqref{eq:linargsforslarge} and only 148 function arguments survive. We find
\begin{align}
\label{eq:finallinargsforslarge}
    \Bigg\{&l_2,-l_2,l_3,l_6,-l_6,l_7,l_8,l_9,-\frac{1}{l_6},\frac{1}{l_5},\frac{1}{l_4},-\frac{1}{l_2},\frac{1}{l_1},-l_1 l_2,l_1 l_3,\frac{l_1}{l_4},l_2^2,l_2
   l_6,-\frac{l_2}{l_4},-\frac{l_2}{l_3},l_3 l_4,\frac{l_3}{l_9},\frac{l_3}{l_8},
   \nonumber\\
   &l_5 l_7,l_6^2,-\frac{l_6}{l_7},\frac{l_7}{l_9},\frac{l_7}{l_8},-\frac{l_{11}}{l_8},\frac{l_{11}}{l_8},-\frac{l_{10}}{l_9},\frac{l_8}{l_9},\frac{l_{11}}{l_7},-\frac{l_{10}}{l_7},-\frac{l_7}{l_6}
   ,-\frac{l_{11}}{l_6},\frac{l_{10}}{l_6},\frac{l_6}{l_5},-\frac{l_{11}}{l_5},\frac{l_8}{l_5},-\frac{l_{11}}{l_3},-\frac{l_3}{l_2},
   \nonumber\\
   &-\frac{l_4}{l_2},\frac{l_{11}}{l_2},\frac{l_2}{l_1},\frac{l_
   {11}}{l_1},\frac{l_8}{l_1},-\frac{1}{l_1 l_2},\frac{l_1 l_2}{l_4},\frac{l_1 l_3}{l_9},-\frac{l_1 l_6}{l_8},-\frac{l_1 l_{11}}{l_9},\frac{l_1 l_6}{l_5},-\frac{l_1}{l_2^2},l_2^3,l_2^2
   l_6,\frac{l_2^2}{l_4},-\frac{l_2 l_5}{l_8},-\frac{l_2 l_6}{l_8},
   \nonumber\\
   &-\frac{l_2 l_7}{l_9},\frac{l_2 l_7}{l_8},\frac{l_2 l_{11}}{l_8},-\frac{l_2 l_{10}}{l_8},\frac{l_2 l_8}{l_9},-\frac{l_2
   l_7}{l_3},-\frac{l_2 l_8}{l_3},\frac{l_3 l_6}{l_8},-\frac{l_3 l_6}{l_7},-\frac{l_6 l_{11}}{l_8},\frac{l_6 l_{10}}{l_9},-\frac{l_{11}}{l_6 l_8},\frac{l_{10}}{l_6 l_9},-\frac{l_{10}}{l_4
   l_7},
   \nonumber\\
   &-\frac{l_7}{l_3 l_6},-\frac{l_{11}}{l_3 l_6},-\frac{l_{11}}{l_3 l_5},-\frac{l_3}{l_2 l_8},-\frac{l_3}{l_2 l_7},\frac{l_{11}}{l_2 l_9},\frac{l_{11}}{l_2 l_8},\frac{l_{11}}{l_2
   l_7},-\frac{l_9}{l_2 l_7},-\frac{l_8}{l_2 l_6},\frac{l_{11}}{l_2 l_5},-\frac{l_8}{l_2 l_5},-\frac{l_{10}}{l_2^2},-\frac{l_2^2}{l_1},
   \nonumber\\
   &\frac{l_2 l_5}{l_1},\frac{l_{10}}{l_1 l_8},\frac{l_9}{l_1
   l_8},\frac{l_{11}}{l_1 l_7},-\frac{l_{11}}{l_1 l_6},-\frac{l_8}{l_1 l_6},\frac{l_{10}}{l_1 l_3},\frac{l_2}{l_1^2},\frac{l_4}{l_1^2},\frac{l_1 l_3 l_6}{l_9},-\frac{l_1 l_3
   l_6}{l_7},\frac{l_1 l_{11}}{l_4 l_7},-\frac{l_1 l_3}{l_2^2},-\frac{l_2^2 l_6}{l_9},\frac{l_2^2 l_7}{l_9},
   \nonumber\\
   &\frac{l_2^2 l_{11}}{l_9},\frac{l_2 l_3 l_6}{l_9},-\frac{l_2 l_3 l_6}{l_8},\frac{l_2
   l_{11}}{l_4 l_8},\frac{l_2 l_{10}}{l_3 l_4},-\frac{l_5 l_7}{l_6^2},\frac{l_7 l_9}{l_8^2},\frac{l_{11}^2}{l_8^2},-\frac{l_{10} l_{11}}{l_8 l_9},-\frac{l_6^2}{l_5 l_7},\frac{l_6 l_{11}}{l_5
   l_7},-\frac{l_6 l_{10}}{l_5 l_7},-\frac{l_{11}}{l_3 l_4 l_6},
   \nonumber\\
   &\frac{l_{10}}{l_3 l_4 l_6},\frac{l_{11}}{l_2 l_5 l_7},-\frac{l_8}{l_2 l_3 l_6},-\frac{l_{10}}{l_2^2 l_8},-\frac{l_{10}}{l_2^2
   l_7},-\frac{l_9}{l_2^2 l_6},-\frac{l_2^2}{l_1 l_3},\frac{l_2 l_7}{l_1 l_8},-\frac{l_2 l_{11}}{l_1 l_9},-\frac{l_2 l_{11}}{l_1 l_3},-\frac{l_7}{l_1 l_3 l_6},-\frac{l_{11}}{l_1 l_3
   l_6},
   \nonumber\\
   &\frac{l_{10}}{l_1 l_3 l_6},\frac{l_{10}}{l_1 l_3 l_5},-\frac{l_1 l_3}{l_2^2 l_7},-\frac{l_2^3}{l_3 l_4},-\frac{l_2^2 l_{11}}{l_3 l_4},\frac{l_3^2 l_6}{l_8^2},-\frac{l_3^2 l_6}{l_7
   l_9},-\frac{l_7 l_9}{l_3^2 l_6},-\frac{l_8 l_{11}}{l_3^2 l_6},\frac{l_9 l_{11}}{l_2 l_7 l_8},\frac{l_8 l_{11}}{l_2 l_7^2},\frac{l_8 l_{11}}{l_2 l_5^2},-\frac{l_{10}}{l_2^2 l_5
   l_7},
   \nonumber\\
   &-\frac{l_3 l_4}{l_2^3},-\frac{l_2^2 l_7}{l_1 l_3},-\frac{l_8 l_{11}}{l_1^2 l_6},\frac{l_{10} l_{11}}{l_3 l_4 l_5 l_7},-\frac{l_9 l_{10}}{l_2^2 l_7^2},-\frac{l_{11}^2}{l_1 l_3 l_5
   l_7},-\frac{l_9 l_{11}}{l_1 l_3^2 l_6},\frac{l_9 l_{10}}{l_1^2 l_3^2 l_6}\Bigg\}
\end{align}
for our final set of preferred ${\rm Li}_n$ function arguments above the two-mass threshold. We can proceed analogously to filter the set of ${\rm Li}_{2,1}$, ${\rm Li}_{3,1}$, and ${\rm Li}_{2,2}$ function argument pairs which survive our first-entry cuts. For the sake of brevity, let us simply state that we find 2496 preferred pairs of function arguments $(f_i, f_j)$ after using {\tt Reduce} to throw away all pairs for which $f_i > 1$ or $f_i f_j > 1$ for $(w,z)$ in $\mathcal{D}_a$. As explained in \cite{vonManteuffel:2017myy}, we can further improve the numerical performance of our functional basis in {\tt GiNaC} by reordering the lists of function arguments obtained to ensure that those which lead to convergent power series expansions are given precedence.

In region $\mathcal{D}_{b_1}$, the first-entry conditions are the same as for $\mathcal{D}_a$ and again lead to \eqref{eq:linargsforslarge}. As mentioned above, we can use {\tt Reduce} to efficiently check which ${\rm Li}_n$ basis functions have suitable analytic properties below the two-mass threshold as well; despite the fact that $w$ and $z$ become complex functions of the original kinematic variables in $\mathcal{D}_{b_1}$ and the remaining below-threshold regions, we can proceed as before provided that we first eliminate $\mathrm{Im}[w]$ and $\mathrm{Im}[z]$ via Eq. \eqref{eq:imasre} and work only with $\mathrm{Re}[w]$ and $\mathrm{Re}[z]$. It turns out that a surprisingly large number of exceptional phase space points and trajectories cause problems for particular basis functions in the $(w,z)$ representation. Specifically, we find that certain ${\rm Li}_n$ arguments assume values on the real axis greater than one
\begin{itemize}
    \item when ~$s = 3\,m^2$ ~for~ $-3\,m^2 < t < 0$.
    \item when ~$s = -\frac{4\, m^2 t}{m^2 - t}$ ~for~ $-3\,m^2 < t < -m^2$.
    \item when ~$s = -\frac{2\,m^2(-2\,t-m^2)}{t}$ ~for~ $-\left(2+\sqrt{2}\right)m^2 < t < -m^2$.
    \item when ~$s = \frac{4\,m^2+t + \sqrt{-t\,(8\,m^2-t)}}{2}$ ~for~ $-\left(2+\sqrt{2}\right)m^2 < t < 0$.
\end{itemize}
We also observe that a number of function arguments have imaginary parts which vanish identically below threshold. Any such ${\rm Li}_n$ function with an argument which can attain values greater than one on $\mathcal{D}_{b_1}$ must be discarded because it will be ill-defined unless small imaginary parts are assigned to $w$ and $z$. 

In the end, we find the 178 preferred ${\rm Li}_n$ function arguments
\begin{align}
\label{eq:linargsforb1}
    \Bigg\{&l_1,l_2,-l_2,l_3,l_4,l_5,l_6,-l_6,l_7,\frac{1}{l_9},\frac{1}{l_8},\frac{1}{l_7},\frac{1}{l_6},-\frac{1}{l_6},\frac{1}{l_5},\frac{1}{l
   _4},\frac{1}{l_3},\frac{1}{l_2},-\frac{1}{l_2},\frac{1}{l_1},-l_1 l_2,l_1 l_3,\frac{l_1}{l_8},\frac{l_1}{l_4},\frac{l_1}{l_2},
   \nonumber \\
   &l_2^2,l_2 l_6,-\frac{l_2}{l_4},-\frac{l_2}{l_3},\frac{l_3}{l_9},\frac{l_3}{l_8},\frac{l_5}{l_8},\frac{l_5}{l_6},l_6^2,-\frac{l_6}{l_7},\frac{l_7}{l_9},\frac{l_7}{l_8},-\frac{l_{11}}{l_8},\frac{l_{11}}{l_8},-\frac{l_{10}}{l_9},\frac{l_8}{l_9},\frac{l_9}{l_8},\frac{l_{11}}{l_7},-\frac{l_{10}}{l_7},\frac{l_8}{l_7},\frac{l_9}{l_7},-\frac{l_7}{l_6},
   \nonumber \\
   &-\frac{l_{11}}{l_6},\frac{l_{10}}{l_6},\frac{1}{l_6^2},\frac{l_6}{l_5},-\frac{l_{11}}{l_5},\frac{l_8}{l_5},\frac{1}{l_5 l_7},-\frac{l_{11}}{l_3},\frac{l_8}{l_3},\frac{l_9}{l_3},\frac{1}{l_3
   l_4},-\frac{l_3}{l_2},-\frac{l_4}{l_2},\frac{l_{11}}{l_2},\frac{1}{l_2
   l_6},\frac{1}{l_2^2},\frac{l_2}{l_1},\frac{l_4}{l_1},\frac{l_{11}}{l_1},\frac{l_8}{l_1},
   \nonumber \\
   &\frac{1}{l_1 l_3},-\frac{1}{l_1 l_2},\frac{l_1
   l_2}{l_4},\frac{l_1 l_3}{l_9},-\frac{l_1 l_6}{l_8},-\frac{l_1 l_{11}}{l_9},\frac{l_1 l_6}{l_5},\frac{l_1}{l_2
   l_5},-\frac{l_1}{l_2^2},l_2^3,l_2^2 l_6,\frac{l_2^2}{l_4},-\frac{l_2 l_5}{l_8},-\frac{l_2 l_6}{l_8},-\frac{l_2 l_7}{l_9},\frac{l_2 l_7}{l_8},
   \nonumber \\
   &\frac{l_2 l_{11}}{l_8},-\frac{l_2 l_{10}}{l_8},\frac{l_2 l_8}{l_9},-\frac{l_2 l_7}{l_3},-\frac{l_2 l_8}{l_3},\frac{l_3
   l_6}{l_8},-\frac{l_3 l_6}{l_7},-\frac{l_6 l_{11}}{l_8},\frac{l_6 l_{10}}{l_9},-\frac{l_{11}}{l_6 l_8},\frac{l_{10}}{l_6
   l_9},-\frac{l_{10}}{l_4 l_7},-\frac{l_7}{l_3 l_6},-\frac{l_{11}}{l_3 l_6},
   \nonumber \\
   &\frac{l_8}{l_3 l_6},-\frac{l_{11}}{l_3 l_5},-\frac{l_3}{l_2
   l_8},-\frac{l_3}{l_2 l_7},\frac{l_{11}}{l_2 l_9},\frac{l_{11}}{l_2 l_8},\frac{l_9}{l_2 l_8},\frac{l_{11}}{l_2 l_7},\frac{l_8}{l_2
   l_7},-\frac{l_9}{l_2 l_7},\frac{l_{11}}{l_2 l_5},-\frac{l_8}{l_2 l_5},\frac{l_4}{l_2^2},-\frac{l_{10}}{l_2^2},\frac{1}{l_2^2
   l_6},\frac{1}{l_2^3},
   \nonumber \\
   &-\frac{l_2^2}{l_1},\frac{l_2 l_5}{l_1},\frac{l_5}{l_1 l_6},\frac{l_{10}}{l_1 l_8},\frac{l_9}{l_1
   l_8},\frac{l_{11}}{l_1 l_7},-\frac{l_{11}}{l_1 l_6},-\frac{l_8}{l_1 l_6},\frac{l_{10}}{l_1 l_3},\frac{l_9}{l_1 l_3},\frac{l_4}{l_1
   l_2},\frac{l_2}{l_1^2},\frac{l_4}{l_1^2},\frac{l_1 l_3 l_6}{l_9},-\frac{l_1 l_3 l_6}{l_7},\frac{l_1 l_{11}}{l_4 l_7},
   \nonumber \\
   &-\frac{l_1 l_3}{l_2^2},-\frac{l_2^2 l_6}{l_9},\frac{l_2^2 l_7}{l_9},\frac{l_2^2 l_{11}}{l_9},\frac{l_2 l_3 l_6}{l_9},-\frac{l_2 l_3
   l_6}{l_8},\frac{l_2 l_{11}}{l_4 l_8},\frac{l_2 l_{10}}{l_3 l_4},\frac{l_7 l_9}{l_8^2},\frac{l_{11}^2}{l_8^2},-\frac{l_{10} l_{11}}{l_8
   l_9},-\frac{l_6^2}{l_5 l_7},\frac{l_6 l_{11}}{l_5 l_7},
   \nonumber \\
   &-\frac{l_6 l_{10}}{l_5 l_7},-\frac{l_{11}}{l_3 l_4 l_6},\frac{l_{10}}{l_3 l_4
   l_6},\frac{l_{11}}{l_2 l_5 l_7},-\frac{l_8}{l_2 l_3 l_6},\frac{l_9}{l_2 l_3 l_6},-\frac{l_{10}}{l_2^2 l_8},-\frac{l_{10}}{l_2^2
   l_7},\frac{l_9}{l_2^2 l_7},-\frac{l_2^2}{l_1 l_3},\frac{l_2 l_7}{l_1 l_8},-\frac{l_2 l_{11}}{l_1 l_9},
   \nonumber \\
   &-\frac{l_2 l_{11}}{l_1
   l_3},-\frac{l_7}{l_1 l_3 l_6},-\frac{l_{11}}{l_1 l_3 l_6},\frac{l_{10}}{l_1 l_3 l_6},\frac{l_9}{l_1 l_3 l_6},\frac{l_{10}}{l_1 l_3
   l_5},-\frac{l_1 l_3}{l_2^2 l_7},-\frac{l_2^3}{l_3 l_4},-\frac{l_2^2 l_{11}}{l_3 l_4},\frac{l_3^2 l_6}{l_8^2},-\frac{l_3^2 l_6}{l_7
   l_9},-\frac{l_7 l_9}{l_3^2 l_6},
   \nonumber \\
   &-\frac{l_8 l_{11}}{l_3^2 l_6},\frac{l_9 l_{11}}{l_2 l_7 l_8},\frac{l_8 l_{11}}{l_2 l_7^2},\frac{l_8
   l_{11}}{l_2 l_5^2},-\frac{l_{10}}{l_2^2 l_5 l_7},-\frac{l_2^2 l_7}{l_1 l_3},-\frac{l_8 l_{11}}{l_1^2 l_6},\frac{l_{10} l_{11}}{l_3 l_4
   l_5 l_7},-\frac{l_9 l_{10}}{l_2^2 l_7^2},-\frac{l_{11}^2}{l_1 l_3 l_5 l_7},-\frac{l_9 l_{11}}{l_1 l_3^2 l_6},
   \nonumber \\
   &\frac{l_9 l_{10}}{l_1^2 l_3^2 l_6}\Bigg\}
\end{align}
for $\mathcal{D}_{b_1}$. In an analogous manner, we find 3314 preferred pairs of ${\rm Li}_{2,1}$, ${\rm Li}_{3,1}$, and ${\rm Li}_{2,2}$ function arguments $(f_i, f_j)$ after using {\tt Reduce} to throw away all pairs for which either $\mathrm{Re}[f_i] > 1$ and $\mathrm{Im}[f_i] = 0$ or, alternatively, $\mathrm{Re}[f_i f_j] > 1$ and  $\mathrm{Im}[f_i f_j] = 0$ for $(w,z)$ in $\mathcal{D}_{b_1}$.

For regions $\mathcal{D}_{b_2}$ and $\mathcal{D}_{b_3}$, the first-entry conditions lead to the preliminary list of 194 ${\rm Li}_n$ function arguments
\begin{align}
\label{eq:linargsforb2b3}
    \Bigg\{&l_1,l_2,-l_2,l_3,l_4,l_5,l_6,-l_6,l_7,l_8,l_9,\frac{1}{l_8},\frac{1}{l_7},\frac{1}{l_6},-\frac{1}{l_6},\frac{1}{l_5},\frac{1}{l_4},\frac{1}{l_3},\frac{1}{l_2},-\frac{1}{l_2},\frac{1}{l_1},-l_1 l_2,l_1 l_3,\frac{l_1}{l_8},\frac{l_1}{l_4},
    \nonumber \\
    &\frac{l_1}{l_2},l_2^2,l_2
   l_6,-\frac{l_2}{l_4},-\frac{l_2}{l_3},l_3 l_4,\frac{l_3}{l_8},l_5
   l_7,\frac{l_5}{l_8},\frac{l_5}{l_6},l_6^2,\frac{l_6}{l_{10}},-\frac{l_6}{l_7},\frac{l_7}{l_8},-\frac{l_7}{l_{10}},-\frac{l_{11}}{l_8},\frac{l_{11}}{l_8},-\frac{l_{11}}{l_{10}},\frac{l_9}{l_8},-\frac{l_9}{l_{10}},
   \nonumber \\
   &\frac{l_{11}}{l_7},-\frac{l_{10}}{l_7},\frac{l_8}{l_7},\frac{l
   _9}{l_7},-\frac{l_7}{l_6},-\frac{l_{11}}{l_6},\frac{l_{10}}{l_6},\frac{1}{l_6^2},\frac{l_6}{l_5},-\frac{l_{11}}{l_5},\frac{l_8}{l_5},\frac{1}{l_5 l_7},-\frac{l_{11}}{l_3},\frac{l_8}{l_3},\frac{l_9}{l_3},\frac{1}{l_3
   l_4},-\frac{l_3}{l_2},-\frac{l_4}{l_2},\frac{l_{11}}{l_2},\frac{1}{l_2
   l_6},
   \nonumber \\
   &\frac{1}{l_2^2},\frac{l_2}{l_1},\frac{l_4}{l_1},\frac{l_{11}}{l_1},\frac{l_8}{l_1},\frac{1}{l_1 l_3},-\frac{1}{l_1
   l_2},\frac{l_1^2}{l_4},\frac{l_1^2}{l_2},\frac{l_1 l_2}{l_4},\frac{l_1 l_3}{l_{10}},\frac{l_1 l_3}{l_7},-\frac{l_1 l_6}{l_8},-\frac{l_1
   l_{11}}{l_{10}},\frac{l_1 l_8}{l_{10}},\frac{l_1 l_6}{l_5},\frac{l_1}{l_2
   l_5},-\frac{l_1}{l_2^2},l_2^3,
   \nonumber \\
   &-\frac{l_2^2}{l_{10}},\frac{l_2^2}{l_6},\frac{l_2^2}{l_4},-\frac{l_2 l_5}{l_8},-\frac{l_2
   l_6}{l_8},\frac{l_2 l_7}{l_8},\frac{l_2 l_{11}}{l_8},-\frac{l_2 l_{11}}{l_{10}},-\frac{l_2 l_{10}}{l_8},-\frac{l_2 l_7}{l_3},\frac{l_3
   l_4}{l_8},\frac{l_3 l_6}{l_8},-\frac{l_3 l_6}{l_7},-\frac{l_4 l_7}{l_{10}},
   \nonumber \\
   &-\frac{l_6 l_{11}}{l_8},\frac{l_6
   l_9}{l_{10}},-\frac{l_{11}}{l_6 l_8},\frac{l_9}{l_6 l_{10}},\frac{l_9}{l_5 l_7},-\frac{l_{10}}{l_4 l_7},-\frac{l_7}{l_3
   l_6},-\frac{l_{11}}{l_3 l_6},\frac{l_8}{l_3 l_6},-\frac{l_{11}}{l_3 l_5},\frac{l_8}{l_3 l_4},-\frac{l_3}{l_2 l_7},\frac{l_{11}}{l_2
   l_8},\frac{l_9}{l_2 l_8},
   \nonumber \\
   &-\frac{l_8}{l_2 l_{10}},\frac{l_{11}}{l_2 l_7},\frac{l_8}{l_2 l_7},-\frac{l_9}{l_2 l_7},-\frac{l_8}{l_2
   l_6},\frac{l_{11}}{l_2 l_5},-\frac{l_8}{l_2 l_5},\frac{l_{11}}{l_2
   l_3},\frac{l_4}{l_2^2},\frac{l_6}{l_2^2},-\frac{l_{10}}{l_2^2},\frac{1}{l_2^3},-\frac{l_2^2}{l_1},\frac{l_2 l_5}{l_1},\frac{l_5}{l_1
   l_6},\frac{l_{10}}{l_1 l_8},\frac{l_9}{l_1 l_8},
   \nonumber \\
   &\frac{l_{11}}{l_1 l_7},-\frac{l_{11}}{l_1 l_6},-\frac{l_8}{l_1 l_6},\frac{l_7}{l_1
   l_3},\frac{l_{10}}{l_1 l_3},\frac{l_9}{l_1 l_3},\frac{l_4}{l_1 l_2},\frac{l_2}{l_1^2},\frac{l_4}{l_1^2},\frac{l_1 l_3
   l_6}{l_{10}},\frac{l_1 l_8}{l_4 l_7},\frac{l_1 l_{11}}{l_2 l_7},-\frac{l_1 l_3}{l_2^2},-\frac{l_2^2 l_7}{l_{10}},-\frac{l_2^2
   l_8}{l_{10}},\frac{l_2 l_{11}}{l_3 l_6},
   \nonumber \\
   &\frac{l_2 l_{10}}{l_3 l_4},\frac{l_3 l_4 l_6}{l_{10}},-\frac{l_5 l_7}{l_6^2},\frac{l_7
   l_9}{l_8^2},\frac{l_{11}^2}{l_8^2},-\frac{l_9 l_{11}}{l_8 l_{10}},-\frac{l_{11}^2}{l_7 l_{10}},-\frac{l_6^2}{l_5 l_7},\frac{l_6
   l_{11}}{l_5 l_7},\frac{l_{10}}{l_3 l_4 l_6},\frac{l_3 l_4}{l_2 l_{10}},\frac{l_4 l_{11}}{l_2 l_8},\frac{l_{11}}{l_2 l_5
   l_7},\frac{l_9}{l_2 l_3 l_6},
   \nonumber \\
   &-\frac{l_{10}}{l_2^2 l_8},-\frac{l_{10}}{l_2^2 l_7},\frac{l_9}{l_2^2 l_7},-\frac{l_9}{l_2^2
   l_6},-\frac{l_2^2}{l_1 l_3},-\frac{l_2 l_{11}}{l_1 l_3},\frac{l_4 l_7}{l_1 l_8},-\frac{l_4 l_{11}}{l_1 l_{10}},-\frac{l_{11}}{l_1 l_3
   l_6},\frac{l_{10}}{l_1 l_3 l_6},\frac{l_9}{l_1 l_3 l_6},\frac{l_9}{l_1 l_3 l_5},-\frac{l_2^3}{l_3 l_4},
   \nonumber \\
   &-\frac{l_2 l_7 l_{11}}{l_8
   l_{10}},-\frac{l_7 l_9}{l_3^2 l_6},\frac{l_{11}^2}{l_3^2 l_6},-\frac{l_8 l_{11}}{l_3^2 l_6},\frac{l_9 l_{11}}{l_2 l_7 l_8},\frac{l_8
   l_{11}}{l_2 l_7^2},\frac{l_8 l_{11}}{l_2 l_5^2},-\frac{l_3 l_4}{l_2^3},-\frac{l_8 l_{11}}{l_1^2 l_6},\frac{l_1 l_3 l_6}{l_2^2
   l_7},\frac{l_2^2 l_8}{l_3 l_4 l_6},-\frac{l_9 l_{11}^2}{l_8^2 l_{10}},
   \nonumber \\
   &\frac{l_9 l_{11}}{l_2 l_3 l_5 l_7},\frac{l_3 l_4 l_6}{l_2^2
   l_8},-\frac{l_9 l_{10}}{l_2^2 l_7^2},\frac{l_6 l_9}{l_2^2 l_5 l_7},\frac{l_2^2 l_7}{l_1 l_3 l_6},-\frac{l_{11}^2}{l_1 l_3 l_5
   l_7},-\frac{l_9 l_{11}}{l_1 l_3^2 l_6},\frac{l_2 l_7 l_{11}}{l_1 l_3^2 l_6},\frac{l_9 l_{10}}{l_1^2 l_3^2 l_6},\frac{l_9 l_{11}^2}{l_1
   l_3^3 l_5 l_6}\Bigg\}.
\end{align}
The rest of the filtering procedure in the remaining regions is closely analogous to what is described above for $\mathcal{D}_{b_1}$, but we nevertheless summarize our findings for the benefit of the reader. 

In $\mathcal{D}_{b_2}$, besides those functions whose arguments have imaginary parts which vanish identically, we find that certain ${\rm Li}_n$ arguments assume values on the real axis greater than one
\begin{itemize}
    \item when ~$s = -\frac{4\, m^2 t}{m^2 - t}$ ~for~ $-m^2 < t < -\frac{m^2}{3}$.
    \item when ~$s = -\frac{2\,m^2(-2\,t-m^2)}{t}$ ~for~ $-m^2 < t < -\frac{2\,m^2}{3}$.
    \item when ~$s = \frac{4\,m^2+t - \sqrt{-t\,(8\,m^2-t)}}{2}$ ~for~ $-\frac{m^2}{3} < t < 0$.
\end{itemize}
After discarding all of the offending ${\rm Li}_n$ function arguments, we find that 184 preferred ones remain:
\begin{align}
\label{eq:linargsforb2}
    \Bigg\{&l_1,l_2,-l_2,l_3,l_4,l_5,l_6,-l_6,l_7,l_9,\frac{1}{l_8},\frac{1}{l_7},\frac{1}{l_6},-\frac{1}{l_6},\frac{1}{l_5},\frac{1}{l_4},\frac{
   1}{l_3},\frac{1}{l_2},-\frac{1}{l_2},\frac{1}{l_1},-l_1 l_2,l_1 l_3,\frac{l_1}{l_8},\frac{l_1}{l_4},\frac{l_1}{l_2},
   \nonumber \\
   &l_2^2,l_2
   l_6,-\frac{l_2}{l_4},-\frac{l_2}{l_3},l_3
   l_4,\frac{l_3}{l_8},\frac{l_5}{l_8},\frac{l_5}{l_6},l_6^2,\frac{l_6}{l_{10}},-\frac{l_6}{l_7},\frac{l_7}{l_8},-\frac{l_7}{l_{10}},-\frac{
   l_{11}}{l_8},\frac{l_{11}}{l_8},-\frac{l_{11}}{l_{10}},\frac{l_9}{l_8},-\frac{l_9}{l_{10}},\frac{l_{11}}{l_7},-\frac{l_{10}}{l_7},
   \nonumber \\
   &\frac{l_8}{l_7},\frac{l_9}{l_7},-\frac{l_7}{l_6},-\frac{l_{11}}{l_6},\frac{1}{l_6^2},\frac{l_6}{l_5},-\frac{l_{11}}{l_5},\frac{l_8}{l_5},\frac{1
   }{l_5 l_7},-\frac{l_{11}}{l_3},\frac{l_8}{l_3},\frac{l_9}{l_3},\frac{1}{l_3
   l_4},-\frac{l_3}{l_2},-\frac{l_4}{l_2},\frac{l_{11}}{l_2},\frac{1}{l_2
   l_6},\frac{1}{l_2^2},\frac{l_2}{l_1},\frac{l_4}{l_1},\frac{l_{11}}{l_1},
   \nonumber \\
   &\frac{l_8}{l_1},\frac{1}{l_1 l_3},-\frac{1}{l_1 l_2},\frac{l_1
   l_2}{l_4},\frac{l_1 l_3}{l_{10}},\frac{l_1 l_3}{l_7},-\frac{l_1 l_6}{l_8},-\frac{l_1 l_{11}}{l_{10}},\frac{l_1 l_6}{l_5},\frac{l_1}{l_2
   l_5},-\frac{l_1}{l_2^2},l_2^3,-\frac{l_2^2}{l_{10}},\frac{l_2^2}{l_6},\frac{l_2^2}{l_4},-\frac{l_2 l_5}{l_8},-\frac{l_2
   l_6}{l_8},
   \nonumber \\
   &\frac{l_2 l_7}{l_8},\frac{l_2 l_{11}}{l_8},-\frac{l_2 l_{11}}{l_{10}},-\frac{l_2 l_{10}}{l_8},-\frac{l_2 l_7}{l_3},\frac{l_3
   l_4}{l_8},\frac{l_3 l_6}{l_8},-\frac{l_3 l_6}{l_7},-\frac{l_4 l_7}{l_{10}},-\frac{l_6 l_{11}}{l_8},\frac{l_6
   l_9}{l_{10}},-\frac{l_{11}}{l_6 l_8},\frac{l_9}{l_6 l_{10}},\frac{l_9}{l_5 l_7},
   \nonumber \\
   &-\frac{l_{10}}{l_4 l_7},-\frac{l_7}{l_3
   l_6},-\frac{l_{11}}{l_3 l_6},\frac{l_8}{l_3 l_6},-\frac{l_{11}}{l_3 l_5},\frac{l_8}{l_3 l_4},-\frac{l_3}{l_2 l_7},\frac{l_{11}}{l_2
   l_8},\frac{l_9}{l_2 l_8},-\frac{l_8}{l_2 l_{10}},\frac{l_{11}}{l_2 l_7},\frac{l_8}{l_2 l_7},-\frac{l_9}{l_2 l_7},\frac{l_{11}}{l_2
   l_5},
   \nonumber \\
   &-\frac{l_8}{l_2 l_5},\frac{l_{11}}{l_2 l_3},\frac{l_4}{l_2^2},\frac{l_6}{l_2^2},\frac{1}{l_2^3},-\frac{l_2^2}{l_1},\frac{l_2
   l_5}{l_1},\frac{l_5}{l_1 l_6},\frac{l_{10}}{l_1 l_8},\frac{l_9}{l_1 l_8},\frac{l_{11}}{l_1 l_7},-\frac{l_{11}}{l_1 l_6},-\frac{l_8}{l_1
   l_6},\frac{l_7}{l_1 l_3},\frac{l_{10}}{l_1 l_3},\frac{l_9}{l_1 l_3},\frac{l_4}{l_1 l_2},\frac{l_2}{l_1^2},
   \nonumber \\
   &\frac{l_4}{l_1^2},\frac{l_1 l_3
   l_6}{l_{10}},\frac{l_1 l_{11}}{l_2 l_7},-\frac{l_1 l_3}{l_2^2},-\frac{l_2^2 l_7}{l_{10}},-\frac{l_2^2 l_8}{l_{10}},\frac{l_2 l_{11}}{l_3
   l_6},\frac{l_2 l_{10}}{l_3 l_4},\frac{l_3 l_4 l_6}{l_{10}},\frac{l_7 l_9}{l_8^2},\frac{l_{11}^2}{l_8^2},-\frac{l_9 l_{11}}{l_8
   l_{10}},-\frac{l_{11}^2}{l_7 l_{10}},-\frac{l_6^2}{l_5 l_7},
   \nonumber \\
   &\frac{l_6 l_{11}}{l_5 l_7},\frac{l_{10}}{l_3 l_4 l_6},\frac{l_3 l_4}{l_2
   l_{10}},\frac{l_4 l_{11}}{l_2 l_8},\frac{l_{11}}{l_2 l_5 l_7},\frac{l_9}{l_2 l_3 l_6},-\frac{l_{10}}{l_2^2 l_8},-\frac{l_{10}}{l_2^2
   l_7},\frac{l_9}{l_2^2 l_7},-\frac{l_9}{l_2^2 l_6},-\frac{l_2^2}{l_1 l_3},-\frac{l_2 l_{11}}{l_1 l_3},\frac{l_4 l_7}{l_1 l_8},
   \nonumber \\
   &-\frac{l_4 l_{11}}{l_1 l_{10}},-\frac{l_{11}}{l_1 l_3 l_6},\frac{l_{10}}{l_1 l_3 l_6},\frac{l_9}{l_1 l_3 l_6},\frac{l_9}{l_1 l_3
   l_5},-\frac{l_2^3}{l_3 l_4},-\frac{l_2 l_7 l_{11}}{l_8 l_{10}},-\frac{l_7 l_9}{l_3^2 l_6},\frac{l_{11}^2}{l_3^2 l_6},-\frac{l_8
   l_{11}}{l_3^2 l_6},\frac{l_9 l_{11}}{l_2 l_7 l_8},\frac{l_8 l_{11}}{l_2 l_7^2},
   \nonumber \\
   &\frac{l_8 l_{11}}{l_2 l_5^2},-\frac{l_3
   l_4}{l_2^3},-\frac{l_8 l_{11}}{l_1^2 l_6},\frac{l_1 l_3 l_6}{l_2^2 l_7},\frac{l_2^2 l_8}{l_3 l_4 l_6},-\frac{l_9 l_{11}^2}{l_8^2
   l_{10}},\frac{l_9 l_{11}}{l_2 l_3 l_5 l_7},\frac{l_3 l_4 l_6}{l_2^2 l_8},-\frac{l_9 l_{10}}{l_2^2 l_7^2},\frac{l_6 l_9}{l_2^2 l_5
   l_7},\frac{l_2^2 l_7}{l_1 l_3 l_6},-\frac{l_{11}^2}{l_1 l_3 l_5 l_7},
   \nonumber \\
   &-\frac{l_9 l_{11}}{l_1 l_3^2 l_6},\frac{l_2 l_7 l_{11}}{l_1 l_3^2
   l_6},\frac{l_9 l_{10}}{l_1^2 l_3^2 l_6},\frac{l_9 l_{11}^2}{l_1 l_3^3 l_5 l_6}\Bigg\}.
\end{align}
In much the same way, we find 3420 preferred pairs of function arguments for ${\rm Li}_{2,1}$, ${\rm Li}_{3,1}$, and ${\rm Li}_{2,2}$ in $\mathcal{D}_{b_2}$.

In $\mathcal{D}_{b_3}$, besides those functions whose arguments have imaginary parts which vanish identically, we find that certain ${\rm Li}_n$ arguments assume values on the real axis greater than one
\begin{itemize}
    \item when ~$s = -\frac{4\, m^2 t}{m^2 - t}$ ~for~ $-\frac{m^2}{3} < t < 0$.
    \item when ~$s = -\frac{2\,m^2(-2\,t-m^2)}{t}$ ~for~ $-\frac{2\,m^2}{3} < t < -\left(2-\sqrt{2}\right)m^2$.
    \item when ~$s = \frac{4\,m^2+t - \sqrt{-t\,(8\,m^2-t)}}{2}$ ~for~ $-\left(2-\sqrt{2}\right)m^2 < t < -\frac{m^2}{3}$.
\end{itemize}
After filtering, we find that 184 preferred ${\rm Li}_n$ function arguments remain:
\begin{align}
\label{eq:linargsforb3}
    \Bigg\{&l_1,l_2,-l_2,l_3,l_4,l_5,l_6,-l_6,l_7,l_9,\frac{1}{l_8},\frac{1}{l_7},\frac{1}{l_6},-\frac{1}{l_6},\frac{1}{l_5},\frac{1}{l_4},\frac{
   1}{l_3},\frac{1}{l_2},-\frac{1}{l_2},\frac{1}{l_1},-l_1 l_2,l_1 l_3,\frac{l_1}{l_8},\frac{l_1}{l_4},\frac{l_1}{l_2},
   \nonumber \\
   &l_2^2,l_2
   l_6,-\frac{l_2}{l_3},l_3
   l_4,\frac{l_3}{l_8},\frac{l_5}{l_8},\frac{l_5}{l_6},l_6^2,\frac{l_6}{l_{10}},-\frac{l_6}{l_7},\frac{l_7}{l_8},-\frac{l_7}{l_{10}},-\frac{
   l_{11}}{l_8},\frac{l_{11}}{l_8},-\frac{l_{11}}{l_{10}},\frac{l_9}{l_8},-\frac{l_9}{l_{10}},\frac{l_{11}}{l_7},-\frac{l_{10}}{l_7},\frac{l
   _8}{l_7},\frac{l_9}{l_7},
   \nonumber \\
   &-\frac{l_7}{l_6},-\frac{l_{11}}{l_6},\frac{1}{l_6^2},\frac{l_6}{l_5},-\frac{l_{11}}{l_5},\frac{l_8}{l_5},\frac{1
   }{l_5 l_7},-\frac{l_{11}}{l_3},\frac{l_8}{l_3},\frac{l_9}{l_3},\frac{1}{l_3
   l_4},-\frac{l_3}{l_2},-\frac{l_4}{l_2},\frac{l_{11}}{l_2},\frac{1}{l_2
   l_6},\frac{1}{l_2^2},\frac{l_2}{l_1},\frac{l_4}{l_1},\frac{l_{11}}{l_1},\frac{l_8}{l_1},
   \nonumber \\
   &\frac{1}{l_1 l_3},-\frac{1}{l_1
   l_2},\frac{l_1^2}{l_4},\frac{l_1^2}{l_2},\frac{l_1 l_2}{l_4},\frac{l_1 l_3}{l_{10}},\frac{l_1 l_3}{l_7},-\frac{l_1 l_6}{l_8},-\frac{l_1
   l_{11}}{l_{10}},\frac{l_1 l_8}{l_{10}},\frac{l_1 l_6}{l_5},\frac{l_1}{l_2
   l_5},-\frac{l_1}{l_2^2},l_2^3,-\frac{l_2^2}{l_{10}},\frac{l_2^2}{l_6},\frac{l_2^2}{l_4},-\frac{l_2 l_5}{l_8},
   \nonumber \\
   &-\frac{l_2 l_6}{l_8},\frac{l_2 l_7}{l_8},\frac{l_2 l_{11}}{l_8},-\frac{l_2 l_{11}}{l_{10}},-\frac{l_2 l_{10}}{l_8},-\frac{l_2 l_7}{l_3},\frac{l_3
   l_4}{l_8},\frac{l_3 l_6}{l_8},-\frac{l_3 l_6}{l_7},-\frac{l_4 l_7}{l_{10}},-\frac{l_6 l_{11}}{l_8},\frac{l_6
   l_9}{l_{10}},-\frac{l_{11}}{l_6 l_8},\frac{l_9}{l_6 l_{10}},
   \nonumber \\
   &\frac{l_9}{l_5 l_7},-\frac{l_7}{l_3 l_6},-\frac{l_{11}}{l_3
   l_6},\frac{l_8}{l_3 l_6},-\frac{l_{11}}{l_3 l_5},\frac{l_8}{l_3 l_4},-\frac{l_3}{l_2 l_7},\frac{l_{11}}{l_2 l_8},\frac{l_9}{l_2
   l_8},-\frac{l_8}{l_2 l_{10}},\frac{l_{11}}{l_2 l_7},\frac{l_8}{l_2 l_7},-\frac{l_9}{l_2 l_7},\frac{l_{11}}{l_2 l_5},-\frac{l_8}{l_2
   l_5},
   \nonumber \\
   &\frac{l_{11}}{l_2 l_3},\frac{l_4}{l_2^2},\frac{l_6}{l_2^2},\frac{1}{l_2^3},-\frac{l_2^2}{l_1},\frac{l_2 l_5}{l_1},\frac{l_5}{l_1
   l_6},\frac{l_9}{l_1 l_8},\frac{l_{11}}{l_1 l_7},-\frac{l_{11}}{l_1 l_6},-\frac{l_8}{l_1 l_6},\frac{l_7}{l_1 l_3},\frac{l_{10}}{l_1
   l_3},\frac{l_9}{l_1 l_3},\frac{l_4}{l_1 l_2},\frac{l_4}{l_1^2},\frac{l_1 l_3 l_6}{l_{10}},\frac{l_1 l_8}{l_4 l_7},
   \nonumber \\
   &\frac{l_1 l_{11}}{l_2
   l_7},-\frac{l_1 l_3}{l_2^2},-\frac{l_2^2 l_7}{l_{10}},-\frac{l_2^2 l_8}{l_{10}},\frac{l_2 l_{11}}{l_3 l_6},\frac{l_2 l_{10}}{l_3
   l_4},\frac{l_3 l_4 l_6}{l_{10}},\frac{l_7 l_9}{l_8^2},\frac{l_{11}^2}{l_8^2},-\frac{l_9 l_{11}}{l_8 l_{10}},-\frac{l_{11}^2}{l_7
   l_{10}},-\frac{l_6^2}{l_5 l_7},\frac{l_6 l_{11}}{l_5 l_7},\frac{l_{10}}{l_3 l_4 l_6},
   \nonumber \\
   &\frac{l_3 l_4}{l_2 l_{10}},\frac{l_4 l_{11}}{l_2
   l_8},\frac{l_{11}}{l_2 l_5 l_7},\frac{l_9}{l_2 l_3 l_6},-\frac{l_{10}}{l_2^2 l_8},-\frac{l_{10}}{l_2^2 l_7},\frac{l_9}{l_2^2
   l_7},-\frac{l_9}{l_2^2 l_6},-\frac{l_2^2}{l_1 l_3},-\frac{l_2 l_{11}}{l_1 l_3},\frac{l_4 l_7}{l_1 l_8},-\frac{l_4 l_{11}}{l_1
   l_{10}},-\frac{l_{11}}{l_1 l_3 l_6},
   \nonumber \\
   &\frac{l_{10}}{l_1 l_3 l_6},\frac{l_9}{l_1 l_3 l_6},\frac{l_9}{l_1 l_3 l_5},-\frac{l_2^3}{l_3
   l_4},-\frac{l_2 l_7 l_{11}}{l_8 l_{10}},-\frac{l_7 l_9}{l_3^2 l_6},\frac{l_{11}^2}{l_3^2 l_6},-\frac{l_8 l_{11}}{l_3^2 l_6},\frac{l_9
   l_{11}}{l_2 l_7 l_8},\frac{l_8 l_{11}}{l_2 l_7^2},\frac{l_8 l_{11}}{l_2 l_5^2},-\frac{l_3 l_4}{l_2^3},-\frac{l_8 l_{11}}{l_1^2
   l_6},
   \nonumber \\
   &\frac{l_1 l_3 l_6}{l_2^2 l_7},\frac{l_2^2 l_8}{l_3 l_4 l_6},-\frac{l_9 l_{11}^2}{l_8^2 l_{10}},\frac{l_9 l_{11}}{l_2 l_3 l_5
   l_7},\frac{l_3 l_4 l_6}{l_2^2 l_8},-\frac{l_9 l_{10}}{l_2^2 l_7^2},\frac{l_6 l_9}{l_2^2 l_5 l_7},\frac{l_2^2 l_7}{l_1 l_3
   l_6},-\frac{l_{11}^2}{l_1 l_3 l_5 l_7},-\frac{l_9 l_{11}}{l_1 l_3^2 l_6},\frac{l_2 l_7 l_{11}}{l_1 l_3^2 l_6},
   \nonumber \\
   &\frac{l_9 l_{10}}{l_1^2
   l_3^2 l_6},\frac{l_9 l_{11}^2}{l_1 l_3^3 l_5 l_6}\Bigg\}.
\end{align}
Note that \eqref{eq:linargsforb2}  and \eqref{eq:linargsforb3} are not identical even though they are of identical length. As before, we find 3420 preferred pairs of function arguments for ${\rm Li}_{2,1}$, ${\rm Li}_{3,1}$, and ${\rm Li}_{2,2}$ in $\mathcal{D}_{b_3}$.

Although it is not {\it a priori} clear that such restrictive choices for the ans\"{a}tze of {\rm Li} functions in each region are actually allowed, it turns out that we can in fact use the sets of function arguments derived from the considerations outlined above to write explicit solutions which satisfy the differential equations for the twelve two-massive-line master integrals with rational symbols in $\mathcal{D}_a$, $\mathcal{D}_{b_1}$, $\mathcal{D}_{b_2}$, and $\mathcal{D}_{b_3}$. We have explicitly checked numerically against the $G$ function representation with kinematic variables in the arguments
at a large number of physical phase space points that, as desired, our results are correct independent of whether explicit positive imaginary parts are assigned to $w$ and $z$. In the next section, we will discuss the more complicated master integrals which involve the intrinsically algebraic part of the full symbol alphabet, $\mathcal{L}_A$, in a similar manner.

\section{Weight-four multiple polylogarithms for Drell-Yan master integrals}
\label{sec:DYresult}

In this section, we apply the techniques introduced in Sections \ref{sec:diffeqint} and \ref{sec:i0} to  explicitly integrate the $\epsilon\,\ud \ln$ form for $\{\mathbf{m}_1,\ldots,\mathbf{m}_{36}\}$, \eqref{eq:dlogform}, in the physical region above the two-mass threshold  ($s > 4\,m^2$) through to weight four in terms of real-valued {\rm Li} functions.
To our knowledge, this is the first time a complete solution
in terms of standard multiple polylogarithms has been found for a multi-loop Feynman integral with an $\epsilon\,\ud \ln$ differential equation containing unrationalizable symbol letters. We fix our integration constants using a variety of established techniques, such as direct evaluation in Feynman parameters, exploiting regular kinematic limits of the differential equations, and exploiting kinematic limits where particular integrals are power-suppressed and must therefore vanish (see Section 5.2 of \cite{Bonciani:2016ypc}).

Due to the significant length of our results,
we refrain from presenting them explicitly in the text. For the benefit of the reader interested in the detailed structure of our solution, we provide the result through to weight four for the six-line normal form integral from the right panel of Figure \ref{fig:algLS}, $\mathbf{m}_{32}$, as an ancillary file included with our arXiv submission. $\mathbf{m}_{32}$ has a number of notable analytic features. We find a basis of multiple polylogarithms comprised of: 14 $\ln$ functions (by construction, $\ln(l_{10})$ and $\ln(l_{11})$ cannot appear), 46 ${\rm Li}_2$ functions, 235 ${\rm Li}_3$ functions, 342 ${\rm Li}_4$ functions, 28 ${\rm Li}_{2,1}$ functions, 742 ${\rm Li}_{3,1}$ functions, and 324 ${\rm Li}_{2,2}$ functions. To give the reader a feeling for the complexity of our result for $\mathbf{m}_{32}$, we present the complete list of 347 ${\rm Li}_n$ function arguments which appear in it:
\begin{align}
\Bigg\{&\frac{1}{l_1},-l_2,l_2,\frac{l_2}{l_1^2},\frac{l_2}{l_1},-l_1
   l_2,-\frac{l_2^2}{l_1},l_2^3,-\frac{l_2}{l_3},-\frac{l_2^2}{l_1
   l_3},l_3,\frac{l_4}{l_1^2},-\frac{l_4}{l_2},\frac{1}{l_5},\frac
   {l_2 l_5}{l_1},-l_6,l_6,l_2 l_6,l_2^2
   l_6,\frac{l_6}{l_5},\frac{l_1 l_6}{l_5},
   \nonumber \\
   &-\frac{l_1 l_3}{l_2^2
   l_7},-\frac{l_3}{l_2 l_7},-\frac{l_6}{l_7},-\frac{l_3
   l_6}{l_7},-\frac{l_1 l_3 l_6}{l_7},-\frac{l_6^2}{l_5
   l_7},l_7,\frac{l_3^2
   l_6}{l_8^2},\frac{l_3}{l_8},-\frac{l_3}{l_2 l_8},-\frac{l_2
   l_5}{l_8},-\frac{l_1 l_6}{l_8},\frac{l_3 l_6}{l_8},-\frac{l_2
   l_3 l_6}{l_8},\frac{l_7}{l_8},
   \nonumber \\
   &\frac{l_2 l_7}{l_8},\frac{l_2
   l_7}{l_1
   l_8},l_8,\frac{l_8}{l_1},\frac{l_8}{l_5},\frac{l_3}{l_9},\frac{
   l_1 l_3}{l_9},-\frac{l_2^2 l_6}{l_9},\frac{l_1 l_3
   l_6}{l_9},\frac{l_2 l_3 l_6}{l_9},-\frac{l_3^2 l_6}{l_7
   l_9},\frac{l_7}{l_9},-\frac{l_2 l_7}{l_9},\frac{l_2^2
   l_7}{l_9},\frac{l_8}{l_9},\frac{l_2 l_8}{l_9},\frac{l_7
   l_9}{l_8^2},\frac{l_9}{l_1 l_8},
   \nonumber \\
   &\frac{l_{10}}{l_1
   l_3},\frac{l_{10}}{l_1 l_3 l_5},\frac{l_{10}}{l_1 l_3
   l_6},-\frac{l_{10}}{l_7},-\frac{l_{10}}{l_2^2
   l_7},-\frac{l_{10}}{l_2^2 l_5 l_7},-\frac{l_6 l_{10}}{l_5
   l_7},\frac{l_{10}}{l_1 l_8},-\frac{l_2
   l_{10}}{l_8},-\frac{l_{10}}{l_9},\frac{l_{10}}{l_6
   l_9},\frac{l_6 l_{10}}{l_9},\frac{l_9 l_{10}}{l_1^2 l_3^2
   l_6},
   \nonumber \\
   &-\frac{l_9 l_{10}}{l_2^2
   l_7^2},\frac{l_{11}}{l_1},\frac{l_{11}}{l_2},-\frac{l_{11}}{l_3
   },-\frac{l_2 l_{11}}{l_1
   l_3},-\frac{l_{11}}{l_5},\frac{l_{11}}{l_2
   l_5},-\frac{l_{11}}{l_3
   l_5},-\frac{l_{11}}{l_6},-\frac{l_{11}}{l_1
   l_6},-\frac{l_{11}}{l_3 l_6},-\frac{l_{11}}{l_1 l_3
   l_6},\frac{l_{11}}{l_7},\frac{l_{11}}{l_1
   l_7},
   \nonumber \\
   &\frac{l_{11}}{l_2 l_7},\frac{l_{11}}{l_2 l_5
   l_7},\frac{l_6 l_{11}}{l_5
   l_7},-\frac{l_{11}}{l_8},\frac{l_{11}}{l_8},\frac{l_{11}}{l_2
   l_8},\frac{l_2 l_{11}}{l_8},-\frac{l_{11}}{l_6 l_8},-\frac{l_6
   l_{11}}{l_8},\frac{l_8 l_{11}}{l_2 l_5^2},-\frac{l_8
   l_{11}}{l_1^2 l_6},-\frac{l_8 l_{11}}{l_3^2 l_6},\frac{l_8
   l_{11}}{l_2 l_7^2},-\frac{l_1 l_{11}}{l_9},
   \nonumber \\
   &-\frac{l_2
   l_{11}}{l_1 l_9},\frac{l_2^2 l_{11}}{l_9},-\frac{l_9
   l_{11}}{l_1 l_3^2 l_6},\frac{l_9 l_{11}}{l_2 l_7
   l_8},-\frac{l_{10} l_{11}}{l_8 l_9},-\frac{l_{11}^2}{l_1 l_3
   l_5 l_7},\frac{l_1 l_7 l_9}{l_{13}^2},\frac{l_1
   l_3}{l_{13}},\frac{l_7}{l_{13}},\frac{l_1
   l_7}{l_{13}},\frac{l_1
   l_8}{l_{13}},\frac{l_9}{l_{13}},\frac{l_9}{l_5
   l_{13}},\frac{l_7 l_9}{l_{13}},
   \nonumber \\
   &\frac{l_7 l_9}{l_8
   l_{13}},\frac{l_1 l_{11}}{l_2 l_{13}},-\frac{l_1 l_7
   l_9}{l_{12}
   l_{13}},\frac{l_{12}}{l_{13}},\frac{l_{13}}{l_1},-\frac{l_1 l_2
   l_7 l_{10}}{l_{14}^2},-\frac{l_{10} l_{13}^2}{l_7
   l_{14}^2},-\frac{l_2 l_{10} l_{13}^2}{l_9 l_{14}^2},-\frac{l_1
   l_3}{l_{14}},\frac{l_1 l_2 l_7}{l_{14}},\frac{l_2^2
   l_7}{l_{14}},-\frac{l_{10}}{l_{14}},
   \nonumber \\
   &-\frac{l_1 l_{10}}{l_2
   l_{14}},-\frac{l_{10}}{l_5 l_{14}},-\frac{l_7
   l_{10}}{l_{14}},\frac{l_2 l_7 l_{10}}{l_3 l_{14}},-\frac{l_7
   l_{10}}{l_8 l_{14}},\frac{l_1 l_{11}}{l_{14}},\frac{l_1 l_2 l_7
   l_{10}}{l_{12} l_{14}},\frac{l_2 l_7 l_{13}}{l_8
   l_{14}},-\frac{l_1 l_3^2 l_6 l_{13}}{l_7 l_9 l_{14}},\frac{l_2
   l_7 l_{13}}{l_9 l_{14}},\frac{l_2^2 l_7 l_{13}}{l_9
   l_{14}},
   \nonumber \\
   &-\frac{l_{10} l_{13}}{l_7 l_{14}},-\frac{l_{10}
   l_{13}}{l_1 l_7 l_{14}},-\frac{l_{10} l_{13}}{l_2 l_7
   l_{14}},-\frac{l_1 l_{10} l_{13}}{l_4 l_7 l_{14}},-\frac{l_{10}
   l_{13}}{l_8 l_{14}},-\frac{l_{10} l_{13}}{l_9
   l_{14}},-\frac{l_2 l_{10} l_{13}}{l_9 l_{14}},-\frac{l_8 l_{10}
   l_{13}}{l_7 l_9 l_{14}},\frac{l_{11} l_{13}}{l_7
   l_{14}},
   \nonumber \\
   &\frac{l_2 l_{11} l_{13}}{l_9 l_{14}},\frac{l_2 l_7
   l_{10} l_{13}}{l_3 l_{12} l_{14}},\frac{l_{14}}{l_1
   l_2},\frac{l_{14}}{l_1 l_2 l_5
   l_7},\frac{l_{14}}{l_{13}},\frac{l_{14}}{l_2 l_{13}},\frac{l_9
   l_{14}}{l_1 l_2 l_7 l_{13}},-\frac{l_7
   l_{10}}{l_{15}^2},\frac{l_2 l_6 l_7 l_{10}}{l_{15}^2},\frac{l_2
   l_6 l_{10} l_{13}^2}{l_1 l_9 l_{15}^2},
   \nonumber \\
   &-\frac{l_2^2 l_6 l_7
   l_{10}^2 l_{13}^2}{l_9 l_{14}^2 l_{15}^2},-\frac{l_6
   l_{14}^2}{l_1 l_{15}^2},-\frac{l_6 l_7 l_9 l_{14}^2}{l_{13}^2
   l_{15}^2},\frac{l_1 l_3
   l_6}{l_{15}},-\frac{l_7}{l_{15}},\frac{l_2
   l_7}{l_{15}},\frac{l_2 l_5 l_7}{l_{15}},\frac{l_2 l_6
   l_7}{l_{15}},\frac{l_2 l_3 l_6 l_7}{l_8
   l_{15}},
   \nonumber \\
   &\frac{l_{10}}{l_{15}},-\frac{l_{10}}{l_2
   l_{15}},\frac{l_6 l_{10}}{l_{15}},-\frac{l_2 l_6
   l_{10}}{l_{15}},\frac{l_6 l_{10}}{l_5 l_{15}},\frac{l_2 l_7
   l_{10}}{l_1 l_3 l_{15}},\frac{l_3 l_6 l_{10}}{l_8
   l_{15}},\frac{l_8 l_{10}}{l_3 l_{15}},-\frac{l_2 l_7
   l_{11}}{l_3 l_{15}},\frac{l_2 l_{10} l_{11}}{l_3 l_4
   l_{15}},-\frac{l_2 l_6 l_7 l_{10}}{l_{12} l_{15}},
   \nonumber \\
   &\frac{l_2
   l_{13}}{l_1 l_{15}},-\frac{l_2^2 l_6 l_{13}}{l_9
   l_{15}},\frac{l_1 l_3 l_6 l_{13}}{l_9 l_{15}},\frac{l_2 l_3 l_6
   l_{13}}{l_9 l_{15}},-\frac{l_{10} l_{13}}{l_1 l_2
   l_{15}},\frac{l_{10} l_{13}}{l_1 l_3 l_{15}},\frac{l_2 l_{10}
   l_{13}}{l_1^2 l_3 l_{15}},\frac{l_2^2 l_{10} l_{13}}{l_1 l_3
   l_4 l_{15}},\frac{l_6 l_{10} l_{13}}{l_1 l_8 l_{15}},\frac{l_6
   l_{10} l_{13}}{l_9 l_{15}},
   \nonumber \\
   &\frac{l_2 l_6 l_{10} l_{13}}{l_1 l_9
   l_{15}},-\frac{l_2 l_6 l_{11} l_{13}}{l_9 l_{15}},-\frac{l_2
   l_6 l_{10} l_{13}}{l_{12} l_{15}},-\frac{l_{10} l_{13}}{l_{14}
   l_{15}},-\frac{l_2 l_6 l_{10} l_{13}}{l_{14} l_{15}},\frac{l_2
   l_7 l_{10} l_{13}}{l_3 l_{14} l_{15}},-\frac{l_2 l_3 l_6 l_{10}
   l_{13}}{l_9 l_{14} l_{15}},\frac{l_2 l_7 l_{10} l_{13}}{l_9
   l_{14} l_{15}},
   \nonumber \\
   &\frac{l_2^2 l_6 l_7 l_{10} l_{13}}{l_9 l_{14}
   l_{15}},-\frac{l_2 l_{10}^2 l_{13}}{l_3 l_4 l_{14}
   l_{15}},\frac{l_1 l_2 l_3 l_6 l_{11} l_{13}}{l_9 l_{14}
   l_{15}},\frac{l_2 l_6 l_{10} l_{11} l_{13}}{l_5 l_9 l_{14}
   l_{15}},\frac{l_2 l_6 l_7 l_{10}^2 l_{13}}{l_8 l_{12} l_{14}
   l_{15}},\frac{l_{14}}{l_1 l_{15}},-\frac{l_6
   l_{14}}{l_{15}},-\frac{l_6 l_{14}}{l_1 l_{15}},
   \nonumber \\
   &-\frac{l_6
   l_{14}}{l_7 l_{15}},-\frac{l_3 l_6 l_{14}}{l_7
   l_{15}},\frac{l_3 l_6 l_{14}}{l_2 l_7 l_{15}},\frac{l_7
   l_{14}}{l_1 l_{15}},-\frac{l_2 l_7 l_{14}}{l_1 l_3
   l_{15}},-\frac{l_2 l_6 l_7 l_{14}}{l_1 l_8 l_{15}},-\frac{l_6
   l_{11} l_{14}}{l_2 l_7 l_{15}},\frac{l_2 l_6 l_7 l_{14}}{l_{12}
   l_{15}},-\frac{l_3 l_6 l_{14}}{l_{13} l_{15}},\frac{l_7
   l_{14}}{l_{13} l_{15}},
   \nonumber \\
   &\frac{l_2 l_6 l_7 l_{14}}{l_{13}
   l_{15}},-\frac{l_2 l_7^2 l_{14}}{l_3 l_{13} l_{15}},\frac{l_3
   l_6 l_9 l_{14}}{l_2 l_7 l_{13} l_{15}},-\frac{l_1 l_3 l_6
   l_{11} l_{14}}{l_2 l_7 l_{13} l_{15}},-\frac{l_6 l_9 l_{11}
   l_{14}}{l_2 l_5 l_7 l_{13} l_{15}},\frac{l_2 l_6 l_7^2 l_9
   l_{14}}{l_8 l_{12} l_{13} l_{15}},-\frac{l_6 l_{11} l_{13}
   l_{14}}{l_2 l_7 l_9 l_{15}},-\frac{l_{15}}{l_2^2 l_6
   l_7},
   \nonumber \\
   &\frac{l_{15}}{l_{13}},-\frac{l_1^2 l_3 l_{15}}{l_2^2 l_7
   l_{14}},\frac{l_2 l_7 l_{15}}{l_8 l_{14}},\frac{l_1 l_2 l_7
   l_{15}}{l_{12} l_{14}},\frac{l_1^2 l_{11} l_{15}}{l_2 l_{13}
   l_{14}},-\frac{l_2 l_{13} l_{15}}{l_9 l_{14}},-\frac{l_{13}
   l_{15}}{l_6 l_9 l_{14}},\frac{l_2 l_7 l_{13} l_{15}}{l_8 l_{12}
   l_{14}},-\frac{l_2 l_6 l_7 l_9}{l_{16}^2},
   \nonumber \\
   &-\frac{l_1 l_2 l_6
   l_7^2 l_9^2}{l_{13}^2 l_{16}^2},\frac{l_1 l_2^2 l_6 l_7^2 l_9
   l_{10}}{l_{14}^2 l_{16}^2},-\frac{l_2^2 l_6^2 l_7^2 l_9
   l_{10}}{l_{15}^2 l_{16}^2},\frac{l_2 l_6^2 l_7^2 l_9^2
   l_{14}^2}{l_{13}^2 l_{15}^2 l_{16}^2},-\frac{l_1^2 l_2 l_7 l_9
   l_{11} l_{15}}{l_{13} l_{14} l_{16}^2},-\frac{l_2 l_5
   l_7}{l_{16}},-\frac{l_2 l_6 l_7}{l_{16}},
   \nonumber \\
   &-\frac{l_6 l_9}{l_2
   l_{16}},-\frac{l_2 l_7 l_9}{l_1 l_3 l_{16}},\frac{l_3 l_6
   l_9}{l_8 l_{16}},-\frac{l_9 l_{11}}{l_3 l_{16}},\frac{l_2 l_6
   l_7 l_9}{l_{12} l_{16}},\frac{l_1 l_3 l_6 l_9}{l_{13}
   l_{16}},-\frac{l_1 l_3 l_6 l_9}{l_2 l_{13} l_{16}},-\frac{l_1
   l_7 l_9}{l_{13} l_{16}},\frac{l_2 l_6 l_7 l_9}{l_{13}
   l_{16}},-\frac{l_2 l_7^2 l_9}{l_3 l_{13} l_{16}},
   \nonumber \\
   &-\frac{l_1 l_9
   l_{11}}{l_{13} l_{16}},\frac{l_1 l_2 l_6 l_7^2 l_9}{l_{12}
   l_{13} l_{16}},\frac{l_1 l_2 l_3 l_6 l_7}{l_{14}
   l_{16}},\frac{l_1^2 l_2 l_3 l_6 l_7}{l_{14} l_{16}},\frac{l_2
   l_6 l_7 l_9}{l_{14} l_{16}},\frac{l_2 l_7 l_{10}}{l_{14}
   l_{16}},\frac{l_1 l_2 l_7 l_{10}}{l_{14} l_{16}},\frac{l_2 l_7
   l_9 l_{10}}{l_{14} l_{16}},\frac{l_2 l_7 l_9 l_{10}}{l_3 l_{14}
   l_{16}},
   \nonumber \\
   &\frac{l_2 l_7 l_9 l_{10}}{l_1 l_3 l_{14}
   l_{16}},\frac{l_2 l_7 l_9 l_{10}}{l_3 l_4 l_{14}
   l_{16}},\frac{l_2 l_7 l_9 l_{10}}{l_8 l_{14} l_{16}},-\frac{l_1
   l_2 l_7 l_{11}}{l_{14} l_{16}},-\frac{l_1 l_6 l_7 l_9
   l_{10}}{l_{12} l_{14} l_{16}},\frac{l_2 l_7^2 l_9 l_{10}}{l_8
   l_{13} l_{14} l_{16}},\frac{l_2 l_7 l_9 l_{10} l_{13}}{l_1 l_3
   l_8 l_{14} l_{16}},\frac{l_2 l_6 l_7^2}{l_{15}
   l_{16}},
   \nonumber \\
   &-\frac{l_2^2 l_6 l_7^2}{l_{15} l_{16}},\frac{l_2^3 l_6
   l_7^2}{l_{15} l_{16}},\frac{l_2 l_6 l_7 l_9}{l_{15}
   l_{16}},-\frac{l_2 l_6 l_7 l_{10}}{l_{15} l_{16}},\frac{l_6 l_9
   l_{10}}{l_{15} l_{16}},-\frac{l_6 l_9 l_{10}}{l_2 l_{15}
   l_{16}},-\frac{l_2 l_6 l_9 l_{10}}{l_{15} l_{16}},-\frac{l_2
   l_6 l_9 l_{10}}{l_4 l_{15} l_{16}},\frac{l_2^2 l_6 l_7^2 l_9
   l_{10}}{l_1 l_3 l_{12} l_{15} l_{16}},
   \nonumber \\
   &\frac{l_2^2 l_6 l_7 l_8
   l_{13}}{l_{14} l_{15} l_{16}},-\frac{l_6 l_7 l_{10}
   l_{13}}{l_{14} l_{15} l_{16}},\frac{l_1 l_2 l_6 l_7 l_{10}
   l_{13}}{l_{14} l_{15} l_{16}},\frac{l_2^2 l_6 l_7 l_{10}
   l_{13}}{l_1 l_{14} l_{15} l_{16}},-\frac{l_2^3 l_6 l_7 l_{10}
   l_{13}}{l_{14} l_{15} l_{16}},\frac{l_2 l_6 l_7 l_{10}^2
   l_{13}}{l_8 l_{14} l_{15} l_{16}},
   \nonumber \\
   &-\frac{l_2^2 l_6 l_7 l_{10}
   l_{11} l_{13}}{l_9 l_{14} l_{15} l_{16}},\frac{l_1 l_3^2 l_6^2
   l_9 l_{14}}{l_2 l_7 l_{13} l_{15} l_{16}},\frac{l_6 l_7 l_9
   l_{14}}{l_{13} l_{15} l_{16}},\frac{l_2 l_6 l_7 l_9
   l_{14}}{l_{13} l_{15} l_{16}},\frac{l_2 l_6 l_7 l_9 l_{14}}{l_1
   l_{13} l_{15} l_{16}},\frac{l_2 l_6 l_7^2 l_9 l_{14}}{l_8
   l_{13} l_{15} l_{16}},\frac{l_6 l_8 l_9 l_{14}}{l_{13} l_{15}
   l_{16}},
   \nonumber \\
   &\frac{l_6 l_9 l_{11} l_{14}}{l_{13} l_{15}
   l_{16}},\frac{l_3 l_6 l_9 l_{12} l_{14}}{l_2 l_7 l_{13} l_{15}
   l_{16}},\frac{l_6 l_9^2 l_{14}^2}{l_2 l_{13}^2 l_{15}
   l_{16}},-\frac{l_6 l_{15}}{l_{16}},\frac{l_6 l_9 l_{15}}{l_5
   l_{13} l_{16}},-\frac{l_1 l_2 l_6 l_{15}}{l_{14}
   l_{16}},-\frac{l_{12} l_{15}}{l_6 l_{14} l_{16}},\frac{l_1 l_2
   l_6 l_7 l_9 l_{15}}{l_{13} l_{14} l_{16}},
   \nonumber \\
   &-\frac{l_7 l_9 l_{12}
   l_{15}}{l_6 l_{13} l_{14} l_{16}},-\frac{l_{16}}{l_2
   l_7},-\frac{l_3 l_{16}}{l_2 l_7
   l_8},\frac{l_{16}}{l_9},-\frac{l_{16}}{l_2
   l_9},\frac{l_{16}}{l_6 l_9},-\frac{l_3 l_4 l_{16}}{l_2 l_8
   l_9},-\frac{l_{11} l_{16}}{l_3 l_6
   l_9},\frac{l_{16}}{l_{12}},\frac{l_{13} l_{16}}{l_1 l_6
   l_9},-\frac{l_{13} l_{16}}{l_2 l_7^2 l_9},
   \nonumber \\
   &-\frac{l_1 l_3 l_{13}
   l_{16}}{l_2 l_7^2 l_9},-\frac{l_{13} l_{16}}{l_2 l_7
   l_9},\frac{l_{13} l_{16}}{l_1 l_6 l_8 l_9},-\frac{l_8 l_{13}
   l_{16}}{l_2 l_7^2 l_9},\frac{l_{13} l_{16}}{l_9
   l_{12}},\frac{l_{14} l_{16}}{l_1 l_2 l_6 l_7},\frac{l_{14}
   l_{16}}{l_1 l_2 l_6 l_7 l_9},\frac{l_{16}}{l_{15}},-\frac{l_6
   l_{16}}{l_{15}},-\frac{l_{10} l_{16}}{l_9 l_{15}},
   \nonumber \\
   &\frac{l_{10}
   l_{16}}{l_6 l_9 l_{15}},\frac{l_6 l_{10} l_{16}}{l_9
   l_{15}},\frac{l_{11} l_{16}}{l_8 l_{15}},-\frac{l_{10} l_{11}
   l_{16}}{l_8 l_9 l_{15}},\frac{l_6 l_{10} l_{16}}{l_5 l_{12}
   l_{15}},\frac{l_{13} l_{16}}{l_9 l_{15}},\frac{l_{13}
   l_{16}}{l_1 l_9 l_{15}},-\frac{l_6 l_{13} l_{16}}{l_7 l_9
   l_{15}},-\frac{l_3 l_6 l_{13} l_{16}}{l_7 l_9
   l_{15}},\frac{l_{10} l_{13} l_{16}}{l_1 l_6 l_9
   l_{15}},
   \nonumber \\
   &\frac{l_{10} l_{13} l_{16}}{l_1 l_3 l_6 l_9
   l_{15}},-\frac{l_{10} l_{13} l_{16}}{l_7 l_9
   l_{15}},-\frac{l_{10} l_{13} l_{16}}{l_1 l_7 l_9
   l_{15}},-\frac{l_6 l_{10} l_{13} l_{16}}{l_1 l_5 l_7 l_9
   l_{15}},-\frac{l_2 l_{10} l_{13} l_{16}}{l_1 l_8 l_9
   l_{15}},\frac{l_{11} l_{13} l_{16}}{l_7 l_9 l_{15}},\frac{l_6
   l_{10} l_{13} l_{16}}{l_9 l_{12} l_{15}},\frac{l_{14}
   l_{16}}{l_1 l_2 l_7 l_{15}},
   \nonumber \\
   &\frac{l_6 l_{14} l_{16}}{l_1 l_2
   l_5 l_7 l_{15}},\frac{l_{14} l_{16}}{l_1 l_8
   l_{15}},\frac{l_{14} l_{16}}{l_9 l_{15}},\frac{l_2 l_{14}
   l_{16}}{l_1 l_9 l_{15}},-\frac{l_6 l_{14} l_{16}}{l_7 l_9
   l_{15}},\frac{l_3 l_6 l_{14} l_{16}}{l_2 l_7 l_9
   l_{15}},\frac{l_8 l_{14} l_{16}}{l_2 l_7 l_9
   l_{15}},\frac{l_{11} l_{14} l_{16}}{l_7 l_9
   l_{15}},\frac{l_{13} l_{14} l_{16}}{l_1 l_7 l_9
   l_{15}},
   \nonumber \\
   &\frac{l_{13} l_{14} l_{16}}{l_1 l_2 l_7 l_9
   l_{15}},\frac{l_2 l_{13} l_{14} l_{16}}{l_1 l_7 l_9
   l_{15}},\frac{l_6 l_{13} l_{14} l_{16}}{l_1 l_7 l_9
   l_{15}},\frac{l_6 l_{13} l_{14} l_{16}}{l_1 l_2 l_7 l_9
   l_{15}},\frac{l_{12} l_{13} l_{14} l_{16}}{l_1^2 l_2 l_6 l_7
   l_9 l_{15}},-\frac{l_3 l_{12} l_{13} l_{15} l_{16}}{l_2^2 l_6
   l_7^3 l_9 l_{14}},-\frac{l_{10} l_{16}^2}{l_9
   l_{15}^2},
   \nonumber \\
   &-\frac{l_{10} l_{13}^2 l_{16}^2}{l_1 l_7 l_9^2
   l_{15}^2},\frac{l_{14}^2 l_{16}^2}{l_1 l_2 l_7 l_9
   l_{15}^2},\frac{l_{13}^2 l_{14}^2 l_{16}^2}{l_1^2 l_2 l_7^2
   l_9^2 l_{15}^2},\frac{l_{11} l_{13} l_{14} l_{16}^2}{l_2 l_6
   l_7^2 l_9^2 l_{15}}\Bigg\}.
\end{align}
It is also worth pointing out that our choice of alphabet, Eq. \eqref{eq:fullalphabet}, produces remarkably simple integration constants. We find
\begin{equation}
\label{eq:intconstwt3}
    \frac{3}{2}\zeta_3  -\frac{3}{2} \pi^3 i
\end{equation}
at weight three and 
\begin{equation}
\label{eq:intconstwt4}
    \frac{1335349}{32}\zeta_2^2 - 116\,\zeta_3\,\pi i
\end{equation}
at weight four.

In lieu of our explicit analytic results, it is straightforward to obtain high-precision numerical results for our master integrals using {\tt GiNaC}.
For the randomly chosen physical phase space point
\begin{equation}
\label{eq:finalPSptDY}
    \left(s,t,m^2\right) = \left(17, -7, 6241/1681\right).
\end{equation}
we find for the most complicated master integrals with two massive internal lines
\newpage
\begin{align}
\label{eq:numresults}
&\mathbf{m}_{32}
\approx \epsilon^3 \big(0.066537984962080530758\ldots - 27.508245870011457529\ldots\, i\big) 
\nonumber \\
&\qquad + \epsilon^4 \big(51.615607433806381131\ldots - 149.06326619542437190\ldots\, i\big) +\mathcal{O}\left(\epsilon^5\right),
\\
&\mathbf{m}_{33}
\approx \epsilon^2 \big(10.163316917366188927\ldots + 6.2974465571355440423\ldots\, i\big)
\nonumber \\
&\qquad + \epsilon^3 \big(33.914009430201406423\ldots + 4.6486595371603574921\ldots\, i\big)
\nonumber \\
&\qquad + \epsilon^4 \big(163.17321004422879959\ldots - 128.72756457117576796\ldots\, i\big)
+\mathcal{O}\left(\epsilon^5\right),
\\
&\mathbf{m}_{34}
\approx \epsilon^2 \big(-9.3166453894096456380\ldots - 4.6722528592943756861\ldots\, i\big)
\nonumber \\
&\qquad + \epsilon^3 \big(-12.274144284891231677\ldots - 11.270075866466130873\ldots\, i\big)
\nonumber \\
&\qquad + \epsilon^4 \big(-51.057330106861359687\ldots + 87.629800828432935443\ldots\, i\big)
+\mathcal{O}\left(\epsilon^5\right),
\\
&\mathbf{m}_{35}
\approx \epsilon^4 \big(6.9039856473317646358\ldots -0.013343873471826080269\ldots\, i\big)
 +\mathcal{O}\left(\epsilon^5\right),
\\
&\mathbf{m}_{36}
\approx \epsilon^3 \big(-4.7564239669560836801\ldots + 4.0753242804814306037\ldots\, i\big)
\nonumber \\
&\qquad + \epsilon^4 \big(-8.5216864119748844907\ldots - 13.318764764536663942\ldots\, i\big)
+\mathcal{O}\left(\epsilon^5\right).
\end{align}
Using the finite integral method \cite{vonManteuffel:2014qoa,vonManteuffel:2015gxa,vonManteuffel:2017myy} and {\tt SecDec 3} \cite{Borowka:2015mxa}, we were able to independently check these numerical results to a few decimal digits.

We find that a double precision evaluation of all 36 master integrals ${m_1,\ldots,m_{36}}$ at the point \eqref{eq:finalPSptDY} using {\tt GiNaC} takes $0.5$ seconds on one Intel E3-1275 CPU core.
The performance in this central point in phase space is therefore not much worse than what is encountered for Feynman integrals involving rational alphabets, see \emph{e.g.}\ the discussion in \cite{Gehrmann:2015ora}.
Therefore, we expect the functional representation discussed in this paper to be well-suited for direct usage in Monte-Carlo programs.

\section{Outlook}
\label{sec:outlook}

In this article, we considered irreducible Feynman integrals
satisfying $\epsilon\,\ud \ln$ differential equations with non-rational symbol letters and described methods for their evaluation in terms of standard multiple polylogarithms.
We show for the first time that a complete solution in terms of standard multiple polylogarithms can be obtained even in the presence of unrationalizable symbol letters.
In particular, new techniques for the construction of an ansatz which matches the symbol in a particular region of phase space allowed us to calculate the two-loop master integrals for the mixed EW-QCD corrections to Drell-Yan lepton pair production in the physical region using ${\rm Li}$ functions through to weight four.
We discussed in detail how to optimize the functional basis to allow for fast and stable numerical evaluations, systematically avoiding arguments on branch cuts which would require an explicit $+i\,0$ prescription.
The presented techniques and results may be applied directly to the calculation of the virtual amplitudes for the full EW-QCD corrections to Drell-Yan production of relative order $\alpha \alpha_s$.

As motivation, we also considered a rather non-trivial master integral for massive Bhabha electron-positron scattering and showed that it may be directly integrated from Feynman parameters to all orders in $\epsilon$ in terms of multiple polylogarithms despite the presence of root-valued symbol letters. Besides the applications discussed in this paper, $\epsilon\,\ud\ln$ differential equations with root-valued symbol letters also appear in other interesting contexts, such as the next-to-leading order QCD corrections to $H$+jet production with full heavy top quark mass dependence.
One may therefore expect that our techniques could be fruitfully applied also to other problems of current phenomenological interest.

\acknowledgments
We gratefully acknowledge Erik Panzer for inspiring and stimulating discussions over the last several years about various aspects of this work. RMS is also very grateful to Francis Brown, Erik Panzer, and All Souls College of Oxford University for their hospitality at an early stage of this work. We thank Stefan M\"{u}ller-Stach, Duco van Straten, and Marco Besier for many enlightening discussions about the theory of rational parametrization. We would also like to thank Hubert Spiesberger for his support and for collaborations on related topics.
We thank Roberto Bonciani, Stefano Di Vita, Pierpaolo Mastrolia, and Ulrich Schubert for a comparison of results.
MH was supported in part by the German Research Foundation (DFG), through the Collaborative Research Center, Project ID 204404729, SFB 1044, and the Cluster of Excellence PRISMA$^+$, Project ID 39083149, EXC 2118/1.
AvM was supported in part by the National Science Foundation under Grant No.\ 1719863.
The authors would like to express a special thanks to the Mainz Institute for Theoretical Physics (MITP) of the Cluster of Excellence PRISMA$^+$ for its hospitality and support.
Our figures were generated using {\tt Jaxodraw} \cite{Binosi:2003yf}, based on {\tt AxoDraw} \cite{Vermaseren:1994je}.

\appendix

\section{Differential equations for the Drell-Yan masters with two massive lines}
\label{sec:dydeq}

In terms of the symbol letters $\{l_1,\ldots,l_{16}\}$  defined in \eqref{eq:fullalphabet}, the system of differential equations for $\mathbf{m}_{1},\ldots,\mathbf{m}_{36}$ in \eqref{eq:normformints} can be cleanly written in differential form: 
\begin{align}
\label{eq:dlogform}
    \epsilon^{-1} &\ud \mathbf{m}_1 = \mathbf{m}_1 \bigg[4 \,\ud \ln \left(l_1\right)-4 \,\ud \ln \left(l_2\right)+4 \,\ud\ln \left(l_3\right)+2 \,\ud\ln \left(l_6\right)-4\,\ud\ln \left(l_7\right)\bigg],
    \nonumber \\
    \epsilon^{-1} &\ud \mathbf{m}_2 = \mathbf{m}_2 \bigg[2\,\ud\ln \left(l_1\right)-\ud\ln \left(l_2\right)\bigg],
    \nonumber \\
    \epsilon^{-1} &\ud \mathbf{m}_3 = \mathbf{m}_3 \bigg[4\,\ud\ln \left(l_1\right)-2\,\ud\ln \left(l_2\right)\bigg],
    \nonumber \\
    \epsilon^{-1} &\ud \mathbf{m}_4 =
    \mathbf{m}_4 \bigg[6 \,\ud\ln \left(l_1\right)-3 \,\ud\ln \left(l_2\right)\bigg]+\mathbf{m}_5 \bigg[\ud\ln \left(l_2\right)-\ud\ln\left(l_4\right)\bigg],
    \nonumber \\
    \epsilon^{-1} &\ud \mathbf{m}_5 = \mathbf{m}_4 \bigg[12 \,\ud\ln \left(l_1\right)-6 \,\ud\ln \left(l_2\right)\bigg]+\mathbf{m}_5 \bigg[4 \,\ud\ln \left(l_1\right)+2 \,\ud\ln\left(l_2\right)-4 \,\ud\ln \left(l_4\right)\bigg],
    \nonumber \\
    \epsilon^{-1} &\ud \mathbf{m}_6 = 0,
    \nonumber \\
    \epsilon^{-1} &\ud \mathbf{m}_7 =
    \mathbf{m}_7 \bigg[4 \,\ud\ln \left(l_1\right)-4 \,\ud\ln \left(l_2\right)+4 \,\ud\ln \left(l_3\right)+2 \,\ud\ln \left(l_6\right)-4
    \,\ud\ln \left(l_7\right)\bigg],
    \nonumber \\
    \epsilon^{-1} &\ud \mathbf{m}_8 =
    \mathbf{m}_1 \bigg[-\ud\ln \left(l_3\right)-\frac{1}{2}\ud\ln \left(l_6\right)+\frac{1}{2}\ud\ln \left(l_8\right)+\frac{1}{2}\ud\ln
    \left(l_{11}\right)\bigg]
    \nonumber \\
    &+\mathbf{m}_3 \bigg[\ud\ln \left(l_3\right)+\frac{1}{2}\ud\ln\left(l_6\right)
    -\frac{1}{2}\ud\ln \left(l_8\right)-\frac{1}{2}\ud\ln \left(l_{11}\right)\bigg]
    \nonumber \\
    &+\mathbf{m}_8
    \bigg[4 \,\ud\ln \left(l_1\right)-\ud\ln \left(l_2\right)+2 \,\ud\ln \left(l_3\right)+\ud\ln \left(l_6\right)
    +2 \,\ud\ln
    \left(l_7\right)-2\, \ud \ln \left(l_8\right)-2 \, \ud\ln \left(l_{11}\right)\bigg],
    \nonumber \\
    \epsilon^{-1} &\ud \mathbf{m}_9 = \mathbf{m}_2\bigg[-\ud\ln \left(l_2\right)\bigg]+\mathbf{m}_9 \bigg[2 \,\ud\ln \left(l_1\right)-2 \,\ud\ln \left(l_3\right)\bigg],
    \nonumber \\
    \epsilon^{-1} &\ud \mathbf{m}_{10} =
     \mathbf{m}_3 \bigg[-\frac{1}{4}\ud\ln \left(l_2\right)\bigg]+ \mathbf{m}_4 \bigg[\frac{3}{2}\ud\ln \left(l_2\right)\bigg]+ \mathbf{m}_5 \bigg[-\frac{3}{4}\ud\ln
   \left(l_2\right)\bigg]
   \nonumber \\
   &+\mathbf{m}_{10} \bigg[4 \,\ud\ln \left(l_1\right)-\ud\ln \left(l_2\right)-2 \,\ud\ln
   \left(l_3\right)\bigg],
   \nonumber \\
   \epsilon^{-1} &\ud \mathbf{m}_{11} =
   \mathbf{m}_{11} \bigg[6 \,\ud \ln \left(l_1\right)-3 \,\ud \ln \left(l_2\right)\bigg]+\mathbf{m}_{12} \bigg[\ud \ln \left(l_2\right)\bigg],
   \nonumber \\
   \epsilon^{-1} &\ud \mathbf{m}_{12} = \mathbf{m}_3 \bigg[-2\,\ud\ln \left(l_2\right)\bigg]+\mathbf{m}_{11} \bigg[3 \,\ud\ln \left(l_2\right)\bigg]+\mathbf{m}_{12} \bigg[4 \,\ud\ln \left(l_1\right)-\ud \ln
   \left(l_2\right)-2 \,\ud\ln \left(l_3\right)\bigg],
   \nonumber \\
   \epsilon^{-1} &\ud \mathbf{m}_{13} = \mathbf{m}_1 \bigg[-3\,\ud \ln \left(l_2\right)-3 \,\ud\ln \left(l_7\right)+\frac{3}{2}\,\ud\ln \left(l_9\right)+\frac{3}{2}\,\ud\ln
   \left(l_{10}\right)\bigg]
   \nonumber \\
   &+\mathbf{m}_6 \bigg[6 \,\ud\ln \left(l_1\right)+6 \,\ud\ln \left(l_3\right)+3 \,\ud\ln\left(l_6\right)-3 \,\ud\ln \left(l_9\right)-3 \,\ud\ln \left(l_{10}\right)\bigg]
   \nonumber \\
   &+\mathbf{m}_{13} \bigg[2 \,\ud\ln
   \left(l_1\right)-4 \,\ud\ln \left(l_2\right)+2 \,\ud\ln \left(l_3\right)+\ud\ln \left(l_6\right)-4 \,\ud\ln
   \left(l_7\right)+\ud\ln \left(l_9\right)+\ud\ln \left(l_{10}\right)\bigg],
   \nonumber \\
   \epsilon^{-1} &\ud \mathbf{m}_{14} = \mathbf{m}_3 \bigg[\ud\ln \left(l_3\right)+\frac{1}{2}\ud\ln \left(l_6\right)-\frac{1}{2}\ud\ln \left(l_8\right)-\frac{1}{2}\ud\ln
   \left(l_{11}\right)\bigg]
   \nonumber \\
   &+\mathbf{m}_7 \bigg[3\,\ud \ln \left(l_2\right)+6\,\ud \ln \left(l_7\right)-3\,\ud \ln
   \left(l_8\right)-3 \,\ud\ln \left(l_{11}\right)\bigg]
   \nonumber \\
   &+\mathbf{m}_{14} \bigg[4 \,\ud\ln \left(l_1\right)-\ud\ln
   \left(l_2\right)+2 \,\ud\ln \left(l_7\right)-\ud\ln \left(l_8\right)-\ud\ln \left(l_{11}\right)\bigg],
   \nonumber \\
   \epsilon^{-1} &\ud \mathbf{m}_{15} = \mathbf{m}_1 \bigg[-\ud\ln \left(l_2\right)-2\,\ud \ln \left(l_7\right)+\ud\ln \left(l_8\right)+\ud\ln
   \left(l_{11}\right)\bigg]
   \nonumber \\
   &+\mathbf{m}_3 \bigg[2 \,\ud\ln \left(l_3\right)+\ud\ln \left(l_6\right)-\ud\ln\left(l_8\right)-\ud\ln \left(l_{11}\right)\bigg]
   \nonumber \\
   &+\mathbf{m}_8 \bigg[4\,\ud \ln \left(l_2\right)+8 \,\ud\ln
   \left(l_7\right)-4 \,\ud\ln \left(l_8\right)-4 \,\ud\ln \left(l_{11}\right)\bigg]+\mathbf{m}_{15} \bigg[4 \,\ud\ln
   \left(l_1\right)-2 \,\ud\ln \left(l_2\right)\bigg],
   \nonumber \\
   \epsilon^{-1} &\ud \mathbf{m}_{16} = \mathbf{m}_2 \bigg[\ud\ln \left(l_2\right)-\ud\ln \left(l_4\right)\bigg]+\mathbf{m}_3 \bigg[\ud\ln \left(l_1\right)-\frac{1}{6}\ud\ln
   \left(l_2\right)-\frac{1}{3}\ud\ln \left(l_4\right)\bigg]
   \nonumber \\
   &+\mathbf{m}_4 \bigg[6 \,\ud\ln \left(l_1\right)-3 \,\ud\ln
   \left(l_2\right)\bigg]+\mathbf{m}_5 \bigg[-\ud\ln \left(l_1\right)+\frac{2}{3}\ud \ln \left(l_2\right)-\frac{1}{6}\ud\ln
   \left(l_4\right)\bigg]
   \nonumber \\
   &+\mathbf{m}_6 \bigg[-2\,\ud \ln \left(l_2\right)+2 \,\ud\ln \left(l_4\right)\bigg]+\mathbf{m}_{16}
   \bigg[6 \,\ud\ln \left(l_1\right)-\frac{7}{3}\ud \ln \left(l_2\right)-\frac{2}{3}\ud\ln\left(l_4\right)\bigg]
   \nonumber \\
   &+\mathbf{m}_{17} \bigg[4 \,\ud\ln \left(l_1\right)-\frac{2}{3}\ud \ln\left(l_2\right)-\frac{4}{3}\ud \ln \left(l_4\right)\bigg],
   \nonumber \\
   \epsilon^{-1} &\ud \mathbf{m}_{17} = \mathbf{m}_2 \bigg[\ud\ln \left(l_2\right)-\ud\ln \left(l_4\right)\bigg]+\mathbf{m}_3 \bigg[\frac{1}{3}\ud\ln\left(l_2\right)-\frac{1}{3}\ud\ln \left(l_4\right)\bigg]
   \nonumber \\
   &+\mathbf{m}_5 \bigg[-\frac{1}{3}\ud\ln \left(l_2\right)+\frac{1}{3}\ud\ln\left(l_4\right)\bigg]+\mathbf{m}_6 \bigg[-2 \,\ud\ln\left(l_2\right)+2 \,\ud\ln \left(l_4\right)\bigg]
   \nonumber \\
   &+\mathbf{m}_{16} \bigg[\frac{2}{3}\,\ud\ln \left(l_2\right)-\frac{2}{3}\,\ud \ln\left(l_4\right)\bigg]+\mathbf{m}_{17} \bigg[4 \,\ud\ln \left(l_1\right)-\frac{2}{3}\,\ud \ln\left(l_2\right)-\frac{4}{3}\,\ud \ln \left(l_4\right)\bigg],
   \nonumber \\
   \epsilon^{-1} &\ud \mathbf{m}_{18} = \mathbf{m}_4 \bigg[4 \,\ud\ln \left(l_1\right)+4\,\ud \ln \left(l_3\right)+2 \,\ud\ln \left(l_6\right)-2 \,\ud\ln \left(l_9\right)-2\,\ud\ln \left(l_{10}\right)\bigg]
   \nonumber \\
   &+\mathbf{m}_5 \bigg[-\ud\ln \left(l_1\right)-\ud \ln \left(l_3\right)-\frac{1}{2}\ud\ln
   \left(l_6\right)+\frac{1}{2}\ud\ln \left(l_9\right)+\frac{1}{2}\ud\ln \left(l_{10}\right)\bigg]
   \nonumber \\
   &+\mathbf{m}_7
   \bigg[-4 \,\ud\ln \left(l_1\right)-4 \,\ud\ln \left(l_3\right)-2 \,\ud\ln \left(l_6\right)+2 \,\ud\ln \left(l_9\right)+2\,\ud
   \ln \left(l_{10}\right)\bigg]
   \nonumber \\
   &+\mathbf{m}_{18} \bigg[\ud\ln \left(l_2\right)+4\,\ud\ln \left(l_7\right)-3 \,\ud\ln
   \left(l_8\right)+\ud\ln \left(l_9\right)+\ud\ln \left(l_{10}\right)-3 \,\ud\ln
   \left(l_{11}\right)\bigg]
   \nonumber \\
   &+\mathbf{m}_{19} \bigg[-2\,\ud \ln \left(l_1\right)-\ud\ln \left(l_8\right)+\ud\ln
   \left(l_9\right)+\ud\ln \left(l_{10}\right)-\ud\ln \left(l_{11}\right)\bigg],
   \nonumber \\
   \epsilon^{-1} &\ud \mathbf{m}_{19} = \mathbf{m}_4 \bigg[-12 \,\ud\ln \left(l_1\right)+3 \,\ud\ln \left(l_2\right)-6 \,\ud\ln \left(l_3\right)+6 \,\ud\ln\left(l_4\right)-3 \,\ud\ln \left(l_6\right)+6 \,\ud\ln \left(l_7\right)\bigg]
   \nonumber \\
   &+\mathbf{m}_5 \bigg[3 \,\ud\ln
   \left(l_1\right)-\ud\ln \left(l_2\right)+2 \,\ud\ln \left(l_3\right)-\frac{3}{2}\ud\ln \left(l_4\right)+\ud\ln
   \left(l_6\right)-2\,\ud \ln \left(l_7\right)\bigg]
   \nonumber \\
   &+\mathbf{m}_7 \bigg[12\,\ud \ln \left(l_1\right)-6 \,\ud\ln
   \left(l_4\right)\bigg]
   \nonumber \\
   &+\mathbf{m}_{18} \bigg[12 \,\ud\ln \left(l_1\right)-6\,\ud \ln \left(l_2\right)-6\,\ud \ln
   \left(l_4\right)-12\,\ud \ln \left(l_7\right)+6\,\ud \ln \left(l_8\right)+6\,\ud \ln
   \left(l_{11}\right)\bigg]
   \nonumber \\
   &+\mathbf{m}_{19} \bigg[10\,\ud \ln \left(l_1\right)-3\,\ud \ln \left(l_2\right)-4\,\ud \ln
   \left(l_4\right)-4\,\ud \ln \left(l_7\right)+2\,\ud \ln \left(l_8\right)+2\,\ud \ln \left(l_{11}\right)\bigg],
   \nonumber \\
   \epsilon^{-1} &\ud \mathbf{m}_{20} = \mathbf{m}_3 \bigg[-\ud\ln \left(l_1\right)+\ud\ln \left(l_2\right)-\ud\ln \left(l_3\right)-\frac{1}{2}\ud\ln\left(l_6\right)+\ud\ln \left(l_7\right)\bigg]
   \nonumber \\
   &+\mathbf{m}_4 \bigg[-4 \,\ud\ln \left(l_1\right)+4 \,\ud\ln\left(l_2\right)-4 \,\ud\ln \left(l_3\right)-2 \,\ud\ln \left(l_6\right)+4 \,\ud\ln \left(l_7\right)\bigg]
   \nonumber \\
   &+\mathbf{m}_5\bigg[\ud\ln \left(l_1\right)-\ud\ln \left(l_2\right)+\ud\ln \left(l_3\right)+\frac{1}{2}\ud\ln\left(l_6\right)-\ud\ln \left(l_7\right)\bigg]
   \nonumber \\
   &+\mathbf{m}_8 \bigg[-2 \,\ud\ln \left(l_1\right)+2 \,\ud\ln\left(l_2\right)-2\,\ud \ln \left(l_3\right)-\ud\ln \left(l_6\right)+2 \,\ud\ln \left(l_7\right)\bigg]+\mathbf{m}_{20}
   \bigg[-4 \,\ud\ln \left(l_2\right)
   \nonumber \\
   &-4\,\ud \ln \left(l_7\right)+2\,\ud \ln \left(l_9\right)+2\,\ud \ln\left(l_{10}\right)\bigg]+\mathbf{m}_{21} \bigg[-2\,\ud \ln \left(l_1\right)+\ud\ln \left(l_2\right)\bigg],
   \nonumber \\
   \epsilon^{-1} &\ud \mathbf{m}_{21} = \mathbf{m}_1 \bigg[-\frac{1}{2}\ud \ln \left(l_2\right)+\ud\ln \left(l_3\right)+\frac{1}{2}\ud\ln
   \left(l_4\right)+\frac{1}{2}\ud\ln \left(l_6\right)-\frac{1}{2}\ud\ln \left(l_8\right)-\frac{1}{2}\ud\ln
   \left(l_{11}\right)\bigg]
   \nonumber \\
   &+\mathbf{m}_3 \bigg[2 \,\ud\ln \left(l_1\right)-\frac{3}{2}\ud\ln \left(l_2\right)+\ud\ln
   \left(l_3\right)-\frac{1}{2}\ud\ln \left(l_4\right)+\frac{1}{2}\ud\ln \left(l_6\right)-2\,\ud \ln\left(l_7\right)+\frac{1}{2}\ud\ln \left(l_8\right)
   \nonumber \\
   &+\frac{1}{2}\ud\ln \left(l_{11}\right)\bigg]+\mathbf{m}_4 \bigg[6\,\ud
   \ln \left(l_1\right)-6\,\ud \ln \left(l_2\right)+6\,\ud \ln \left(l_3\right)+3\,\ud \ln \left(l_6\right)-6\,\ud\ln\left(l_7\right)\bigg]
   \nonumber \\
   &+\mathbf{m}_5 \bigg[-2\,\ud \ln \left(l_1\right)+2\,\ud \ln \left(l_2\right)-2\,\ud \ln\left(l_3\right)-\ud\ln \left(l_6\right)+2 \,\ud\ln \left(l_7\right)\bigg]
   \nonumber \\
   &+\mathbf{m}_8 \bigg[6\,\ud \ln\left(l_1\right)-4 \,\ud\ln \left(l_2\right)+2\,\ud \ln \left(l_3\right)-2\,\ud \ln \left(l_4\right)+\ud\ln\left(l_6\right)-6\,\ud \ln \left(l_7\right)+2\,\ud \ln \left(l_8\right)
   \nonumber \\
   &+2\,\ud \ln\left(l_{11}\right)\bigg]+\mathbf{m}_{20} \bigg[12\,\ud \ln \left(l_1\right)-6 \,\ud\ln \left(l_4\right)\bigg]
   \nonumber \\
   &+\mathbf{m}_{21}
   \bigg[10 \,\ud\ln \left(l_1\right)-5\,\ud \ln \left(l_2\right)-4\,\ud \ln \left(l_4\right)-4\,\ud \ln \left(l_7\right)+2\,\ud\ln \left(l_9\right)+2\,\ud \ln \left(l_{10}\right)\bigg],
   \nonumber \\
   \epsilon^{-1} &\ud \mathbf{m}_{22} = \mathbf{m}_9 \bigg[-2 \, \ud\ln \left(l_2\right)\bigg]+\mathbf{m}_{22} \bigg[4\, \ud \ln \left(l_1\right)-2 \, \ud\ln \left(l_2\right)\bigg],
   \nonumber \\
   \epsilon^{-1} &\ud \mathbf{m}_{23} = \mathbf{m}_3 \bigg[\ud\ln \left(l_1\right)-\frac{1}{2}\ud\ln \left(l_2\right)\bigg]+\mathbf{m}_4 \bigg[4 \, \ud\ln \left(l_1\right)-2
   \, \ud\ln \left(l_2\right)\bigg]
   \nonumber \\
   &+\mathbf{m}_5 \bigg[-\ud\ln \left(l_1\right)+\frac{1}{2}\ud\ln \left(l_2\right)\bigg]+\mathbf{m}_9 \bigg[2\,\ud\ln \left(l_2\right)\bigg]+\mathbf{m}_{10}\bigg[2\, \ud \ln \left(l_2\right)\bigg]
   \nonumber \\
   &+\mathbf{m}_{23} \bigg[4 \, \ud\ln \left(l_1\right)-2 \, \ud\ln
   \left(l_2\right)\bigg],
   \nonumber \\
   \epsilon^{-1} &\ud \mathbf{m}_{24} = \mathbf{m}_{10}\bigg[ -2\, \ud \ln \left(l_2\right)\bigg] +\mathbf{m}_{24} \bigg[6 \, \ud\ln \left(l_1\right)-3 \, \ud\ln \left(l_2\right)\bigg],
   \nonumber \\
   \epsilon^{-1} &\ud \mathbf{m}_{25} = \mathbf{m}_3 \bigg[-\frac{1}{2}\ud\ln \left(l_1\right)+\frac{1}{4}\ud\ln \left(l_2\right)\bigg]+\mathbf{m}_4 \bigg[-2\, \ud \ln \left(l_1\right)+\ud\ln\left(l_2\right)\bigg]
   \nonumber \\
   &+\mathbf{m}_5 \bigg[\frac{1}{2}\ud\ln \left(l_1\right)-\frac{1}{4}\ud\ln\left(l_2\right)\bigg]+\mathbf{m}_{10} \bigg[\ud\ln \left(l_2\right)\bigg]+\mathbf{m}_{11} \bigg[\ud\ln \left(l_1\right)-\frac{1}{2}\ud\ln\left(l_2\right)\bigg]
   \nonumber \\
   &+ \mathbf{m}_{12} \bigg[-\frac{1}{2}\ln \left(l_2\right)\bigg]+\mathbf{m}_{25} \bigg[4\,\ud \ln\left(l_1\right)-2 \,\ud\ln \left(l_2\right)\bigg],
   \nonumber \\
   \epsilon^{-1} &\ud \mathbf{m}_{26} = \mathbf{m}_8 \bigg[-2\,\ud \ln \left(l_2\right)-2\,\ud \ln \left(l_7\right)+\ud\ln \left(l_9\right)+\ud\ln
   \left(l_{10}\right)\bigg]
   \nonumber \\
   &+\mathbf{m}_{11} \bigg[-\ud\ln \left(l_2\right)-\ud\ln \left(l_7\right)+\frac{1}{2}\ud\ln
   \left(l_9\right)+\frac{1}{2}\ud\ln \left(l_{10}\right)\bigg]+\mathbf{m}_{12}\bigg[-\ud\ln \left(l_2\right)\bigg]
   \nonumber \\
   &+\mathbf{m}_{26}
   \bigg[8 \,\ud\ln \left(l_1\right)-4 \,\ud\ln \left(l_2\right)+2\,\ud\ln \left(l_3\right)+\ud\ln \left(l_6\right)-\ud\ln
   \left(l_7\right)-\frac{1}{2}\ud\ln \left(l_9\right)-\frac{1}{2}\ud\ln \left(l_{10}\right)\bigg]
   \nonumber \\
   &+\mathbf{m}_{27}
   \bigg[-\frac{1}{2}\ud\ln \left(l_9\right)+\frac{1}{2}\ud\ln \left(l_{10}\right)\bigg],
   \nonumber \\
   \epsilon^{-1} &\ud \mathbf{m}_{27} = \mathbf{m}_1 \bigg[-\ud\ln \left(l_8\right)+\ud\ln \left(l_{11}\right)\bigg]+\mathbf{m}_3 \bigg[\ud\ln \left(l_8\right)-\ud\ln
   \left(l_{11}\right)\bigg]+\mathbf{m}_8 \bigg[4 \,\ud\ln \left(l_8\right)
   \nonumber \\
   &-3 \,\ud\ln \left(l_9\right)+3 \,\ud\ln
   \left(l_{10}\right)-4 \,\ud\ln \left(l_{11}\right)\bigg]+\mathbf{m}_{11} \bigg[-\frac{3}{2}\ud\ln \left(l_9\right)+\frac{3}{2}\ud\ln\left(l_{10}\right)\bigg]
   \nonumber \\
   &+\mathbf{m}_{12} \bigg[-\ud\ln \left(l_6\right)\bigg]+\mathbf{m}_{26}
   \bigg[\frac{3}{2}\ud\ln \left(l_9\right)-\frac{3}{2}\ud\ln \left(l_{10}\right)\bigg]+\mathbf{m}_{27} \bigg[4 \,\ud\ln\left(l_1\right)-4 \,\ud\ln \left(l_2\right)
   \nonumber \\
   &-2 \,\ud\ln \left(l_3\right)-4 \,\ud\ln \left(l_5\right)+\ud\ln\left(l_6\right)-\ud\ln \left(l_7\right)+\frac{3}{2}\ud\ln \left(l_9\right)+\frac{3}{2}\ud\ln\left(l_{10}\right)\bigg],
   \nonumber \\
   \epsilon^{-1} &\ud \mathbf{m}_{28} = \mathbf{m}_1 \bigg[-\frac{1}{2}\ud\ln \left(l_1\right)-\frac{1}{2}\ud\ln \left(l_3\right)-\frac{1}{4}\ud\ln
   \left(l_6\right)+\frac{1}{4}\ud\ln \left(l_9\right)+\frac{1}{4}\ud\ln \left(l_{10}\right)\bigg]
   \nonumber \\
   &+\mathbf{m}_3
   \bigg[\frac{1}{2}\ud\ln \left(l_1\right)+\frac{1}{2}\ud\ln \left(l_3\right)+\frac{1}{4}\ud\ln\left(l_6\right)-\frac{1}{4}\ud\ln \left(l_9\right)-\frac{1}{4}\ud\ln \left(l_{10}\right)\bigg]
   \nonumber \\
   &+\mathbf{m}_4
   \bigg[3\,\ud \ln \left(l_1\right)+3\,\ud \ln \left(l_3\right)+\frac{3}{2}\ud \ln \left(l_6\right)-\frac{3}{2}\ud \ln
   \left(l_9\right)-\frac{3}{2}\ud \ln \left(l_{10}\right)\bigg]
   \nonumber \\
   &+\mathbf{m}_5 \bigg[-\frac{1}{2}\ud \ln
   \left(l_1\right)-\frac{1}{2}\ud\ln \left(l_3\right)-\frac{1}{4}\ud\ln \left(l_6\right)+\frac{1}{4}\ud\ln\left(l_9\right)+\frac{1}{4}\ud\ln \left(l_{10}\right)\bigg]
   \nonumber \\
   &+\mathbf{m}_8 \bigg[-4 \,\ud\ln \left(l_1\right)-4\,\ud
   \ln \left(l_3\right)-2\,\ud \ln \left(l_6\right)+2\,\ud \ln \left(l_9\right)+2 \,\ud\ln\left(l_{10}\right)\bigg]
   \nonumber \\
   &+\mathbf{m}_{15} \bigg[\ud\ln \left(l_1\right)+\ud\ln \left(l_3\right)+\frac{1}{2}\ud\ln\left(l_6\right)-\frac{1}{2}\ud\ln \left(l_9\right)-\frac{1}{2}\ud\ln \left(l_{10}\right)\bigg]
   \nonumber \\
   &+\mathbf{m}_{17}\bigg[2 \,\ud\ln \left(l_1\right)+2\,\ud \ln \left(l_3\right)+\ud\ln \left(l_6\right)-\ud\ln \left(l_9\right)-\ud\ln\left(l_{10}\right)\bigg]
   \nonumber \\
   &+\mathbf{m}_{20} \bigg[-6 \,\ud\ln \left(l_1\right)-6\,\ud \ln \left(l_3\right)-3 \,\ud\ln\left(l_6\right)+3 \,\ud\ln \left(l_9\right)+3 \,\ud\ln \left(l_{10}\right)\bigg]
   \nonumber \\
   &+\mathbf{m}_{21} \bigg[-2\,\ud \ln
   \left(l_1\right)-2\,\ud \ln \left(l_3\right)-\ud\ln \left(l_6\right)+\ud\ln \left(l_9\right)+\ud\ln\left(l_{10}\right)\bigg]
   \nonumber \\
   &+\mathbf{m}_{28} \bigg[2 \,\ud\ln \left(l_1\right)-2 \,\ud\ln \left(l_2\right)+2\,\ud \ln
   \left(l_3\right)+\ud\ln \left(l_6\right)-2\,\ud \ln \left(l_8\right)+\ud\ln \left(l_9\right)+\ud\ln\left(l_{10}\right)
   \nonumber \\
   &-2 \,\ud\ln \left(l_{11}\right)\bigg]+\mathbf{m}_{29} \bigg[2 \,\ud\ln \left(l_3\right)+\ud\ln
   \left(l_6\right)-\ud\ln \left(l_8\right)-\ud\ln \left(l_{11}\right)\bigg],
   \nonumber \\
   \epsilon^{-1} &\ud \mathbf{m}_{29} = \mathbf{m}_1 \bigg[\ud\ln \left(l_1\right)+\frac{3}{4}\ud\ln \left(l_2\right)+\frac{1}{2}\ud\ln
   \left(l_3\right)+\frac{1}{2}\ud\ln \left(l_4\right)+\frac{1}{4}\ud\ln \left(l_6\right)+\frac{3}{2}\ud\ln
   \left(l_7\right)
   \nonumber \\
   &-\ud\ln \left(l_9\right)-\ud\ln \left(l_{10}\right)\bigg]+\mathbf{m}_2 \bigg[-\ud\ln
   \left(l_2\right)+4 \,\ud\ln \left(l_3\right)+\ud\ln \left(l_4\right)+2\, \ud\ln \left(l_6\right)-2 \,\ud\ln
   \left(l_7\right)
   \nonumber \\
   &-\ud\ln \left(l_9\right)-\ud\ln \left(l_{10}\right)\bigg]+\mathbf{m}_3 \bigg[-\ud\ln
   \left(l_1\right)+\frac{1}{12}\ud\ln \left(l_2\right)-\frac{3}{2} \ud\ln \left(l_3\right)-\frac{1}{6}\ud\ln
   \left(l_4\right)
   \nonumber \\
   &-\frac{3}{4} \ud\ln \left(l_6\right)+\frac{1}{6}\ud\ln \left(l_7\right)+\frac{2}{3} \ud\ln
   \left(l_9\right)+\frac{2}{3} \ud\ln \left(l_{10}\right)\bigg]+\mathbf{m}_4 \bigg[-6 \,\ud\ln
   \left(l_1\right)+\frac{3}{2} \ud\ln \left(l_2\right)
   \nonumber \\
   &-9\, \ud\ln \left(l_3\right)-\frac{9}{2} \ud\ln
   \left(l_6\right)+3\, \ud\ln \left(l_7\right)+3\, \ud\ln \left(l_9\right)+3\, \ud\ln
   \left(l_{10}\right)\bigg]+\mathbf{m}_5 \bigg[\ud\ln \left(l_1\right)
   \nonumber \\
   &-\frac{1}{12}\ud\ln \left(l_2\right)+\frac{3}{2}
   \ud\ln \left(l_3\right)+\frac{1}{6}\ud\ln \left(l_4\right)+\frac{3}{4} \ud\ln \left(l_6\right)-\frac{1}{6}\ud\ln
   \left(l_7\right)-\frac{2}{3} \ud\ln \left(l_9\right)-\frac{2}{3} \ud\ln \left(l_{10}\right)\bigg]
   \nonumber \\
   &+\mathbf{m}_6
   \bigg[\ud\ln \left(l_2\right)-6\, \ud\ln \left(l_3\right)-2\, \ud\ln \left(l_4\right)-3\, \ud\ln \left(l_6\right)+2\,
   \ud\ln \left(l_7\right)+2\, \ud\ln \left(l_9\right)
   \nonumber \\
   &+2\, \ud\ln \left(l_{10}\right)\bigg]+\mathbf{m}_8 \bigg[8\, \ud\ln
   \left(l_1\right)+4\, \ud\ln \left(l_3\right)-2\,\ud\ln \left(l_4\right)+2\, \ud\ln \left(l_6\right)-2\, \ud\ln
   \left(l_9\right)
   \nonumber \\
   &-2\, \ud\ln \left(l_{10}\right)\bigg]+\mathbf{m}_{13} \bigg[\ud\ln \left(l_2\right)-2\, \ud\ln
   \left(l_3\right)-\ud\ln \left(l_6\right)+2\, \ud\ln \left(l_7\right)\bigg]
   \nonumber \\
   &+\mathbf{m}_{15} \bigg[-2\, \ud\ln
   \left(l_1\right)-\frac{1}{2}\ud\ln \left(l_2\right)-\ud\ln \left(l_3\right)-\frac{1}{2}\ud\ln
   \left(l_6\right)-\ud\ln \left(l_7\right)+\ud\ln \left(l_9\right)+\ud\ln \left(l_{10}\right)\bigg]
   \nonumber \\
   &+\mathbf{m}_{16}
   \bigg[\frac{2}{3}\ud\ln \left(l_2\right)+\frac{2}{3} \ud\ln \left(l_4\right)+\frac{4}{3} \ud\ln
   \left(l_7\right)-\frac{2}{3} \ud\ln \left(l_9\right)-\frac{2}{3} \ud\ln
   \left(l_{10}\right)\bigg]
   \nonumber \\
   &+\mathbf{m}_{17} \bigg[-4\, \ud\ln \left(l_1\right)+\frac{1}{3}\ud\ln \left(l_2\right)-2\,
   \ud\ln \left(l_3\right)+\frac{4}{3} \ud\ln \left(l_4\right)-\ud\ln \left(l_6\right)+\frac{2}{3} \ud\ln
   \left(l_7\right)
   \nonumber \\
   &+\frac{2}{3} \ud\ln \left(l_9\right)+\frac{2}{3} \ud\ln
   \left(l_{10}\right)\bigg]+\mathbf{m}_{20} \bigg[12\, \ud\ln \left(l_1\right)-3 \,\ud\ln \left(l_2\right)+6 \,\ud\ln
   \left(l_3\right)-6\, \ud\ln \left(l_4\right)
   \nonumber \\
   &+3\, \ud\ln \left(l_6\right)-6\, \ud\ln \left(l_7\right)\bigg]+\mathbf{m}_{21}
   \bigg[4\, \ud\ln \left(l_1\right)-3\, \ud\ln \left(l_2\right)+2\, \ud\ln \left(l_3\right)-4\, \ud\ln
   \left(l_4\right)
   \nonumber \\
   &+\ud\ln \left(l_6\right)-6\, \ud\ln \left(l_7\right)+2\, \ud\ln \left(l_9\right)+2\, \ud\ln
   \left(l_{10}\right)\bigg]+\mathbf{m}_{28} \bigg[3\, \ud\ln \left(l_2\right)-2\, \ud\ln \left(l_3\right)-\ud\ln
   \left(l_6\right)
   \nonumber \\
   &+6\, \ud\ln \left(l_7\right)-2\, \ud\ln \left(l_8\right)-2\, \ud\ln
   \left(l_{11}\right)\bigg]+\mathbf{m}_{29} \bigg[2\, \ud\ln \left(l_1\right)-\ud\ln \left(l_2\right)-2\, \ud\ln
   \left(l_3\right)-\ud\ln \left(l_6\right)
   \nonumber \\
   &+2\, \ud\ln \left(l_7\right)-\ud\ln \left(l_8\right)+\ud\ln
   \left(l_9\right)+\ud\ln \left(l_{10}\right)-\ud\ln \left(l_{11}\right)\bigg],
   \nonumber \\
   \epsilon^{-1} &\ud \mathbf{m}_{30} = \mathbf{m}_3 \bigg[\ud\ln \left(l_2\right)\bigg]+\mathbf{m}_4 \bigg[6\,\ud\ln \left(l_2\right)\bigg]+\mathbf{m}_5 \bigg[-\ud\ln \left(l_2\right)\bigg]+\mathbf{m}_{16} \bigg[2\,\ud\ln\left(l_2\right)\bigg]
   \nonumber \\
   &+\mathbf{m}_{17} \bigg[4\,\ud\ln \left(l_2\right)\bigg]+\mathbf{m}_{22} \bigg[-2\,\ud\ln \left(l_2\right)\bigg]+\mathbf{m}_{23} \bigg[2\,\ud\ln
   \left(l_2\right)\bigg]+\mathbf{m}_{24}\bigg[-2\,\ud \ln \left(l_2\right)\bigg]
   \nonumber \\
   &+\mathbf{m}_{30} \bigg[6\,\ud \ln \left(l_1\right)-4\,\ud \ln
   \left(l_2\right)+2\,\ud \ln \left(l_3\right)\bigg],
   \nonumber \\
   \epsilon^{-1} &\ud \mathbf{m}_{31} = \mathbf{m}_3 \bigg[-\frac{1}{2}\ud\ln \left(l_8\right)+\frac{1}{2}\ud\ln \left(l_{11}\right)\bigg]+\mathbf{m}_4\bigg[-2 \,\ud\ln \left(l_9\right)+2 \,\ud\ln\left(l_{10}\right)\bigg]
   \nonumber \\
   &+\mathbf{m}_5 \bigg[\frac{1}{2}\ud\ln \left(l_9\right)-\frac{1}{2}\ud\ln
   \left(l_{10}\right)\bigg]+\mathbf{m}_7 \bigg[-3\, \ud\ln \left(l_8\right)+3 \,\ud\ln \left(l_9\right)-3 \,\ud\ln
   \left(l_{10}\right)
   \nonumber \\
   &+3 \,\ud\ln \left(l_{11}\right)\bigg]+\mathbf{m}_{10}\bigg[2\, \ud\ln \left(l_6\right)\bigg]+\mathbf{m}_{14} \bigg[-\ud\ln
   \left(l_8\right)+\ud\ln \left(l_9\right)-\ud\ln \left(l_{10}\right)+\ud\ln \left(l_{11}\right)\bigg]
   \nonumber \\
   &+\mathbf{m}_{18}
   \bigg[3\, \ud\ln \left(l_8\right)-3\, \ud\ln \left(l_{11}\right)\bigg]+\mathbf{m}_{19} \bigg[\ud\ln \left(l_8\right)+\ud\ln
   \left(l_9\right)-\ud\ln \left(l_{10}\right)-\ud\ln \left(l_{11}\right)\bigg]
   \nonumber \\
   &+\mathbf{m}_{24} \bigg[-2\, \ud\ln \left(l_9\right)+2 \,\ud\ln
   \left(l_{10}\right)\bigg]+\mathbf{m}_{31} \bigg[4\, \ud\ln \left(l_1\right)-3 \,\ud\ln\left(l_2\right)-2\, \ud\ln \left(l_3\right)-2\, \ud\ln \left(l_5\right)
   \nonumber \\
   &+\ud\ln \left(l_9\right)+\ud\ln
   \left(l_{10}\right)\bigg],
   \nonumber \\
   \epsilon^{-1} &\ud \mathbf{m}_{32} = \mathbf{m}_1 \bigg[-\frac{1}{4}\ud\ln \left(l_2\right)+\frac{1}{4}\ud\ln \left(l_9\right)-\frac{1}{4}\ud\ln
   \left(l_{10}\right)-\frac{1}{2}\ud\ln \left(l_{13}\right)+\frac{1}{2}\ud\ln
   \left(l_{14}\right)\bigg]
   \nonumber \\
   &+\mathbf{m}_3 \bigg[-\frac{1}{4}\ud\ln \left(l_2\right)-\frac{1}{2}\ud\ln
   \left(l_6\right)-\frac{1}{2}\ud\ln \left(l_7\right)-\frac{1}{2}\ud\ln \left(l_9\right)+\frac{1}{2}\ud\ln
   \left(l_{13}\right)-\frac{1}{2}\ud\ln \left(l_{14}\right)
   \nonumber \\
   &+\frac{1}{2}\ud\ln\left(l_{15}\right)+\frac{1}{2}\ud\ln \left(l_{16}\right)\bigg]+\mathbf{m}_4 \bigg[-\ud\ln \left(l_2\right)-2\,
   \ud\ln \left(l_6\right)-2\, \ud\ln \left(l_7\right)-2\, \ud\ln \left(l_9\right)
   \nonumber \\
   &+2\, \ud\ln \left(l_{13}\right)-2\, \ud\ln
   \left(l_{14}\right)+2\, \ud\ln \left(l_{15}\right)+2\, \ud\ln \left(l_{16}\right)\bigg]+\mathbf{m}_5
   \bigg[\frac{1}{4} \ud\ln \left(l_2\right)+\frac{1}{2}\ud\ln \left(l_6\right)
   \nonumber \\
   &+\frac{1}{2}\ud\ln\left(l_7\right)+\frac{1}{2}\ud\ln \left(l_9\right)-\frac{1}{2}\ud\ln \left(l_{13}\right)+\frac{1}{2}\ud\ln
   \left(l_{14}\right)-\frac{1}{2}\ud\ln \left(l_{15}\right)-\frac{1}{2}\ud\ln
   \left(l_{16}\right)\bigg]
   \nonumber \\
   &+\mathbf{m}_8 \bigg[-2 \,\ud\ln \left(l_2\right)+2 \,\ud\ln \left(l_9\right)-2 \,\ud\ln
   \left(l_{10}\right)-4\, \ud\ln \left(l_{13}\right)+4\, \ud\ln \left(l_{14}\right)\bigg]
   \nonumber \\
   &+\mathbf{m}_{10} \bigg[2\, \ud\ln
   \left(l_1\right)+\ud\ln \left(l_2\right)+2\, \ud\ln \left(l_7\right)+2\, \ud\ln \left(l_9\right)-2\, \ud\ln
   \left(l_{13}\right)-2\, \ud\ln \left(l_{14}\right)
   \nonumber \\
   &+2\, \ud\ln \left(l_{15}\right)-2\, \ud\ln
   \left(l_{16}\right)\bigg]+\mathbf{m}_{11} \bigg[-\frac{1}{2}\ud\ln \left(l_2\right)+\frac{1}{2}\ud\ln\left(l_9\right)-\frac{1}{2}\ud\ln \left(l_{10}\right)-\ud\ln \left(l_{13}\right)
   \nonumber \\
   &+\ud\ln
   \left(l_{14}\right)\bigg]+\mathbf{m}_{12} \bigg[-\ud\ln \left(l_1\right)-\frac{1}{2}\ud\ln \left(l_2\right)-\ud\ln
   \left(l_7\right)-\ud\ln \left(l_9\right)+\ud\ln \left(l_{13}\right)
   \nonumber \\
   &+\ud\ln \left(l_{14}\right)-\ud\ln
   \left(l_{15}\right)+\ud\ln \left(l_{16}\right)\bigg]
   \nonumber \\
   &+\mathbf{m}_{15} \bigg[\frac{1}{2} \ud\ln
   \left(l_2\right)-\frac{1}{2}\ud\ln \left(l_9\right)+\frac{1}{2}\ud\ln \left(l_{10}\right)+\ud\ln\left(l_{13}\right)-\ud\ln \left(l_{14}\right)\bigg]
   \nonumber \\
   &+\mathbf{m}_{20} \bigg[-3\, \ud\ln \left(l_2\right)+3\, \ud\ln
   \left(l_9\right)-3\, \ud\ln \left(l_{10}\right)-6\, \ud\ln \left(l_{13}\right)+6\, \ud\ln
   \left(l_{14}\right)\bigg]
   \nonumber \\
   &+\mathbf{m}_{21} \bigg[-3\, \ud\ln \left(l_2\right)-2\, \ud\ln \left(l_6\right)-2\, \ud\ln
   \left(l_7\right)-2\, \ud\ln \left(l_{10}\right)-2\, \ud\ln \left(l_{13}\right)+2\, \ud\ln \left(l_{14}\right)
   \nonumber \\
   &+2\, \ud\ln\left(l_{15}\right)+2\, \ud\ln \left(l_{16}\right)\bigg]+\mathbf{m}_{24} \bigg[-2\, \ud\ln \left(l_2\right)-2\, \ud\ln
   \left(l_6\right)-2\, \ud\ln \left(l_7\right)-\ud\ln \left(l_9\right)
   \nonumber \\
   &-\ud\ln \left(l_{10}\right)+2\, \ud\ln
   \left(l_{15}\right)+2\, \ud\ln \left(l_{16}\right)\bigg]
   \nonumber \\
   &+\mathbf{m}_{25} \bigg[\ud\ln \left(l_2\right)-\ud\ln
   \left(l_9\right)+\ud\ln \left(l_{10}\right)+2\, \ud\ln \left(l_{13}\right)-2\, \ud\ln\left(l_{14}\right)\bigg]+\mathbf{m}_{26} \bigg[3 \,\ud\ln \left(l_2\right)
   \nonumber \\
   &+3 \,\ud\ln \left(l_6\right)+3 \,\ud\ln
   \left(l_7\right)+\frac{3}{2} \ud\ln \left(l_9\right)+\frac{3}{2}\ud\ln \left(l_{10}\right)-3 \,\ud\ln
   \left(l_{15}\right)-3 \,\ud\ln \left(l_{16}\right)\bigg]
   \nonumber \\
   &+\mathbf{m}_{27} \bigg[\frac{1}{2} \ud\ln
   \left(l_9\right)-\frac{1}{2}\ud\ln \left(l_{10}\right)+\ud\ln \left(l_{15}\right)-\ud\ln
   \left(l_{16}\right)\bigg]
   \nonumber \\
   &+\mathbf{m}_{32} \bigg[4 \,\ud\ln \left(l_1\right)-\ud\ln \left(l_2\right)-2\, \ud\ln
   \left(l_3\right)+2 \,\ud\ln \left(l_7\right)+\ud\ln \left(l_9\right)+\ud\ln \left(l_{10}\right)-2 \,\ud\ln
   \left(l_{12}\right)\bigg],
   \nonumber \\
   \epsilon^{-1} &\ud \mathbf{m}_{33} = \mathbf{m}_1 \bigg[\frac{3}{4} \ud\ln \left(l_9\right)-\frac{3}{4} \ud\ln \left(l_{10}\right)\bigg]+\mathbf{m}_2 \bigg[\ud\ln\left(l_9\right)-\ud\ln \left(l_{10}\right)\bigg]+\mathbf{m}_3 \bigg[-\frac{13}{12} \ud\ln \left(l_9\right)
   \nonumber \\
   &+\frac{13}{12}\ud\ln\left(l_{10}\right)\bigg]+\mathbf{m}_4 \bigg[-\frac{11}{2} \ud\ln \left(l_9\right)+\frac{11}{2} \ud\ln
   \left(l_{10}\right)\bigg]+\mathbf{m}_5 \bigg[\frac{13}{12} \ud\ln\left(l_9\right)-\frac{13}{12} \ud\ln \left(l_{10}\right)\bigg]
   \nonumber \\
   &+\mathbf{m}_6 \bigg[-\ud\ln \left(l_9\right)+\ud\ln\left(l_{10}\right)\bigg]+\mathbf{m}_8 \bigg[6\, \ud\ln \left(l_9\right)-6\, \ud\ln
   \left(l_{10}\right)\bigg]+\mathbf{m}_9\bigg[-2\, \ud\ln \left(l_6\right)\bigg]
   \nonumber \\
   &+\mathbf{m}_{10}\bigg[-4 \,\ud\ln \left(l_6\right)\bigg]+\mathbf{m}_{11} \bigg[2\,\ud\ln \left(l_9\right)-2\, \ud\ln \left(l_{10}\right)\bigg]+ \mathbf{m}_{12} \bigg[2\,\ud\ln \left(l_6\right)\bigg]
   \nonumber \\
   &+\mathbf{m}_{13}
   \bigg[-\frac{1}{3}\ud\ln \left(l_9\right)+\frac{1}{3}\ud\ln \left(l_{10}\right)\bigg]+\mathbf{m}_{15} \bigg[-\frac{3}{2} \ud\ln \left(l_9\right)+\frac{3}{2}\ud\ln \left(l_{10}\right)\bigg]
   \nonumber \\
   &+\mathbf{m}_{16} \bigg[-\frac{2}{3} \ud\ln \left(l_9\right)+\frac{2}{3} \ud\ln\left(l_{10}\right)\bigg]+\mathbf{m}_{17} \bigg[-\frac{7}{3} \ud\ln \left(l_9\right)+\frac{7}{3} \ud\ln
   \left(l_{10}\right)\bigg]
   \nonumber \\
   &+\mathbf{m}_{20} \bigg[5 \,\ud\ln
   \left(l_9\right)-5 \,\ud\ln \left(l_{10}\right)\bigg]+\mathbf{m}_{21} \bigg[3\, \ud\ln \left(l_9\right)-3\, \ud\ln
   \left(l_{10}\right)\bigg]
   \nonumber \\
   &+\mathbf{m}_{22} \bigg[2\, \ud\ln \left(l_9\right)-2\, \ud\ln
   \left(l_{10}\right)\bigg]+\mathbf{m}_{23} \bigg[-\ud\ln \left(l_9\right)+\ud\ln \left(l_{10}\right)\bigg]
   \nonumber \\
   &+\mathbf{m}_{24}
   \bigg[2\, \ud\ln \left(l_9\right)-2\, \ud\ln \left(l_{10}\right)\bigg]+\mathbf{m}_{25} \bigg[-2\, \ud\ln \left(l_9\right)+2\, \ud\ln
   \left(l_{10}\right)\bigg]
   \nonumber \\
   &+\mathbf{m}_{26} \bigg[-2\, \ud\ln
   \left(l_9\right)+2 \,\ud\ln \left(l_{10}\right)\bigg]+\mathbf{m}_{28} \bigg[4\, \ud\ln \left(l_8\right)-\ud\ln \left(l_9\right)+\ud\ln
   \left(l_{10}\right)-4\, \ud\ln \left(l_{11}\right)\bigg]
   \nonumber \\
   &+\mathbf{m}_{29} \bigg[2\, \ud\ln \left(l_8\right)-\ud\ln
   \left(l_9\right)+\ud\ln \left(l_{10}\right)-2\, \ud\ln \left(l_{11}\right)\bigg]
   \nonumber \\
   &+\mathbf{m}_{32} \bigg[2 \,\ud\ln
   \left(l_9\right)-2 \,\ud\ln \left(l_{10}\right)+4 \,\ud\ln \left(l_{15}\right)-4 \,\ud\ln\left(l_{16}\right)\bigg]
   \nonumber \\
   &+\mathbf{m}_{33} \bigg[2\, \ud\ln \left(l_1\right)-4\, \ud\ln \left(l_2\right)-2 \,\ud\ln
   \left(l_3\right)-6\, \ud\ln \left(l_5\right)+2 \,\ud\ln \left(l_6\right)-2 \,\ud\ln \left(l_7\right)
   \nonumber \\
   &+2 \,\ud\ln\left(l_9\right)+2 \,\ud\ln \left(l_{10}\right)\bigg]+\mathbf{m}_{35} \bigg[-2\, \ud\ln \left(l_6\right)\bigg]+\mathbf{m}_{36} \bigg[2 \,\ud\ln\left(l_9\right)-2 \,\ud\ln \left(l_{10}\right)\bigg],
   \nonumber \\
   \epsilon^{-1} &\ud \mathbf{m}_{34} = \mathbf{m}_1 \bigg[-\frac{1}{2} \ud\ln \left(l_2\right)-\frac{1}{2}\ud\ln \left(l_7\right)+\frac{1}{4}\ud\ln
   \left(l_9\right)+\frac{1}{4}\ud\ln \left(l_{10}\right)\bigg]
   \nonumber \\
   &+\mathbf{m}_2 \bigg[-2 \,\ud\ln \left(l_2\right)-2\,
   \ud\ln \left(l_7\right)+\ud\ln \left(l_9\right)+\ud\ln \left(l_{10}\right)\bigg]+\mathbf{m}_3 \bigg[\frac{7}{6} \ud\ln
   \left(l_2\right)+\frac{7}{6} \ud\ln \left(l_7\right)
   \nonumber \\
   &-\frac{7}{12} \ud\ln \left(l_9\right)-\frac{7}{12} \ud\ln
   \left(l_{10}\right)\bigg]
   +\mathbf{m}_4 \bigg[5 \,\ud\ln \left(l_2\right)+5 \,\ud\ln \left(l_7\right)-\frac{5}{2} \ud\ln
   \left(l_9\right)-\frac{5}{2} \ud\ln \left(l_{10}\right)\bigg]
   \nonumber \\
   &+\mathbf{m}_5 \bigg[-\frac{7}{6} \ud\ln
   \left(l_2\right)-\frac{7}{6} \ud\ln \left(l_7\right)+\frac{7}{12} \ud\ln \left(l_9\right)+\frac{7}{12} \ud\ln
   \left(l_{10}\right)\bigg]
   \nonumber \\
   &+\mathbf{m}_6 \bigg[2\, \ud\ln \left(l_2\right)+2\, \ud\ln \left(l_7\right)-\ud\ln
   \left(l_9\right)-\ud\ln \left(l_{10}\right)\bigg]+\mathbf{m}_8 \bigg[-4\, \ud\ln \left(l_2\right)-4\, \ud\ln
   \left(l_7\right)
   \nonumber \\
   &+2\, \ud\ln \left(l_9\right)+2\, \ud\ln \left(l_{10}\right)\bigg]+\mathbf{m}_9 \bigg[2\,\ud\ln
   \left(l_2\right)\bigg]+\mathbf{m}_{10} \bigg[4\,\ud\ln \left(l_2\right)\bigg]
   \nonumber \\
   &+\mathbf{m}_{11} \bigg[-2 \,\ud\ln \left(l_2\right)-2 \,\ud\ln
   \left(l_7\right)+\ud\ln \left(l_9\right)+\ud\ln \left(l_{10}\right)\bigg]+ \mathbf{m}_{12} \bigg[-2\,\ud\ln
   \left(l_2\right)\bigg]
   \nonumber \\
   &+\mathbf{m}_{13} \bigg[\frac{2}{3} \ud\ln \left(l_2\right)+\frac{2}{3} \ud\ln
   \left(l_7\right)-\frac{1}{3}\ud\ln \left(l_9\right)-\frac{1}{3}\ud\ln \left(l_{10}\right)\bigg]+\mathbf{m}_{15}
   \bigg[\ud\ln \left(l_2\right)+\ud\ln \left(l_7\right)
   \nonumber \\
   &-\frac{1}{2}\ud\ln \left(l_9\right)-\frac{1}{2}\ud\ln
   \left(l_{10}\right)\bigg]
   +\mathbf{m}_{16} \bigg[\frac{4}{3} \ud\ln \left(l_2\right)+\frac{4}{3} \ud\ln
   \left(l_7\right)-\frac{2}{3} \ud\ln \left(l_9\right)-\frac{2}{3} \ud\ln
   \left(l_{10}\right)\bigg]
   \nonumber \\
   &+\mathbf{m}_{17} \bigg[\frac{2}{3} \ud\ln \left(l_2\right)+\frac{2}{3} \ud\ln
   \left(l_7\right)-\frac{1}{3}\ud\ln \left(l_9\right)-\frac{1}{3}\ud\ln \left(l_{10}\right)\bigg]+\mathbf{m}_{20}
   \bigg[2\, \ud\ln \left(l_2\right)+2\, \ud\ln \left(l_7\right)
   \nonumber \\
   &-\ud\ln \left(l_9\right)-\ud\ln
   \left(l_{10}\right)\bigg]+\mathbf{m}_{21} \bigg[-2 \,\ud\ln \left(l_2\right)-2\, \ud\ln \left(l_7\right)+\ud\ln
   \left(l_9\right)+\ud\ln \left(l_{10}\right)\bigg]
   \nonumber \\
   &+\mathbf{m}_{22} \bigg[-4 \,\ud\ln \left(l_1\right)-4\, \ud\ln
   \left(l_3\right)-2 \,\ud\ln \left(l_6\right)+2 \,\ud\ln \left(l_9\right)+2\, \ud\ln\left(l_{10}\right)\bigg]
   \nonumber \\
   &+\mathbf{m}_{24} \bigg[-4\, \ud\ln \left(l_2\right)-4\, \ud\ln \left(l_7\right)+2\, \ud\ln
   \left(l_9\right)+2\, \ud\ln \left(l_{10}\right)\bigg]+\mathbf{m}_{26} \bigg[4\, \ud\ln \left(l_2\right)+4\, \ud\ln
   \left(l_7\right)
   \nonumber \\
   &-2\, \ud\ln \left(l_9\right)-2\, \ud\ln \left(l_{10}\right)\bigg]+\mathbf{m}_{28} \bigg[2\, \ud\ln
   \left(l_2\right)+2\, \ud\ln \left(l_7\right)-\ud\ln \left(l_9\right)-\ud\ln \left(l_{10}\right)\bigg]
   \nonumber \\
   &+\mathbf{m}_{29}
   \bigg[2 \,\ud\ln \left(l_2\right)+2\, \ud\ln \left(l_7\right)-\ud\ln \left(l_9\right)-\ud\ln
   \left(l_{10}\right)\bigg]+\mathbf{m}_{30} \bigg[-2\,\ud\ln \left(l_2\right)\bigg]
   \nonumber \\
   &+\mathbf{m}_{32} \bigg[4\, \ud\ln \left(l_2\right)+4 \,\ud\ln
   \left(l_6\right)+4 \,\ud\ln \left(l_7\right)+2 \,\ud\ln \left(l_9\right)+2 \,\ud\ln \left(l_{10}\right)
   \nonumber \\
   &-4 \,\ud\ln\left(l_{15}\right)-4\, \ud\ln \left(l_{16}\right)\bigg]+\mathbf{m}_{33} \bigg[\ud\ln \left(l_9\right)-\ud\ln
   \left(l_{10}\right)\bigg]
   \nonumber \\
   &+\mathbf{m}_{34} \bigg[4 \,\ud\ln \left(l_1\right)-4\, \ud\ln \left(l_2\right)-2 \,\ud\ln
   \left(l_7\right)+\ud\ln \left(l_9\right)+\ud\ln \left(l_{10}\right)\bigg]+\mathbf{m}_{35} \bigg[2\,\ud\ln
   \left(l_2\right)\bigg]
   \nonumber \\
   &+\mathbf{m}_{36} \bigg[-4\, \ud\ln \left(l_2\right)-4\, \ud\ln \left(l_7\right)+2\, \ud\ln
   \left(l_9\right)+2\, \ud\ln \left(l_{10}\right)\bigg],
   \nonumber \\
   \epsilon^{-1} &\ud \mathbf{m}_{35} = \mathbf{m}_1 \bigg[-\frac{3}{4}\ud\ln \left(l_2\right)\bigg]+\mathbf{m}_2 \bigg[\ud\ln \left(l_2\right)\bigg]+ \mathbf{m}_3 \bigg[\frac{5}{12}\ud\ln
   \left(l_2\right)\bigg]+\mathbf{m}_4 \bigg[\frac{7}{2}\ud\ln \left(l_2\right)\bigg]
   \nonumber \\
   &+ \mathbf{m}_5 \bigg[-\frac{5}{12}\ud\ln \left(l_2\right)\bigg]+\mathbf{m}_6 \bigg[-\ud\ln\left(l_2\right)\bigg]+ \mathbf{m}_8 \bigg[-6\,\ud\ln \left(l_2\right)\bigg]+ \mathbf{m}_{13} \bigg[-\frac{1}{3}\ud\ln \left(l_2\right)\bigg]
   \nonumber \\
   &+\mathbf{m}_{15} \bigg[\frac{3}{2}\ud\ln \left(l_2\right)\bigg]+ \mathbf{m}_{16} \bigg[\frac{4}{3}\ud\ln \left(l_2\right)\bigg]+ \mathbf{m}_{17}\bigg[\frac{11}{3} \ud\ln\left(l_2\right)\bigg]+\mathbf{m}_{20} \bigg[-\ud\ln \left(l_2\right)\bigg]
   \nonumber \\
   &+\mathbf{m}_{21} \bigg[\ud\ln \left(l_2\right)\bigg]+\mathbf{m}_{22} \bigg[-2\,\ud\ln\left(l_2\right)\bigg]+\mathbf{m}_{25} \bigg[2\, \ud\ln \left(l_2\right)\bigg]+\mathbf{m}_{26} \bigg[-2\,\ud\ln \left(l_2\right)\bigg]
   \nonumber \\
   &+\mathbf{m}_{27} \bigg[2\,\ud\ln
   \left(l_6\right)\bigg]+\mathbf{m}_{28} \bigg[-3\, \ud\ln \left(l_2\right)\bigg]+\mathbf{m}_{29} \bigg[-\ud\ln \left(l_2\right)\bigg]
   \nonumber \\
   &+\mathbf{m}_{30} \bigg[-2\, \ud\ln\left(l_2\right)-2\, \ud\ln \left(l_7\right)+\ud\ln \left(l_9\right)+\ud\ln \left(l_{10}\right)\bigg]+\mathbf{m}_{32}
   \bigg[4 \,\ud\ln \left(l_1\right)+2\, \ud\ln \left(l_2\right)
   \nonumber \\
   &+4\, \ud\ln \left(l_7\right)+4\, \ud\ln \left(l_9\right)-4\,
   \ud\ln \left(l_{13}\right)-4\, \ud\ln \left(l_{14}\right)+4\, \ud\ln \left(l_{15}\right)-4\, \ud\ln
   \left(l_{16}\right)\bigg]
   \nonumber \\
   &+\mathbf{m}_{33} \bigg[2\,\ud\ln \left(l_6\right)\bigg]
   +\mathbf{m}_{35} \bigg[2 \,\ud\ln \left(l_1\right)-2\, \ud\ln
   \left(l_2\right)-2\, \ud\ln \left(l_3\right)-2\, \ud\ln \left(l_6\right)+2\, \ud\ln \left(l_7\right)
   \nonumber \\
   &+\ud\ln\left(l_9\right)+\ud\ln \left(l_{10}\right)\bigg]+\mathbf{m}_{36} \bigg[-4\,\ud\ln \left(l_2\right)\bigg],
   \nonumber \\
   \epsilon^{-1} &\ud \mathbf{m}_{36} = \mathbf{m}_1\bigg[\frac{5}{4} \ud\ln \left(l_1\right)-\frac{3}{8} \ud\ln \left(l_2\right)+\frac{1}{2}\ud\ln
   \left(l_3\right)+\frac{1}{4}\ud\ln \left(l_6\right)-\frac{1}{4}\ud\ln \left(l_9\right)-\frac{1}{4}\ud\ln\left(l_{10}\right)\bigg]
   \nonumber \\
   &+\mathbf{m}_2\bigg[\ud\ln \left(l_1\right)+\frac{1}{2}\ud\ln \left(l_2\right)+2\, \ud\ln
   \left(l_3\right)+\ud\ln \left(l_6\right)-\ud\ln \left(l_9\right)-\ud\ln \left(l_{10}\right)\bigg]
   \nonumber \\
   &+\mathbf{m}_3\bigg[-\frac{17}{12} \ud\ln \left(l_1\right)-\frac{1}{24}\ud\ln \left(l_2\right)-\frac{3}{2} \ud\ln\left(l_3\right)-\frac{3}{4} \ud\ln \left(l_6\right)+\frac{3}{4} \ud\ln \left(l_9\right)+\frac{3}{4} \ud\ln
   \left(l_{10}\right)\bigg]
   \nonumber \\
   &+\mathbf{m}_4\bigg[-\frac{13}{2} \ud\ln \left(l_1\right)-\frac{1}{4}\ud\ln\left(l_2\right)-7\, \ud\ln \left(l_3\right)-\frac{7}{2} \ud\ln \left(l_6\right)+\frac{7}{2} \ud\ln\left(l_9\right)+\frac{7}{2} \ud\ln \left(l_{10}\right)\bigg]
   \nonumber \\
   &+\mathbf{m}_5\bigg[\frac{17}{12} \ud\ln
   \left(l_1\right)+\frac{1}{24}\ud\ln \left(l_2\right)+\frac{3}{2} \ud\ln \left(l_3\right)+\frac{3}{4} \ud\ln
   \left(l_6\right)-\frac{3}{4} \ud\ln \left(l_9\right)-\frac{3}{4} \ud\ln \left(l_{10}\right)\bigg]
   \nonumber \\
   &+\mathbf{m}_6\bigg[-\ud\ln \left(l_1\right)-\frac{1}{2}\ud\ln \left(l_2\right)-2\, \ud\ln \left(l_3\right)-\ud\ln
   \left(l_6\right)+\ud\ln \left(l_9\right)+\ud\ln \left(l_{10}\right)\bigg]
   \nonumber \\
   &+\mathbf{m}_8\bigg[6\, \ud\ln
   \left(l_1\right)-3 \,\ud\ln \left(l_2\right)\bigg] +\mathbf{m}_9\bigg[-\ud\ln \left(l_2\right)\bigg] +\mathbf{m}_{10}\bigg[-\ud\ln \left(l_2\right)\bigg]
   \nonumber \\
   &+\mathbf{m}_{13}\bigg[-\frac{1}{3} \ud\ln \left(l_1\right)-\frac{1}{6}\ud\ln \left(l_2\right)-\frac{2}{3} \ud\ln
   \left(l_3\right)-\frac{1}{3}\ud\ln \left(l_6\right)+\frac{1}{3}\ud\ln \left(l_9\right)+\frac{1}{3}\ud\ln\left(l_{10}\right)\bigg]
   \nonumber \\
   &+\mathbf{m}_{15}\bigg[-\frac{5}{2} \ud\ln \left(l_1\right)+\frac{3}{4} \ud\ln
   \left(l_2\right)-\ud\ln \left(l_3\right)-\frac{1}{2}\ud\ln \left(l_6\right)+\frac{1}{2}\ud\ln\left(l_9\right)+\frac{1}{2}\ud\ln \left(l_{10}\right)\bigg]
   \nonumber \\
   &+\mathbf{m}_{16}\bigg[\frac{2}{3} \ud\ln
   \left(l_1\right)-\frac{1}{3}\ud\ln \left(l_2\right)\bigg]+\mathbf{m}_{17}\bigg[-\frac{5}{3} \ud\ln
   \left(l_1\right)-\frac{1}{6}\ud\ln \left(l_2\right)-2\, \ud\ln \left(l_3\right)-\ud\ln \left(l_6\right)
   \nonumber \\
   &+\ud\ln\left(l_9\right)+\ud\ln \left(l_{10}\right)\bigg]+\mathbf{m}_{20}\bigg[7\, \ud\ln \left(l_1\right)-\frac{1}{2}\ud\ln
   \left(l_2\right)+6 \,\ud\ln \left(l_3\right)+3\, \ud\ln \left(l_6\right)-3\, \ud\ln \left(l_9\right)
   \nonumber \\
   &-3\, \ud\ln
   \left(l_{10}\right)\bigg]+\mathbf{m}_{21}\bigg[\ud\ln \left(l_1\right)+\frac{1}{2}\ud\ln \left(l_2\right)+2\, \ud\ln
   \left(l_3\right)+\ud\ln \left(l_6\right)-\ud\ln \left(l_9\right)-\ud\ln \left(l_{10}\right)\bigg]
   \nonumber \\
   &+\mathbf{m}_{22}\bigg[2\, \ud\ln \left(l_1\right)+2\, \ud\ln \left(l_3\right)+\ud\ln \left(l_6\right)-\ud\ln
   \left(l_9\right)-\ud\ln \left(l_{10}\right)\bigg]+\mathbf{m}_{23}\bigg[-2 \,\ud\ln \left(l_1\right)
   \nonumber \\
   &+\ud\ln\left(l_2\right)-2\, \ud\ln \left(l_3\right)-\ud\ln \left(l_6\right)+\ud\ln \left(l_7\right)+\frac{1}{2}\ud\ln
   \left(l_9\right)+\frac{1}{2}\ud\ln \left(l_{10}\right)\bigg] 
   \nonumber \\
   &+\mathbf{m}_{24}\bigg[-2\,\ud\ln \left(l_1\right)+\ud\ln \left(l_2\right)\bigg]+\mathbf{m}_{25}\bigg[-2 \ud\ln \left(l_1\right)+\ud\ln \left(l_2\right)\bigg]+\mathbf{m}_{26}\bigg[4 \,\ud\ln \left(l_1\right)
   \nonumber \\
   &-\ud\ln \left(l_2\right)+2\, \ud\ln \left(l_3\right)+\ud\ln
   \left(l_6\right)-\ud\ln \left(l_9\right)-\ud\ln \left(l_{10}\right)\bigg] +\mathbf{m}_{27}\bigg[-\ud\ln \left(l_9\right)+\ud\ln
   \left(l_{10}\right)\bigg]
   \nonumber \\
   &+\mathbf{m}_{28}\bigg[\ud\ln \left(l_1\right)-\frac{3}{2}\ud\ln
   \left(l_2\right)-2\, \ud\ln \left(l_3\right)-\ud\ln \left(l_6\right)+\ud\ln \left(l_9\right)+\ud\ln
   \left(l_{10}\right)\bigg]
   \nonumber \\
   &+\mathbf{m}_{29}\bigg[-\ud\ln \left(l_1\right)-\frac{1}{2}\ud\ln \left(l_2\right)-2 \,\ud\ln
   \left(l_3\right)-\ud\ln \left(l_6\right)+\ud\ln \left(l_9\right)+\ud\ln \left(l_{10}\right)\bigg]
   \nonumber \\
   &+\mathbf{m}_{30}\bigg[2\,
   \ud\ln \left(l_2\right) \bigg]+\mathbf{m}_{32}\bigg[-3\, \ud\ln \left(l_2\right)-2\, \ud\ln \left(l_6\right)-2 \,\ud\ln
   \left(l_7\right)-2\, \ud\ln \left(l_{10}\right)-2\, \ud\ln \left(l_{13}\right)
   \nonumber \\
   &+2\, \ud\ln \left(l_{14}\right)+2\, \ud\ln
   \left(l_{15}\right)+2\, \ud\ln \left(l_{16}\right)\bigg]+\mathbf{m}_{33}\bigg[-\ud\ln\left(l_9\right)+\ud\ln \left(l_{10}\right)\bigg]
   \nonumber \\
   &+\mathbf{m}_{35}\bigg[-2\, \ud\ln \left(l_2\right)\bigg]+\mathbf{m}_{36}\bigg[4\, \ud\ln \left(l_1\right)+2\, \ud\ln
   \left(l_7\right)-\ud\ln \left(l_9\right)-\ud\ln \left(l_{10}\right)\bigg].
\end{align}
\vspace*{2ex}

\section{A symbol alphabet with several independent roots}
\label{sec:multiroot}

\begin{figure}[b!]
    \centering
    \includegraphics[width=0.25\textwidth]{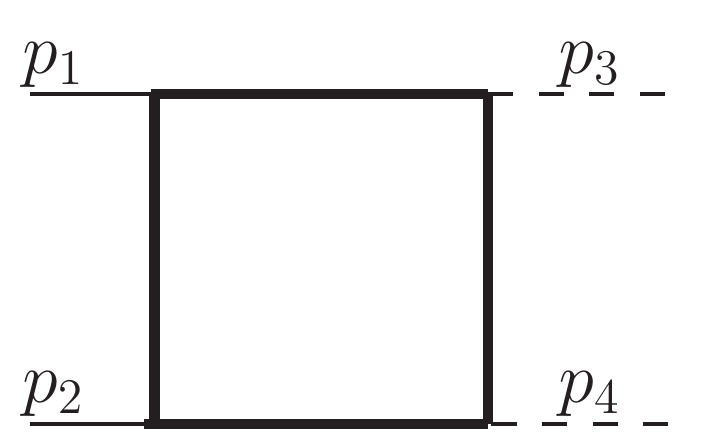}
    \caption{Feynman integral with symbol alphabet depending on several roots.}
    \label{fig:hhdiag}
\end{figure}

In this appendix, we consider the symbol alphabet of the one-loop
box integral shown in Figure\ \ref{fig:hhdiag}.
This integral occurs in processes like $gg\to hh$ or $gg \to ZZ$
with a top quark running in the loop.

We use the kinematical definitions
\begin{align}
   s &= (p_1 + p_2)^2, &
   t &= (p_1 - p_3)^2, &
   p_1^2 = p_2^2 &= 0, &
   p_3^2 = p_4^2 &= m_h^2,
\end{align}
and $m_t^2$ for the internal mass squared.

\newpage
It is straightforward to write down a normal form basis,
\begin{align}
    \mathbf{n}_1 &= \epsilon ~~
    \includegraphics[valign=m,raise=.2cm,height=.07\linewidth,width=.07\linewidth,keepaspectratio]{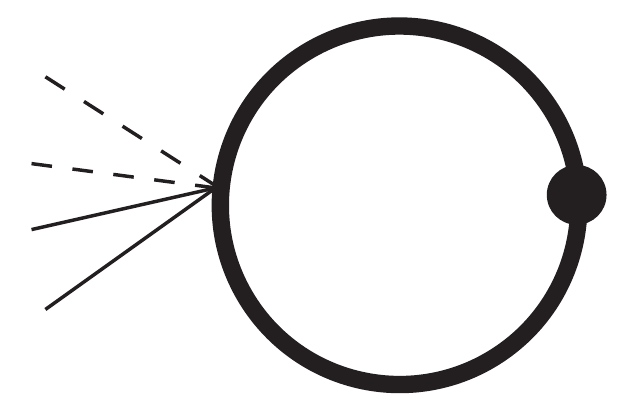} \\
    \mathbf{n}_2 &= \sqrt{s(s - 4\, m_t^2)}~ \epsilon ~~
    \includegraphics[valign=m,raise=.4cm,height=.08\linewidth,width=.08\linewidth,keepaspectratio]{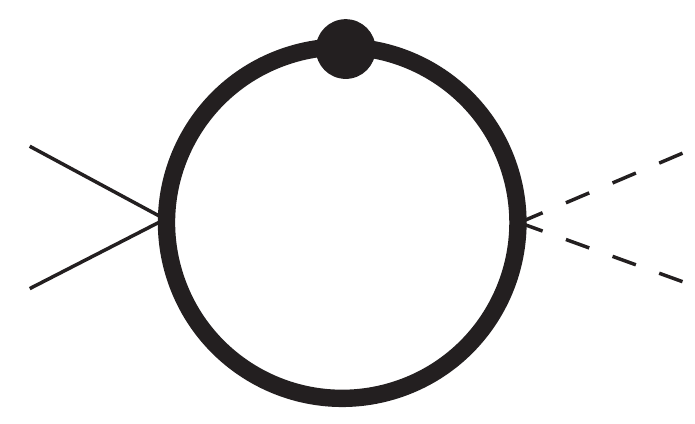}
    \\
    \mathbf{n}_3 &= \sqrt{m_h^2(m_h^2 - 4\, m_t^2)}~ \epsilon ~~
    \includegraphics[valign=m,raise=.4cm,height=.08\linewidth,width=.08\linewidth,keepaspectratio]{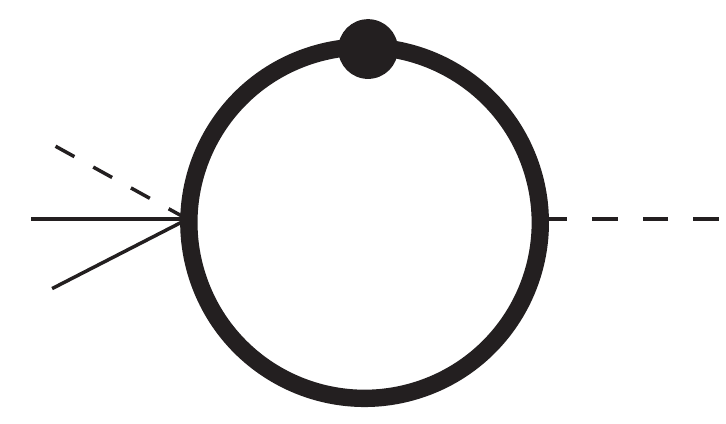}
    \\
    \mathbf{n}_4 &= \sqrt{t(t - 4\, m_t^2)}~ \epsilon ~~
    \includegraphics[valign=m,raise=.4cm,height=.08\linewidth,width=.08\linewidth,keepaspectratio]{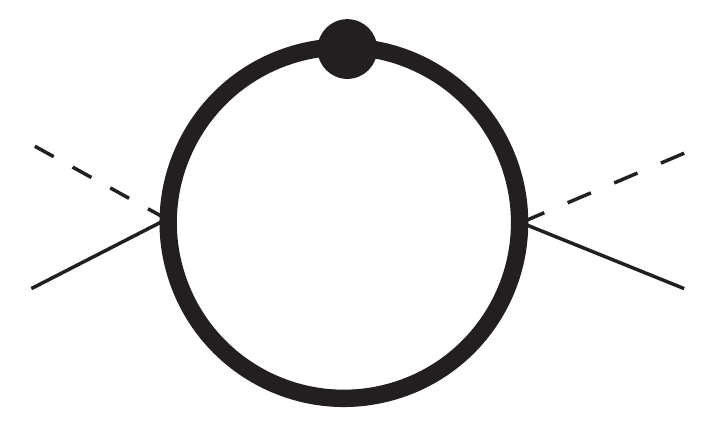}
    \\
    \mathbf{n}_5 &= s\, \epsilon^2 ~~
    \includegraphics[valign=m,raise=.3cm,height=.09\linewidth,width=.09\linewidth,keepaspectratio]{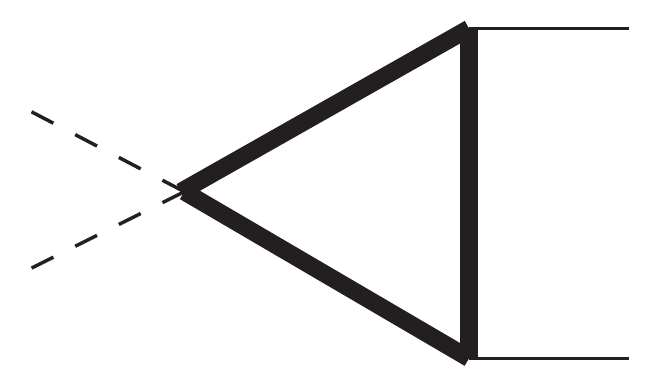}
    \\
    \mathbf{n}_6 &= (m_h^2-t) \epsilon^2 ~~
    \includegraphics[valign=m,raise=.3cm,height=.09\linewidth,width=.09\linewidth,keepaspectratio]{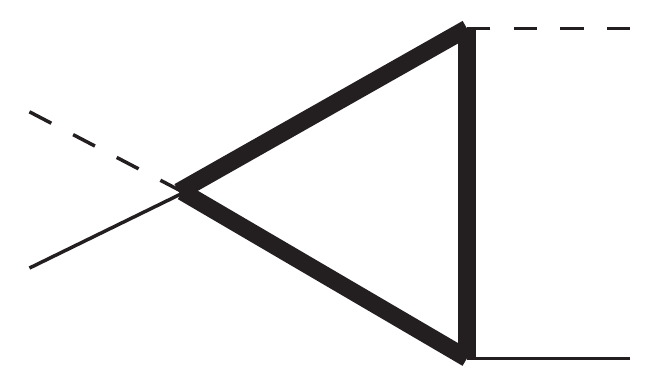}
    \\
    \mathbf{n}_7 &= \sqrt{s(s - 4\, m_h^2)}~ \epsilon^2 ~~
    \includegraphics[valign=m,raise=.4cm,height=.09\linewidth,width=.09\linewidth,keepaspectratio]{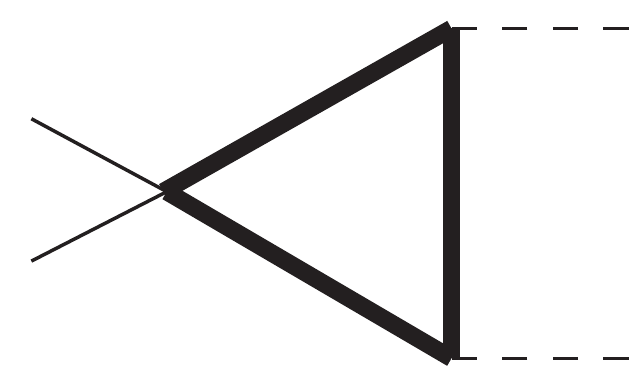}
    \\
    \mathbf{n}_8 &= \sqrt{s (s\, t (t-4\, m_t^2) - 4\, m_t^2 (m_h^2-t)^2)}~ \epsilon^2
    ~
    \includegraphics[valign=m,raise=.4cm,height=.08\linewidth,width=.08\linewidth,keepaspectratio]{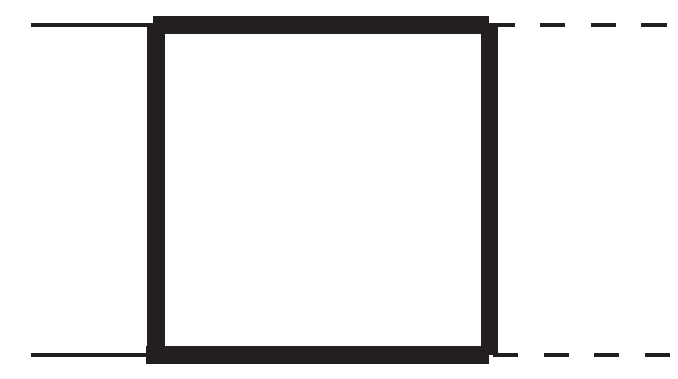}
\end{align}
which fulfils a differential equation of the form
\begin{equation} \label{eq:dlogformhh}
\ud\mathbf{n}_i = \epsilon\; \sum_{j,k} \ud \ln(l_k) \big(A^{(k)}\big)_{ij} \, \mathbf{n}_j\,,
\end{equation}
with $i,j = 1,\ldots,8$.
It is not straightforward, however, to actually arrive at
\eqref{eq:dlogformhh} with a simple form for the letters $l_k$
due to the presence of the five different square roots.
Moreover, while it is is not difficult to choose a set of independent letters, it is not obvious how to minimize the number of independent letters.

While one could try to rationalize some of the roots with a suitable reparametrization, we want to take a different approach here and just directly work with the original kinematic invariants.
Our primary motivation is to extend the construction
of symbol letters presented in Section~\ref{sec:diffeqint} to the case
of several root-valued leading singularities.

Employing the integration measure
\begin{equation}
    \frac{\Gamma(1-\epsilon) m_t^{2\epsilon}}{i \pi^{2-\epsilon}},
\end{equation}
the integrals $m_i$ depend only on the dimensionless ratios of scales
\begin{align}
w_1 &= s/m_t^2, \\
w_2 &= t/m_t^2, \\
w_3 &= m_h^2/m_t^2\,.
\end{align}
In a conventional approach, one might try to directly integrate the terms in the partial differential equations with {\it e.g.}\  {\tt Mathematica} to arrive at the form \eqref{eq:dlogformhh}.
In this approach, we obtained 24 independent letters, many of which are very complicated and involve high powers of the variables.

Instead, the following construction leads to much simpler
and fewer independent letters. Starting from the rational part:
\begin{align}
    \left\{w_1,w_2,w_3,w_3-w_2,w_1+w_3^2-4\, w_3,w_1 w_2+w_2^2-2\, w_2 w_3+w_3^2\right\}
\end{align}
and the five square roots
\begin{align}
    &\left\{\sqrt{(w_1-4) w_1},\sqrt{(w_3-4) w_3},\sqrt{(w_2-4) w_2},\sqrt{w_1 (w_1-4\, w_3)},\nonumber\right.\\
    &\left.\sqrt{w_1 \left(w_1 w_2^2-4\,
   w_1 w_2-4\, w_2^2+8\, w_2 w_3-4\, w_3^2\right)}\right\},
\end{align}
we used the algorithm described in Section \ref{sec:diffeqint} for each of these square roots independently. Next, we formed pairs and triples of the roots and used the algorithm on their products. 
We find a total of 22 independent candidates for letters in this way,
which are much simpler than the original set of letters and involve
only low powers of the variables:
\begin{align}
\tilde{\mathcal{L}} &= \{\tilde{l}_1\,\ldots,\tilde{l}_{22}\} \nonumber\\
&=
\big\{w_1, w_2, w_3, w_2 - w_3, w_1 - 4\, w_3 + w_3^2, w_1 w_2 + w_2^2 - 2\, w_2 w_3 + w_3^2, \sqrt{(-4 + w_1)w_1}, 
\nonumber\\
&\quad
 \sqrt{(-4 + w_2)w_2}, \sqrt{w_3(-4 + w_3)}, \sqrt{w_1(w_1 - 4 w_3)}, (\sqrt{w_1(-4 + w_1)} + w_1)/2,
\nonumber\\
&\quad
  (\sqrt{w_2(-4 + w_2)} + w_2)/2,(\sqrt{w_3(-4 + w_3)} + w_3)/2,
\nonumber\\
&\quad
 \sqrt{w_1(w_1 w_2 ( w_2-4)  - 4 (w_2 - w_3)^2)},(\sqrt{w_1(w_1 - 4 w_3)} \sqrt{w_3(-4 + w_3)} + w_1 w_3)/2,
 \nonumber\\
&\quad
 (w_1^2 + \sqrt{w_1(-4 + w_1)} \sqrt{w_1(w_1 - 4 w_3)} - 2 w_3 w_1)/2,
 \nonumber\\
&\quad
\Big(w_1 w_2 + \sqrt{w_1(w_1 w_2 ( w_2-4)  - 4 (w_2 - w_3)^2)}\Big)/2, 
 \nonumber\\
&\quad
 (w_1^2 w_2 + \sqrt{w_1(w_1 w_2 ( w_2-4)  - 4 (w_2 - w_3)^2)} \sqrt{w_1(w_1 - 4 w_3)} - 2 w_2 w_3 w_1 + 2 w_3^2 w_1)/2, 
 \nonumber\\
&\quad
\Big(-2 w_1^2 - 4 w_2w_1 + w_1^2 w_2 + 4 w_3 w_1
 \nonumber\\
&\quad\quad
 + \sqrt{w_1(-4 + w_1)} \sqrt{w_1(w_1 w_2 ( w_2-4)  - 4 (w_2 - w_3)^2)}\,\Big)/2,
 \nonumber\\
&\quad
\Big(-2 w_1 w_2  - 2 w_1 w_3 + w_1 w_2 w_3 
 \nonumber\\
&\quad\quad
+ \sqrt{w_3(-4 + w_3)}\sqrt{w_1(w_1 w_2 ( w_2-4)  - 4 (w_2 - w_3)^2)}  
 \,\Big)/2,
 \nonumber\\
&\quad
\Big(-4 w_1 w_2 - 2 w_2^2 + w_1 w_2^2  + 4 w_2 w_3 - 2 w_3^2
 \nonumber\\
&\quad\quad
+  \sqrt{w_2(-4 + w_2)}
\sqrt{w_1(w_1 w_2 ( w_2-4)  - 4 (w_2 - w_3)^2)}\,\Big)/2
\big\}.
\end{align}
The factor $1/2$ appears for the same reason as in Section \ref{sec:diffeqint} to prevent any factor of $2$ in the factorization of the product of a letter with its conjugate over the rational part.
All of these letters are power products of the 24 letters in the differential equation obtained in the first attempt.

To go to even lower degrees, one can do a further factorization of the letters by factorizing the square roots further, i.e. by splitting them in a similar way to:
\begin{equation}
    \sqrt{(w_1-4)w_1}=\sqrt{w_1}\sqrt{w_1-4},
\end{equation}
and dividing them out in all letters, if possible.
Note, that this leads to a representation with a possible sum of different square roots in each letter. If one would like to derive this representation along the lines described in Section \ref{sec:diffeqint}, one would need to allow for more general structures than
\begin{equation}
    l_a=q_a+r
\end{equation}
with rational $q_a$.

Interestingly, in our new representation even more letters drop out of the differential equation, and one finds only $19$ letters in the end:
\begin{align}
\mathcal{L} &= \{l_1\,\ldots,l_{19}\} \nonumber\\
=&\Big\{\frac{1}{2}\Big(\sqrt{w_1-4}+\sqrt{w_1}\Big),\sqrt{w_1-4},\sqrt{w_1},\frac{1}{2}\Big(\sqrt{w_2-4}+\sqrt{w_2}\Big),\sqrt{w_2-4},\nonumber\\
&\frac{1}{2}\Big(\sqrt{w_3-4}+\sqrt{w_3}\Big),\frac{1}{2}\Big(\sqrt{w_3-4} \sqrt{w_1-4\, w_3}+\sqrt{w_1}\sqrt{w_3}\Big),\sqrt{w_1-4\, w_3},\nonumber\\
&\frac{1}{2}(\sqrt{w_1-4} \sqrt{w_1-4\, w_3}+w_1-2\,w_3),w_2-w_3,\sqrt{w_3-4},\nonumber\\
&\sqrt{w_1 w_2 ( w_2-4)  - 4 (w_2 - w_3)^2},w_1+w_3^2-4\, w_3,w_1 w_2+w_2^2-2\,w_2 w_3+w_3^2,\nonumber\\
&\frac{1}{2}\Big(\sqrt{w_1 w_2 ( w_2-4)  - 4 (w_2 - w_3)^2}+\sqrt{w_1} \sqrt{w_2-4}\sqrt{w_2}\Big),\nonumber\\
&\frac{1}{2}(\sqrt{w_1 w_2 ( w_2-4)  - 4 (w_2 - w_3)^2}+\sqrt{w_1} w_2),\nonumber\\
&\frac{1}{2}(\sqrt{w_1-4}\sqrt{w_1 w_2 ( w_2-4)  - 4 (w_2 - w_3)^2}+w_1 w_2-2\, w_1-4\, w_2+4\,
w_3),\nonumber\\
&\frac{1}{2}(\sqrt{w_1-4\, w_3} \sqrt{w_1 w_2 ( w_2-4)  - 4 (w_2 - w_3)^2}+w_1 w_2-2\,w_2 w_3+2\, w_3^2),\nonumber\\
&\frac{1}{2}(\sqrt{w_3-4} \sqrt{w_3} \sqrt{w_1 w_2 ( w_2-4)  - 4 (w_2 - w_3)^2}+\sqrt{w_1} w_2 w_3-2 \sqrt{w_1} w_2-2
\sqrt{w_1} w_3)\Big\},
\end{align}
with the differential equation given by:
\begin{align}
\epsilon^{-1}\ud \mathbf{n}_1 &= 0, \nonumber\\
\epsilon^{-1}\ud \mathbf{n}_2 &= \mathbf{n}_1~\ud\ln\left({l_{1}^2}\right) + 
  \mathbf{n}_2~ \ud\ln\left(1/{l_{2}^2}\right), \nonumber\\
\epsilon^{-1}\ud \mathbf{n}_3 &=
 \mathbf{n}_1~ \ud\ln\left({l_{6}^2}\right) + 
  \mathbf{n}_3~ \ud\ln\left(1/{l_{11}^2}\right), \nonumber\\
\epsilon^{-1}\ud \mathbf{n}_4 &=
 \mathbf{n}_1~ \ud\ln\left({l_{4}^2}\right) + 
  \mathbf{n}_4 ~\ud\ln\left(1/{l_{5}^2}\right), \nonumber\\
\epsilon^{-1}\ud \mathbf{n}_5 &=
 \mathbf{n}_2~ \ud\ln\left({l_{1}^2}\right), \nonumber\\
\epsilon^{-1}\ud \mathbf{n}_6 &=
 \mathbf{n}_3~ \ud\ln\left({l_{6}^2}\right)
 + \mathbf{n}_4~ \ud\ln\left({1}/{l_{4}^2}\right), \nonumber\\
\epsilon^{-1}\ud \mathbf{n}_7 &=
 \mathbf{n}_2~ \ud\ln\left({l_{9}^2}/{l_{13}}\right) + 
  \mathbf{n}_3~ \ud\ln\left({l_{13}^2 / l_{7}^4}\right) \
+ \mathbf{n}_7~ \ud\ln\left({l_{8}^2}/{l_{13}}\right), \nonumber\\
\epsilon^{-1}\ud \mathbf{n}_8 &=
  \mathbf{n}_2~ \ud\ln\left({l_{17}^2 l_{3}^2}/{l_{13}}\right)
  + \mathbf{n}_3~ \ud\ln\left({l_{10}^4 l_{13}^2}/{l_{19}^4}\right)
  + \mathbf{n}_4~ \ud\ln\left({l_{15}^4}/{l_{10}^4}\right)
  + \mathbf{n}_5~ \ud\ln\left({l_{16}^2}/{l_{14}}\right) 
 \nonumber\\
 & 
 + \mathbf{n}_6~ \ud\ln\left({l_{14}^2}/{l_{16}^4}\right)
 + \mathbf{n}_7~ \ud\ln\left({l_{18}^2}/{(l_{13} l_{14})}\right)
 + \mathbf{n}_8~ \ud\ln\left({l_{14}}/{l_{12}^2}\right)\,.
\end{align}


\bibliographystyle{JHEP}
\bibliography{dymasters2m}

\end{document}